\documentclass[twocolumn,twocolappendix]{aastex631}
\usepackage{amsmath}

\newcommand\ppxf{\texttt{pPXF}}
\newcommand\Alf{\texttt{alf}}

\shorttitle{Stellar and gas-phase abundances in star-forming galaxies}
\shortauthors{Z. Zhuang et al.}

\graphicspath{{./}{figures/}}

\begin{document}

\title{Metals in Star-Forming Galaxies with KCWI. I. Methodology and First Results on the  Abundances of Iron, Magnesium, and Oxygen}

\correspondingauthor{Zhuyun Zhuang}
\email{zzhuang@astro.caltech.edu}

\author[0000-0002-1945-2299]{Zhuyun Zhuang}
\affiliation{Department of Astronomy, California Institute of Technology, 1200 E. California Blvd., MC 249-17, Pasadena, CA 91125, USA}

\author[0000-0001-6196-5162]{Evan N.\ Kirby}
\affiliation{Department of Physics and Astronomy, University of Notre Dame, 225 Nieuwland Science Hall, Notre Dame, IN 46556, USA}

\author[0000-0002-4834-7260]{Charles C.\ Steidel}
\affiliation{Department of Astronomy, California Institute of Technology, 1200 E. California Blvd., MC 249-17, Pasadena, CA 91125, USA}

\author[0000-0002-4739-046X]{Mithi A.C. de los Reyes}
\affiliation{Department of Physics \& Astronomy, Amherst College, 25 East Drive, Amherst, MA, 01002, USA}

\author[0000-0001-5847-7934]{Nikolaus Z.\ Prusinski}
\affiliation{Department of Astronomy, California Institute of Technology, 1200 E. California Blvd., MC 249-17, Pasadena, CA 91125, USA}

\author[0000-0003-4570-3159]{N.\ Leethochawalit}
\affiliation{National Astronomical Research Institute of Thailand (NARIT), Mae Rim, Chiang Mai, 50180, Thailand}

\author[0000-0002-8435-9402]{Minjung Park}
\affiliation{Center for Astrophysics | Harvard \& Smithsonian, Cambridge, MA, 02138, USA}

\author[0000-0002-1590-8551]{Charlie Conroy}
\affiliation{Center for Astrophysics | Harvard \& Smithsonian, Cambridge, MA, 02138, USA}

\author[0000-0001-5595-757X]{Evan H. Nu\~{n}ez}
\affiliation{Department of Astronomy, California Institute of Technology, 1200 E. California Blvd., MC 249-17, Pasadena, CA 91125, USA}


\begin{abstract}

Understanding the chemical enrichment of different elements is crucial to gaining a complete picture of galaxy chemical evolution. In this study, we present a new sample of 46 low-redshift, low-mass star-forming galaxies at $M_*\sim 10^{8-10}M_{\odot}$ along with two quiescent galaxies at $M_*\sim 10^{8.8}M_{\odot}$ observed with the Keck Cosmic Web Imager (KCWI), aiming to investigate the chemical evolution of galaxies in the transition zone between Local Group satellites and massive field galaxies. We develop a novel method to simultaneously determine stellar abundances of iron and magnesium in star-forming galaxies. With the gas-phase oxygen abundance (O/H)$_{\rm g}$ measured using the strong line method, we are able to make the first-ever apples-to-apples comparison of $\alpha$ elements in the stars and the ISM. We find that the [Mg/H]$_*$--[O/H]$_{\rm g}$ relation is much tighter than the [Fe/H]$_*$--[O/H]$_{\rm g}$ relation, which can be explained by the similar production processes of $\alpha$ elements. Most galaxies in our sample exhibit higher [O/H]$_{\rm g}$ than [Fe/H]$_*$ and [Mg/H]$_*$\@. In addition, we construct mass--metallicity relations (MZRs) measured as three different elements (Fe$_*$, Mg$_*$, O$_{\rm g}$). Compared to the gas O-MZR, the stellar Fe- and Mg-MZRs show larger scatter driven by variations in specific star formation rates (sSFR), with star-forming galaxies exhibiting higher sSFR and lower stellar abundances at fixed mass. The excess of [O/H]$_{\rm g}$ compared to stellar abundances as well as the anti-correlation between sSFR and stellar abundance suggests that galaxy quenching of intermediate-mass galaxies at $M_*\sim 10^{8-10}M_{\odot}$ is primarily driven by starvation.

\end{abstract}


\section{Introduction} \label{sec:intro}
Heavy elements are invaluable tracers of the evolutionary history of galaxies. The amount of metals (i.e., metallicity) residing in the interstellar medium (ISM) and stellar population of a galaxy is governed by the interplay between the gravitational potential well \cite[e.g.,][]{Dekel1986}, nucleosynthesis during star formation \citep[e.g.,][]{Kobayashi20}, efficiency of gas mixing in the ISM \citep[e.g.,][]{Veilleux05}, galactic outflows driven by stellar and AGN feedback \citep[e.g.,][]{Murray05, Faucher-Giguere13, Hopkins14}, and pristine inflows from the circumgalactic and intergalactic medium (CGM and IGM) over cosmic time \citep[e.g.,][]{Keres05}. 

Over the past five decades, significant progress has been made in measuring metallicities in galaxies from the Local Group to the high-redshift universe. Using emission line diagnostics that trace gas metallicity of galaxies, \citet{Lequeux79} first discovered a correlation between stellar masses and gas metallicities of the ISM in star-forming galaxies, which is known as the mass--metallicity relation (MZR)\@. Later, \citet{Tremonti04} measured the gas metallicities of $\sim 53,000$ star-forming galaxies in the Sloan Digital Sky Survey (SDSS) from emission lines and confirmed that the gas metallicity is strongly correlated with stellar mass in all SDSS star-forming galaxies. \citet{Gallazzi05} also discovered a similar correlation between stellar mass and stellar metallicity measured from stellar absorption features for a sample of 44,254 galaxies drawn from the SDSS survey. The trend was later confirmed by other works using various methods to measure metallicities in the SDSS galaxies \citep[e.g.,][]{Andrews13, Zahid17}. Previous studies have also found that both the stellar MZR and the gas MZRs exist in the Local Group satellite galaxies \citep[e.g.][]{Lee2006, Berg12, Kirby13}. More recently, astronomers have shown that the MZR exists at all different redshifts, establishing the stellar MZR of quiescent galaxies up to $z\sim 2$ \citep[][]{Choi14, Kriek19, Leethochawalit19, Zhuang23, Beverage23b} and the gas MZR of star-forming galaxies up to $z\sim 10$ \citep[e.g.,][]{Erb06,Steidel14,Maiolino19,Sanders20, Strom22,Curti23}. All these studies suggest that at fixed stellar mass, high-$z$ galaxies tend to have lower metallicities than their local counterparts. These trends are also captured by large cosmological simulations \citep[e.g.,][]{Ma16, DeRossi17, Torrey19}. 


Generally, measurements of the MZR are based on the abundance of a single element: iron abundance [Fe/H]$_*$ for stars and oxygen abundance [O/H]$_{\rm g}$ for ionized gas in the ISM\@. The gas-phase MZR indicates the chemical enrichment in the current ISM as galaxies grow in mass, while the stellar MZR represents the metals locked in stars averaged over the entire star formation history (SFH)\@. While most work to date has focused on measuring either the stellar or the gas-phase MZR, usually for passively evolving \citep[e.g.,][]{Leethochawalit19} or star-forming galaxies \citep[e.g.,][]{Curti20} respectively, recent work has attempted to relate the gas metallicity measured from nebular emission lines to the stellar metallicity determined from the stellar continuum for nearby \citep[e.g.,][from the rest-optical continuum]{Lian18, Fraser-McKelvie22} and high-$z$ star-forming galaxies \citep[e.g.,][from the rest-FUV continuum to which young OB stars primarily contribute the observed features]{Steidel16, Topping20, Strom22}. Such work has found that the two abundance measurements do not yield the same ``metallicity'' for galaxies with significant ongoing star formation, where gas metallicity measured as oxygen abundance is significantly higher than stellar metallicity that uses iron abundance as a proxy. Moreover, these comparison studies have shown that gas-phase and stellar metallicities scale with stellar masses differently, resulting in different shapes (slopes and normalizations) of the gas-phase and stellar MZRs in the nearby and high-$z$ universe \citep[e.g.,][]{Fraser-McKelvie22, Strom22}.

The mismatch between the stellar and gas-phase metallicity is likely to result from the different recycling times of iron and oxygen. Oxygen is an $\alpha$ element produced primarily by core collapse supernovae (CCSNe) on characteristic timescales of $< 10$ Myr, while the production of Fe is dominated by Type Ia supernovae (SNe) with characteristic delay times $\gtrsim 400$ Myr \citep{Burbidge57,Tinsley80,Chen21}. If a galaxy has formed a substantial fraction of its stars in the relatively recent past, the abundance of $\alpha$ elements relative to iron, [$\alpha$/Fe], will be much higher than the solar ratio. In addition, if [Fe/H]$_*$ is measured from the rest-frame optical instead of UV spectrum, the result reflects the chemical enrichment of relatively old, low-mass stars because the stellar photospheric features of young OB stars reside primarily in the rest-frame UV\@. The oxygen abundance measured from emission lines instead indicates the amount of metals that reside in the current ISM, which better represents the level of chemical enrichment of the youngest stars. The combined effects of the distinct recycling times of iron and oxygen and the stars at different ages to which stellar and gas-phase metallicity are sensitive make it less straightforward to interpret the relation between metallicities in different phases. 

A better way to approach this problem would be to relate two elements for which the abundance ratio is less sensitive to the short-timescale details of the SFH\@.
Magnesium in the stars is useful here. Because oxygen and magnesium are both $\alpha$ elements, their abundance ratio should be much less affected by the varying delayed timescale of Type Ia SNe than [O/Fe] derived from the optical spectrum. Comparing [Mg/H]$_*$ of the stellar population with [O/H]$_{\rm g}$ can therefore inform us about the degree to which metal-poor gas accretion or metal-rich outflows have affected the overall galactic metallicity. Relating the abundances of gas-phase oxygen, stellar iron, and stellar magnesium allows us to correlate chemical enrichment at different timescales. In addition, [Mg/Fe]$_*$ provides unique insights on the chemical enrichment history of a galaxy. 

In this work, we present a novel technique to determine [Fe/H]$_*$ and [Mg/Fe]$_*$ of star-forming galaxies via full-spectrum fitting of the optical stellar continuum. We apply this technique to a new sample of low-redshift, low-mass star-forming galaxies between $10^8M_{\odot}$ and $10^{10}M_{\odot}$ observed with the Keck Cosmic Web Imager \citep[KCWI;][]{Morrissey18}, yielding the first-ever apples-to-apples comparison of $\alpha$ elements in different phases. 

The KCWI sample consists of star-forming galaxies in the transition zone between the Local Group dwarf satellite galaxies and massive field galaxies that have been well-studied by large spectroscopic surveys. The sample's purpose is to fully understand the shape of the stellar MZRs across the full range of galaxy masses. A simple extrapolation of the stellar MZR of satellite galaxies in the Local Group \citep[$M_* \lesssim 10^8 M_{\odot}$,][]{Kirby13} appears to disagree with measurements of more massive, quiescent galaxies in the field \citep[$M_* \gtrsim 10^{9.5} M_{\odot}$,][]{Leethochawalit19}. At $M_*\sim 10^9M_{\odot}$, the high-mass and low-mass MZRs are discrepant in [Fe/H]$_*$ by $\sim 0.6$~dex, much larger than the scatter of either MZR (see Figure~7 in \citealt{Zhuang2021}). In our previous work \citep{Zhuang2021}, we ruled out the possibility that the large discrepancy originates from systematic differences in the techniques used to estimate [Fe/H]$_*$ of galaxies at different masses, and we suggested that there is a transition mass in the stellar MZR in the local universe. Constraining the shape of the MZR, especially the stellar one, in the transition mass is essential to unveil how the physical processes regulating metal retention and interaction with the surrounding environment vary as galaxies grow in mass.

The main goals of this paper are (1) to compare stellar and gas-phase abundances of a new sample of low-mass star-forming galaxies observed with KCWI; (2) to construct the MZR of our sample measured from three different elements (O$_{\rm g}$, Fe$_*$ and Mg$_*$); and (3) to quantify the shape of the stellar MZR for galaxies beyond the Local Group in order to put constraints on the chemical evolution of galaxies. The structure of this paper is as follows. We describe the basic properties of the sample, the observations, and data reductions in Section~\ref{sec:data}. In Section~\ref{sec:emline_measurements}, we present the methods used for emission line measurements and gas-phase metallicity determination. In Section~\ref{sec:stellar_abund_measurement}, we explain the new technique used in this work to measure [Fe/H]$_*$ and [Mg/Fe]$_*$ of star-forming galaxies. We present our measurements from the KCWI data in Section~\ref{sec:results} and discuss their implications when combining the abundance measurements of more massive SDSS galaxies in Section~\ref{sec:discussion}. Finally, we summarize our findings in Section~\ref{sec:summary}. Throughout this work we assume a flat $\Lambda$CDM cosmology with $\Omega_m=0.3$, $\Omega_{\Lambda}=0.7$ and $H_0=70$~km s$^{-1}$~Mpc$^{-1}$.


\section{Data}\label{sec:data}
We use KCWI mounted at Nasmyth focus on the Keck~II telescope to obtain the spectroscopic data for our analysis. We describe the sample selection in Section~\ref{subsec:sample}, global properties of galaxies estimated from broadband photometry in Section~\ref{subsec:sed}, and summarize the KCWI observations and data reduction in Section~\ref{subsec:kcwi_obs}.

\subsection{Target Selection}\label{subsec:sample}
Our sample consists of 46 star-forming galaxies at $z < 0.06$ in a stellar mass range of $10^{8} - 10^{10} M_{\odot}$ (Table~\ref{tab:galaxy_prop}). We chose this mass range because it represents the transition from the Local Group dwarf galaxies to the well-studied massive galaxies in the Hubble flow. The weak stellar absorption features of heavily star-forming galaxies dominant in this mass range has prevented observers from obtaining reliable measurements of not only stellar metallicity but also [Mg/Fe]$_*$ of individual star-forming dwarfs from large spectroscopic optical surveys. Aiming to measure stellar and gas-phase abundances simultaneously of the same star-forming galaxies, we acquired very high-quality integral field spectroscopic data for this small sample to reveal their faint stellar features.

19 galaxies were selected from the clumpy galaxy catalog compiled by \citet{Mehta21} that made use of the classification provided by the citizen science-powered \textit{Galaxy Zoo} based on the SDSS Stripe82 images, while the other 27 galaxies were selected from the SDSS MPA-JHU catalog \citep{Brinchmann04, Tremonti04}. All these targets were randomly selected from the parent catalogs based on their stellar masses. We restricted our sample to galaxies with sSFR between $\rm 10^{-9}-10^{-11}$~yr$^{-1}$ based on the SDSS MPA-JHU catalog \citep[\textit{galSpecLine}, SDSS DR17;][]{Brinchmann04, Tremonti04}, a range that is representative of typical SDSS star-forming galaxies.

Additionally, two quiescent galaxies, 0241-0810 and 0125-0024 at $\log{(M_*/M_{\odot})} \sim 8.8$, were added to the sample as filler targets to investigate whether low-mass quiescent galaxies exhibit stellar abundances distinct from star-forming galaxies at similar masses. 0241-0810 is in the vicinity of the massive elliptical galaxy NGC~1052 with a similar radial velocity.  The dwarf galaxy likely belongs to a virialized group structure in the environment of NGC 1052 \citep{Roman21}. However, 0125-0024 is identified as an isolated dwarf galaxy at least 1~Mpc away from the nearest massive galaxy ($\log{(M_*/M_{\odot})} > 10$) by \citet{Kado-Fong20} from the SDSS\@.

\subsection{Photometry, Broadband SED Fitting and Size Measurements}\label{subsec:sed}
Although the SDSS catalog provides stellar masses and SFRs measured from SDSS $ugriz$ photometry, we performed our own spectral energy distribution (SED) fitting with BAGPIPES \citep{Carnall18, Carnall19} using photometry from far-UV to mid-IR wavelengths to obtain more reliable estimates of galaxies' global properties. 

We retrieved the publicly-available coadded science images from the \textit{Galaxy Evolution Explorer} \citep[\textit{GALEX};][]{Martin05}, the Sloan Digital Sky Survey \citep[SDSS;][]{York00}, the Panoramic Survey Telescope and Rapid Response System \citep[PS1;][]{Flewelling20}, and the Wide-field Infrared Survey Explorer \citep[WISE;][]{Wright10} through the unblurred coadds of the WISE imaging (unWISE) archive \citep{Lang14}. For each galaxy, we first calculated the flux densities ($f_{k}$) in each band $k$ with the same elliptical aperture defined as the \citet{Kron80} radius of the PS1-$i$ image. To account for the resolution mismatch in different filters, each PS1-$i$ detection image was downgraded to the resolutions in other photometric bands, from which we estimated the flux densities in the same aperture as $f'_{\rm ps1-i, k}$. The final photometry $f_{k, \rm final}$ in the band $k$ was calculated as:
\begin{equation}
    f_{k, \rm final} = \frac{f'_{\rm ps1-i, k}}{f_{\rm ps1-i}}f_{k}
\end{equation}
All the photometry was then corrected for Galactic reddening using the $E(B-V)$ values measured by \citet{Schlafly11}.

The SFRs and stellar masses were derived from BAGPIPES, a Bayesian-SED fitting code. We assumed a delayed exponentially declining (delayed-$\tau$) SFH for each galaxy. We fixed the redshifts at the values in the SDSS catalog and adopted the \citet{Calzetti2000} dust attenuation relation with a uniform prior of $0 < A_V < 2$. Nebular emission models based on the CLOUDY photoionization code \citep{Ferland17} with a varying ionization parameter $-4 < \log{U} < -2$ were included to account for the contribution from emission lines and nebular continuum. The other free parameters and their priors were stellar mass formed ($7 < \log{(M_*/M_{\odot})} < 12$), stellar metallicity ($-2 < \log{(Z/Z_{\odot})} < 0.4$), time since the onset of star formation ($1$~Myr $ < T_0 < 14$~Gyr), and the $e$-folding SFR timescale ($1~\rm Myr < \tau < 100~Gyr$). We assumed uniform priors on the logarithmic scale for all the free parameters. The best-fit SFRs and stellar masses were taken as the $50^{\rm th}$ percentile of the marginalized 1D posterior distribution. 
We experimented with four other SFHs: single exponentially declining, log-normal, double-power-law, and constant SFHs. The root-mean-square (RMS) errors of the best-fit values (i.e., the median of the posteriors) between different SFH models are quoted as the reported uncertainties. 

\startlongtable
\begin{deluxetable*}{lcccccr}
\tablecaption{General Properties of the KCWI Galaxy Sample\label{tab:galaxy_prop}}
\tablehead{\colhead{Galaxy} & \colhead{RA} & \colhead{Dec} & \colhead{$z^{a}$} & \colhead{$\rm \log{M_*}^{b}$} & \colhead{$\log{\rm SFR}^{b}$} & \colhead{$R_e$} \\ 
\colhead{} & \colhead{(J2000)} & \colhead{(J2000)} & \colhead{} & \colhead{[$\rm M_{\odot}$]} & \colhead{[$ \rm M_{\odot}\  yr^{-1}$]} & \colhead{(kpc)}}
\tablewidth{1.8\columnwidth} 
\startdata
2225$-$0046 & 22 25 07.92 & $-$00 46 06.03 & 0.016 & 8.09$\pm$0.10 & $-1.84\pm0.08$ & 0.76$\pm$0.01 \\
0231$-$0043 & 02 31 32.17 & $-$00 43 36.63 & 0.009 & 8.09$\pm$0.10 & $-1.67\pm0.08$ & 1.05$\pm$0.02 \\
0133$-$0109 & 01 33 41.39 & $-$01 09 29.76 & 0.016 & 8.11$\pm$0.14 & $-1.36\pm0.13$ & 1.07$\pm$0.02 \\
2329$+$1552 & 23 29 46.17 & $+$15 52 38.96 & 0.009 & 8.23$\pm$0.10 & $-1.67\pm0.08$ & 0.53$\pm$0.01 \\
2152$-$0110 & 21 52 22.46 & $-$01 10 15.95 & 0.016 & 8.40$\pm$0.10 & $-1.72\pm0.13$ & 0.97$\pm$0.01 \\
0845$+$5145 & 08 45 54.52 & $+$51 45 11.12 & 0.014 & 8.42$\pm$0.10 & $-1.80\pm0.14$ & 0.69$\pm$0.01 \\
2334$+$0029 & 23 34 14.80 & $+$00 29 07.28 & 0.024 & 8.42$\pm$0.10 & $-1.19\pm0.07$ & 1.31$\pm$0.03 \\
0121$-$0853 & 01 21 51.93 & $-$08 53 10.63 & 0.018 & 8.43$\pm$0.10 & $-1.55\pm0.10$ & 0.85$\pm$0.01 \\
0013$-$0956 & 00 13 45.42 & $-$09 56 03.67 & 0.013 & 8.51$\pm$0.10 & $-1.45\pm0.21$ & 0.83$\pm$0.01 \\
0155$-$0037 & 01 55 01.92 & $-$00 37 35.27 & 0.023 & 8.56$\pm$0.10 & $-1.50\pm0.12$ & 1.32$\pm$0.03 \\
0140$-$0012 & 01 40 52.90 & $-$00 12 49.52 & 0.018 & 8.59$\pm$0.10 & $-1.74\pm0.16$ & 1.19$\pm$0.03 \\
0023$-$0939 & 00 23 48.32 & $-$09 39 09.85 & 0.020 & 8.59$\pm$0.10 & $-1.10\pm0.06$ & 1.03$\pm$0.01 \\
0233$+$0038 & 02 33 03.92 & $+$00 38 41.59 & 0.024 & 8.67$\pm$0.10 & $-0.93\pm0.08$ & 1.20$\pm$0.01 \\
0256$-$0010 & 02 56 32.90 & $-$00 10 51.13 & 0.029 & 8.69$\pm$0.10 & $-0.89\pm0.08$ & 2.04$\pm$0.01 \\
2259$-$0019 & 22 59 03.19 & $-$00 19 45.82 & 0.016 & 8.71$\pm$0.10 & $-1.21\pm0.09$ & 1.32$\pm$0.01 \\
2300$-$0832 & 23 00 51.88 & $-$08 32 13.36 & 0.025 & 8.75$\pm$0.10 & $-1.58\pm0.15$ & 1.26$\pm$0.06 \\
0233$-$0819 & 02 33 05.84 & $-$08 19 08.79 & 0.019 & 8.77$\pm$0.10 & $-1.61\pm0.17$ & 1.08$\pm$0.01 \\
0125$-$0024 & 01 25 06.09 & $-$00 24 31.22 & 0.016 & 8.78$\pm$0.10 & $-4.67^c$ & 1.24$\pm$0.03 \\
0241$-$0810 & 02 41 35.06 & $-$08 10 24.52 & 0.005 & 8.79$\pm$0.10 & $-2.77^c$ & 0.79$\pm$0.01 \\
2301$+$1451 & 23 01 06.26 & $+$14 51 16.74 & 0.024 & 8.81$\pm$0.11 & $-0.74\pm0.12$ & 1.73$\pm$0.12 \\
0332$-$0600 & 03 32 00.36 & $-$06 00 42.77 & 0.018 & 8.82$\pm$0.10 & $-1.71\pm0.17$ & 1.17$\pm$0.03 \\
2131$-$0613 & 21 31 03.64 & $-$06 13 58.27 & 0.025 & 8.82$\pm$0.10 & $-0.73\pm0.10$ & 1.84$\pm$0.01 \\
0850$+$6145 & 08 50 39.67 & $+$61 45 44.42 & 0.024 & 8.88$\pm$0.10 & $-1.05\pm0.14$ & 1.23$\pm$0.01 \\
0256$-$0707 & 02 56 34.76 & $-$07 07 11.12 & 0.017 & 8.96$\pm$0.10 & $-1.08\pm0.12$ & 1.84$\pm$0.04 \\
0033$+$0012 & 00 33 14.83 & $+$00 12 03.66 & 0.014 & 8.97$\pm$0.10 & $-0.57\pm0.09$ & 2.04$\pm$0.02 \\
2137$+$1243 & 21 37 52.51 & $+$12 43 42.63 & 0.031 & 8.98$\pm$0.11 & $-0.91\pm0.12$ & 1.75$\pm$0.01 \\
0137$-$0942 & 01 37 43.57 & $-$09 42 36.36 & 0.019 & 9.02$\pm$0.10 & $-0.88\pm0.12$ & 2.49$\pm$0.01 \\
0040$-$0017 & 00 40 35.09 & $-$00 17 37.19 & 0.019 & 9.02$\pm$0.10 & $-0.95\pm0.12$ & 2.39$\pm$0.05 \\
0136$+$1356 & 01 36 15.18 & $+$13 56 43.26 & 0.024 & 9.03$\pm$0.10 & $-0.49\pm0.20$ & 1.88$\pm$0.05 \\
2304$-$0933 & 23 04 30.86 & $-$09 33 03.58 & 0.032 & 9.04$\pm$0.10 & $-0.94\pm0.12$ & 1.85$\pm$0.01 \\
0309$-$0041 & 03 09 58.06 & $-$00 41 39.58 & 0.037 & 9.05$\pm$0.10 & $-0.51\pm0.09$ & 3.57$\pm$0.10 \\
0019$-$0004 & 00 19 17.04 & $-$00 04 11.50 & 0.040 & 9.15$\pm$0.10 & $-0.45\pm0.08$ & 2.95$\pm$0.09 \\
0208$+$0048 & 02 08 48.30 & $+$00 48 01.41 & 0.020 & 9.15$\pm$0.10 & $-0.66\pm0.05$ & 3.84$\pm$0.05 \\
0306$-$0033 & 03 06 24.96 & $-$00 33 04.38 & 0.025 & 9.16$\pm$0.10 & $-0.98\pm0.15$ & 1.58$\pm$0.04 \\
0346$+$0023 & 03 46 06.86 & $+$00 23 24.93 & 0.031 & 9.18$\pm$0.10 & $-0.32\pm0.08$ & 2.56$\pm$0.05 \\
0243$+$0032 & 02 43 33.19 & $+$00 32 45.29 & 0.028 & 9.34$\pm$0.10 & $-0.68\pm0.13$ & 3.13$\pm$0.05 \\
2318$+$0105 & 23 18 01.34 & $+$01 05 13.31 & 0.030 & 9.35$\pm$0.10 & $-0.59\pm0.11$ & 3.97$\pm$0.05 \\
2206$-$0041 & 22 06 46.12 & $-$00 41 17.40 & 0.029 & 9.37$\pm$0.16 & $-0.11\pm0.14$ & 1.64$\pm$0.01 \\
2243$+$0040 & 22 43 58.53 & $+$00 40 25.70 & 0.039 & 9.37$\pm$0.10 & $-0.48\pm0.11$ & 4.07$\pm$0.06 \\
0334$+$0106 & 03 34 06.02 & $+$01 06 15.19 & 0.049 & 9.44$\pm$0.10 & $-0.03\pm0.09$ & 3.87$\pm$0.01 \\
0317$-$0004 & 03 17 20.20 & $-$00 04 37.04 & 0.022 & 9.46$\pm$0.10 & $-0.08\pm0.10$ & 3.86$\pm$0.03 \\
2211$+$0006 & 22 11 47.47 & $+$00 06 49.79 & 0.057 & 9.58$\pm$0.10 & $-0.20\pm0.06$ & 4.00$\pm$0.19 \\
0245$-$0045 & 02 45 57.45 & $-$00 45 12.27 & 0.054 & 9.60$\pm$0.10 & $0.04\pm0.10$ & 5.05$\pm$0.10 \\
0030$-$0042 & 00 30 13.43 & $-$00 42 28.45 & 0.041 & 9.63$\pm$0.10 & $0.13\pm0.10$ & 2.74$\pm$0.01 \\
2139$+$0019 & 21 39 56.73 & $+$00 19 23.69 & 0.050 & 9.72$\pm$0.10 & $0.23\pm0.09$ & 4.60$\pm$0.06 \\
0407$-$0634 & 04 07 04.64 & $-$06 34 02.15 & 0.038 & 9.88$\pm$0.10 & $0.01\pm0.06$ & 4.75$\pm$0.06 \\
0254$+$0103 & 02 54 50.52 & $+$01 03 26.97 & 0.043 & 9.90$\pm$0.10 & $0.43\pm0.06$ & 6.48$\pm$0.02 \\
0140$-$0013 & 01 40 47.46 & $-$00 13 05.74 & 0.058 & 10.02$\pm$0.10 & $-0.02\pm0.12$ & 6.28$\pm$0.12
\enddata
\tablenotetext{a}{Redshifts are taken from SDSS DR17 \citep{Abdurrouf22}}
\tablenotetext{b}{The stellar masses and SFRs are derived from broadband SED fitting with BAGPIPES.}
\tablenotetext{c}{Galaxies 0125-0024 and 0241-0810 are quiescent. The SED-based SFRs may not reflect the true level of star formation activity and thus only serve as a proxy of their quiescent nature. }
\end{deluxetable*}

\begin{figure}[htb!]
    \centering
    \includegraphics[width=0.98\columnwidth]{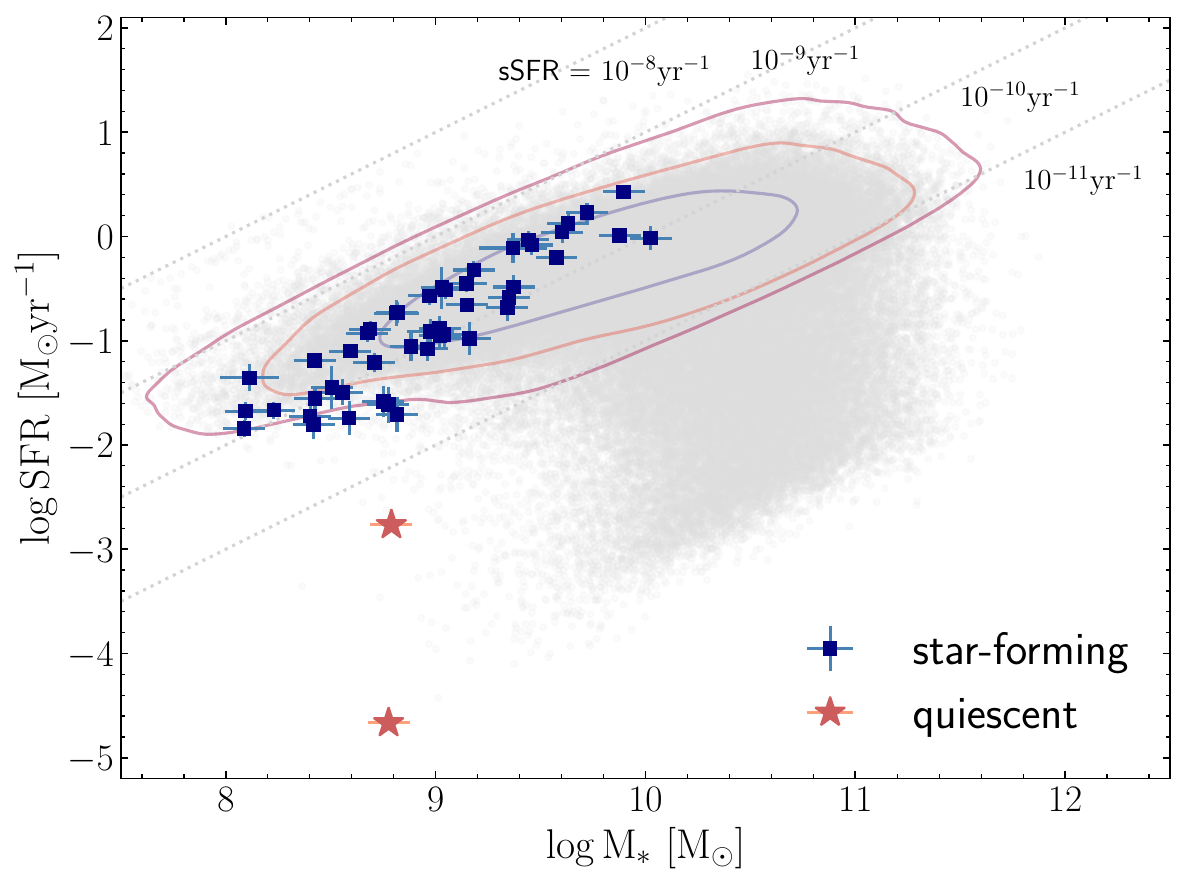}
    \caption{SFR as a function of stellar mass of the star-forming (blue squares) and quiescent (red stars) galaxies in our sample. 
    The contours show the SDSS SFMS at a similar redshift range ($z < 0.06$) taken from the GSWLC catalog \citep{Salim16} at 1-3 $\sigma$. The grey dots indicate all the GSWLC galaxies at the same redshift range as our sample.  }
    \label{fig:logM_SFR}
\end{figure}

All star-forming galaxies in our sample have typical SFRs and stellar masses like those in the SDSS ``star-forming main sequence'' (SFMS), as shown in Figure~\ref{fig:logM_SFR}, except for the two quiescent galaxies. For 38 out of 48 galaxies in our sample, SFR and mass measurements are available from the GALEX-SDSS-WISE Legacy Catalog (GSWLC) \citep{Salim16, Salim18}. For 26 of these galaxies, the stellar masses and the SFRs we measured are consistent with the GSWLC estimates within $2\sigma$ uncertainties. However, we emphasize that the major conclusions in this work remain unchanged when using the mass and SFR estimates from the GSWLC catalog. The derived masses and SFRs are listed in Table~\ref{tab:galaxy_prop}. The SED-based SFR estimates of two spectroscopically confirmed quiescent galaxies show large uncertainties due to their faint UV fluxes, so their best-fit SFRs are only a proxy of their quiescent nature and should be taken with caution. 

We decided to re-measure the sizes of the galaxies in our sample in spite of the existing size measurements in the SDSS catalog, because some of them with multiple star-forming clumps are mis-labeled as several individual galaxies instead of one galaxy in the SDSS catalog.
Because of their complex light profiles, we chose to fit the \citet{Petrosian76} half-light radius ($R_e$) independent of the assumption on galaxy light profiles. We estimated the Petrosian half-light radii of our sample from the PS1-$i$ image using the \texttt{PetroFit} package \citep{Geda22}. The size measurements and the associated errors are listed in Table~\ref{tab:galaxy_prop}.

\subsection{KCWI Observations}\label{subsec:kcwi_obs}

The KCWI data were taken over the course of five nights in October 2021 and September to November 2023. The seeing varied between $\sim 1.0$\arcsec\ and $\sim 2.0$\arcsec\ throughout the five nights. Table~\ref{tab:kcwi_obs} describes the observations of each galaxy. The total exposure time for each galaxy ranged from 5 minutes to 100 minutes. We used the BL grating with the Medium Slicer centered at 4500~\AA\@, which gives a field of view (FoV) of 16.5\arcsec$\times$20.4\arcsec\@ and wavelength coverage of 3500--5600~\AA\@\footnote{For the galaxies observed in 2023, we also have KCWI red channel data covering a spectral range of 6300--9500~\AA\@. For consistency between data taken in different nights, we consider only the KCWI blue spectra.  We are preparing a separate article (Z. Zhuang et al. in prep.) that will discuss the red optical data.}. From the arc spectra, we determined a spectral resolution FWHM of 2.5~\AA\ ($\sigma \sim$ 70 km s$^{-1}$) for both nights. We observed most of the galaxies at multiple position angles (the orientation of the KCWI slices on the sky) to maximize the spatial resolution and minimize covariance during stacking. To perform sky subtraction for the galaxies with their angular sizes comparable or even larger than the IFU FoV, we either placed multiple pointings on the target to obtain the in-field sky or took off-field sky frames adjacent to the science frames with half of the science exposure time. 

\startlongtable
\begin{deluxetable*}{lccccccc}
\tablecaption{KCWI observations of the Galaxy Sample \label{tab:kcwi_obs}}
\tablehead{\colhead{Galaxy} & \colhead{$z$} & \colhead{Exposure} & \colhead{Position Angles} & \colhead{Date} & \colhead{Airmass} & \colhead{Off-field Sky} & \colhead{Equivalent Extraction Radius$^{a}$} \\
\colhead{} & \colhead{} & \colhead{(s)} & \colhead{(deg)} & \colhead{} & \colhead{} & \colhead{} & \colhead{($R_e$)} }
\startdata
2225$-$0046 & 0.016 & 2320 & 0, 90 & 2023-09-23 & 1.12 &  & 1.1 \\
0231$-$0043 & 0.009 & 5160 & 70, 160 & 2021-10-02 & 1.07 & \checkmark & 1.6 \\
0133$-$0109 & 0.016 & 4200 & 50, 140 & 2021-10-02 & 1.16 &  & 1.7 \\
2329$+$1552 & 0.009 & 950 & 90 & 2023-09-23 & 1.04 & \checkmark & 1.4 \\
2152$-$0110 & 0.016 & 2710 & 0,90 & 2023-09-23 & 1.14 &  & 1.6 \\
0845$+$5145 & 0.014 & 1985 & 0, 90 & 2023-11-05 & 1.22 &  & 2.1 \\
2334$+$0029 & 0.024 & 2350 & 0, 90 & 2023-10-13 & 1.06 & \checkmark & 1.6 \\
0121$-$0853 & 0.018 & 2000 & 0, 90 & 2023-10-13 & 1.15 &  & 1.8 \\
0013$-$0956 & 0.013 & 1265 & 0, 90 & 2023-09-23 & 1.21 & \checkmark & 1.2 \\
0155$-$0037 & 0.023 & 1900 & 0, 90 & 2023-11-05 & 1.27 &  & 1.2 \\
0140$-$0012 & 0.018 & 2000 & 0, 90 & 2023-10-13 & 1.18 &  & 1.5 \\
0023$-$0939 & 0.02 & 1000 & 0 & 2023-10-13 & 1.15 & \checkmark & 1.2 \\
0233$+$0038 & 0.024 & 2000 & 0, 90 & 2023-11-05 & 1.24 &  & 1.4 \\
0256$-$0010 & 0.029 & 2965 & 0, 120, 240 & 2023-11-05 & 1.07 &  & 1.3 \\
2259$-$0019 & 0.016 & 3670 & 30, 90, 150 & 2023-09-23 & 1.10 &  & 1.0 \\
2300$-$0832 & 0.025 & 2057 & 0 & 2023-09-23 & 1.14 &  & 1.5 \\
0233$-$0819 & 0.019 & 2000 & 0, 90 & 2023-11-05 & 1.22 &  & 1.5 \\
0125$-$0024 & 0.016 & 2336 & 0, 90 & 2023-10-13 & 1.07 & \checkmark & 0.9 \\
0241$-$0810 & 0.005 & 1000 & 0 & 2023-11-05 & 1.15 & \checkmark & 1.2 \\
2301$+$1451 & 0.024 & 2340 & 0, 90 & 2023-10-13 & 1.12 &  & 1.5 \\
0332$-$0600 & 0.018 & 2000 & 0, 90 & 2023-11-05 & 1.24 &  & 1.5 \\
2131$-$0613 & 0.025 & 2000 & 0, 90 & 2023-10-13 & 1.15 &  & 1.3 \\
0850$+$6145 & 0.024 & 1000 & 90 & 2023-11-05 & 1.35 &  & 1.4 \\
0256$-$0707 & 0.017 & 2000 & 0, 90 & 2023-11-05 & 1.13 & \checkmark & 1.4 \\
0033$+$0012 & 0.014 & 1200 & 45 & 2021-10-02 & 1.14 & \checkmark & 1.8 \\
2137$+$1243 & 0.031 & 2250 & 0,90 & 2023-09-23 & 1.15 &  & 1.7 \\
0137$-$0942 & 0.019 & 2000 & 80, 170 & 2023-09-23 & 1.16 & \checkmark & 1.1 \\
0040$-$0017 & 0.019 & 2400 & 110 & 2021-10-10 & 1.11 & \checkmark & 1.9 \\
0136$+$1356 & 0.024 & 1547 & 0, 90 & 2023-10-13 & 1.04 &  & 1.9 \\
2304$-$0933 & 0.032 & 2000 & 0, 90 & 2023-10-13 & 1.16 &  & 1.7 \\
0309$-$0041 & 0.037 & 3900 & 170 & 2021-10-10 & 1.24 & \checkmark & 1.2 \\
0019$-$0004 & 0.04 & 4800 & 0 & 2021-10-02 & 1.43 &  & 1.8 \\
0208$+$0048 & 0.02 & 2700 & 90 & 2021-10-10 & 1.06 & \checkmark & 1.5 \\
0306$-$0033 & 0.025 & 2000 & 0, 90 & 2023-11-05 & 1.14 &  & 1.6 \\
0346$+$0023 & 0.031 & 1800 & 70, 160 & 2021-10-02 & 1.06 & \checkmark & 1.6 \\
0243$+$0032 & 0.028 & 2400 & 90 & 2021-10-10 & 1.07 & \checkmark & 1.7 \\
2318$+$0105 & 0.03 & 3600 & 60, 150 & 2021-10-10 & 1.11 & \checkmark & 1.5 \\
2206$-$0041 & 0.029 & 1200 & 0, 90 & 2021-10-02 & 1.13 &  & 2.2 \\
2243$+$0040 & 0.039 & 3600 & 50 & 2021-10-10 & 1.18 &  & 1.8 \\
0334$+$0106 & 0.049 & 1985 & 0, 90 & 2023-11-05 & 1.28 &  & 0.6 \\
0317$-$0004 & 0.022 & 600 & 140 & 2021-10-10 & 1.14 & \checkmark & 1.4 \\
2211$+$0006 & 0.057 & 2400 & 0, 90 & 2021-10-10 & 1.24 &  & 1.5 \\
0245$-$0045 & 0.054 & 1800 & 90 & 2021-10-10 & 1.14 & \checkmark & 1.5 \\
0030$-$0042 & 0.041 & 1200 & 45 & 2021-10-02 & 1.19 &  & 1.5 \\
2139$+$0019 & 0.05 & 1200 & 0, 90 & 2021-10-02 & 1.13 & \checkmark & 1.3 \\
0407$-$0634 & 0.038 & 2000 & 0, 90 & 2023-11-05 & 1.39 & \checkmark & 0.9 \\
0254$+$0103 & 0.043 & 2400 & 20, 110 & 2021-10-02 & 1.20 & \checkmark & 1.6 \\
0140$-$0013 & 0.058 & 2700 & 80, 170 & 2021-10-02 & 1.07 &  & 1.4
\enddata
\tablenotetext{a}{Because we used an irregular aperture to extract the spectrum for each galaxy, we estimate the equivalent binning radius as the fraction of the effective radius using the square root of the ratio of the area used for KCWI spectrum extraction and the area within 1$R_e$.  }
\end{deluxetable*}

\subsubsection{Data Reduction}
We reduced the KCWI data using a modified version of the publicly available KCWI data reduction pipeline with the updates detailed in Prusinski et al. (in prep.)\footnote{\url{https://github.com/prusinski/KCWI_DRP}}. This pipeline converts raw, 2D science frames into flux-calibrated, 3D data cubes with a spatial sampling of 0.29\arcsec $\times$ 0.68\arcsec\ and a spectral dispersion of 1~\AA\ per pixel. It also allows us to scale the off-field sky frames for sky subtraction. The reduced data cubes were aligned, stacked, and resampled to a new spatial grid with spatial sampling of 0.3\arcsec $\times$ 0.3\arcsec\@ and dispersion of 1~\AA\ per spectral channel  using the KCWIKit package\footnote{\url{https://github.com/yuguangchen1/KcwiKit}} \citep{Chen2021,Prusinski24}. 

\subsubsection{Binning and Covariance Correction}
In this work, we focus only on the spatially integrated properties of the entire galaxies; results from a spatially-resolved analysis will be presented in future work (Zhuang et al., in prep). 
The integrated spectra for each galaxy were obtained by summing the spectra of spaxels with S/N $>$ 1 per spectral pixel in the stellar continuum between 4200~\AA\ and 4400~\AA\@. As listed in Table~\ref{tab:kcwi_obs}, the extraction regions are equivalent to the area within 0.6--2.2~$R_e$ for the galaxies in our sample.
We selected regions based the S/N threshold instead of adopting a uniform aperture for spectrum extraction in order to maximize the S/N from the galaxies. As we further demonstrate in Appendix~\ref{appendix:aper_effect}, any aperture bias that results from the irregular extraction region affects our main results negligibly. 

The IFU data underestimate the variance of the spatially-binned data due to covariance between adjacent spaxels, so we need to correct the variance of the binned spectra.
This issue originates from the redistribution of the flux in the same pixel onto the new spatial grid when stacking the data. If we adopted standard error propagation to calculate the variances of the stacked data cubes assuming the adjacent pixels are independent of each other, the variances would be underestimated.

As suggested by \citet{OSullivan20} the error of a KCWI data cube can be corrected as 
\begin{equation}
    \sigma_{\rm measured}= C[1+\alpha \log{(N_b)}]\sigma_{\rm nocov},
    \label{eq:sn_calibration1}
\end{equation}
for $N_b \lesssim N_{\rm thresh}$, and as
\begin{equation}
    \sigma_{\rm measured}= C[1+\alpha \log{(N_{\rm thresh})}]\sigma_{\rm nocov},
    \label{eq:sn_calibration2}
\end{equation}
for $N_b > N_{\rm thresh}$, where $N_{\rm thresh}$ represents the kernels beyond which the pixels become uncorrelated. 

To measure the correction parameters $C$, $\alpha$, and $N_{\rm thresh}$, we followed the method developed by \citet{delosReyes23}. In brief, we first simulated mock sky cubes as Gaussian noise centered at zero with variances determined from the variance cubes of the science frames. 
The variance values of mock sky cubes are the same as those of science cubes. The data and variance cubes of the mock sky were passed through the same stacking procedures as the science cubes. The $\sigma_{\rm measured}$ were calculated as the standard deviation of the mock sky data, while the $\sigma_{\rm nocov}$ were estimated as the median of the variance for the stacked sky cube. We then fit the $\sigma_{\rm measured}$ and $\sigma_{\rm nocov}$ to Equation~(\ref{eq:sn_calibration1}) and (\ref{eq:sn_calibration2}) to derive $\alpha$ and $N_{\rm thresh}$.

Finally, the covariance correction factor from Equation~(\ref{eq:sn_calibration1}) and (\ref{eq:sn_calibration2}) was multiplied into the flux uncertainties of the stacked spectra, which we previously calculated from error propagation without accounting for covariance.

\section{Emission-Line Measurements}\label{sec:emline_measurements}
The following section details the spectral analysis for gas-phase abundance measurements. Prior to the spectral analysis, the spectra were first corrected for Galactic extinction using the Galactic reddening maps of \citet{Schlafly11}. We adopted the Galactic extinction law by \citet{O'Donnell94} to calculate the de-reddening correction. The spectra are converted to the rest frame using the spectroscopic redshifts in the SDSS DR17 catalog \citep{Abdurrouf22}.

\subsection{Disentangling stellar continuum and emission lines}\label{subsec:stellar_gas_fit}
The first step in the spectral analysis is to separate the stellar continuum and emission lines in order to measure properties of the stellar population and the nebular gas. To do this, we performed a simultaneous fit of the stellar continuum and emission lines of the integrated spectrum using the Penalized Pixel-Fitting method \citep[\ppxf\@;][]{Cappellari04, Cappellari17, Cappellari23}, which is a $\chi^2$-minimization code that fits for a linear combination of simple stellar population (SSP) models at different ages and metallicities to account for complex SFHs of star-forming galaxies.

For the stellar population synthesis (SPS) templates, we generated SSP models from the Flexible Stellar Population Synthesis code \citep[FSPS;][]{Conroy09, Conroy10} version 3.2, assuming a \citet{Kroupa01} IMF, the MIST isochrones \citep{Choi16}, and the MILES stellar spectral libraries \citep{Sanchez06}. The SSP models accessed through the PYTHON-FSPS bindings \citep{Foreman-Mackey14} have metallicities spanning the range $\rm [Fe/H] = -2.0$ to $+0.5$ with an interval of 0.25~dex, corresponding to the grid of the MIST isochrones. The ages range from $\rm \log(Age)[yr] = 5.0$ to 10.2 with an interval of 0.2~dex. 

Given that the instrumental resolution of KCWI ($\sim 2.5$~\AA\ in the observed frame) is slightly higher than that of the FSPS SSP templates (2.54~\AA\ in the rest frame), the observed spectra were convolved with a Gaussian kernel to match the spectral resolution of the SSP templates. We also restricted the spectral analysis to rest-frame wavelengths between 3650~\AA\ and 5500~\AA\@.

Inspired by the MaNGA data-analysis pipeline \citep{Westfall19}, we ran two iterations of \ppxf\ for each spectrum to better constrain the stellar kinematics and stellar continuum. 
In the first iteration, the spectral regions around potential emission lines (Table~\ref{tab:emline_list}) in the velocity range of $\pm 400$~km s$^{-1}$ were masked to determine the stellar kinematics. We included an eighth-order additive Legendre polynomial and an eighth-order multiplicative Legendre polynomial in the fit of the stellar continuum, motivated by the experience of the MaNGA team \citep{Westfall19, Belfiore19} to improve the quality of kinematics results. The two polynomials allow the overall shape of the model spectra to better match that of the data, which may suffer from imperfect flux calibration and small inaccuracies in the dust extinction correction. Each fit ran \ppxf\@ twice. After the first run with the default emission-line mask, we used a moving boxcar with a wavelength width of 100~\AA\ ($\sim$ 7000 km s$^{-1}$) to determine the local mean and standard deviation of the fit residual. We then re-ran the fit with the updated mask excluding the $>3\sigma$ outliers.  

\begin{figure*}[htb!]
    \centering
    \includegraphics[width=0.95\textwidth]{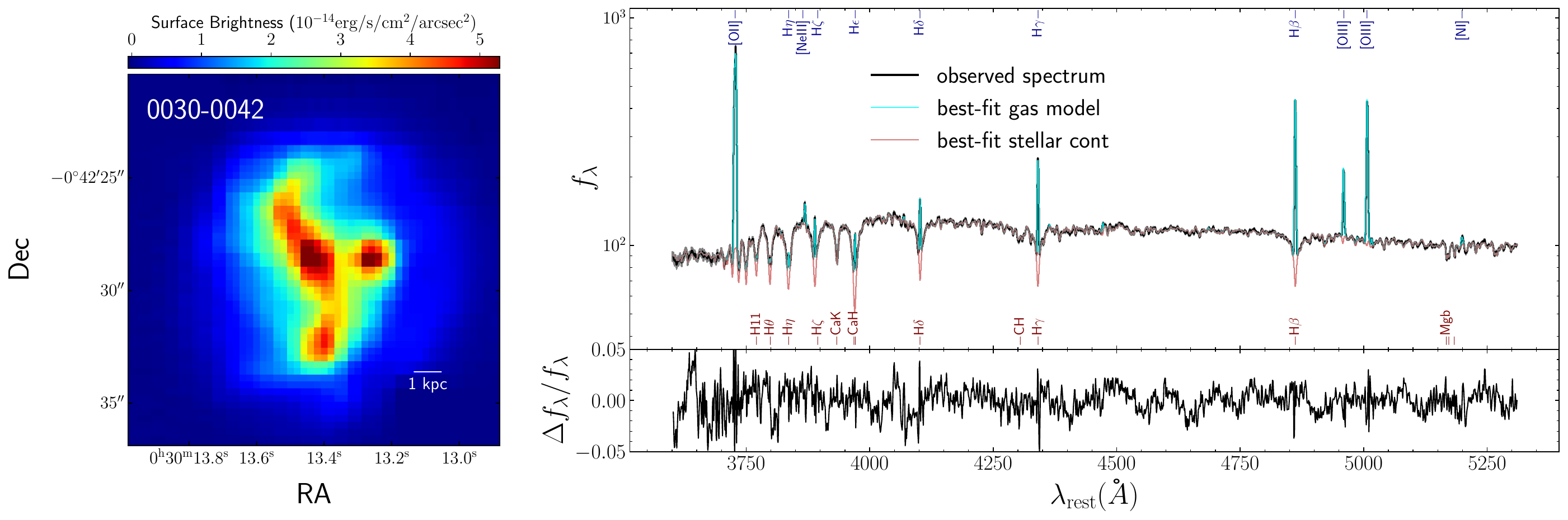}
    \caption{\textbf{Left:} Example KCWI white-light image computed by integrating the spectrum in the wavelength direction for each spatial pixel. \textbf{Right:} Example integrated KCWI spectrum of the same galaxy (top) and the model residual (bottom). In the top panel, the observed integrated spectrum is shown in black, while the best-fit model of the stellar continuum and the emission lines are shown in cyan and red, respectively. The $y$-axis is on a logarithmic scale. Prominent emission lines (blue) and stellar absorption features (red) are marked in the top and middle panels. The full set of the emission lines included in the fit is listed in Table~\ref{tab:emline_list}.  }
    \label{fig:example_kcwi_jointfit}
\end{figure*}

In the second iteration, we fixed the stellar kinematics to the values determined from the first iteration and performed a simultaneous fit of emission lines and stellar continuum. The emission lines and their constraints used in the fits are listed in Table~\ref{tab:emline_list}. As suggested by \citet{Sarzi06} and \citet{Oh11}, this approach is preferable to fitting the emission lines separately on a continuum-subtracted spectrum, which tends to overestimate the emission line fluxes and widths when a small mismatch exists between the stellar continuum and SPS models. Figure~\ref{fig:example_kcwi_jointfit} shows an example of a best-fit model obtained with the joint fitting from the second iteration along with the observed spectrum. 

This method provides measurements of integrated emission line fluxes that properly account for the stellar continuum. 
Additionally, we obtain a emission-free stellar continuum spectrum by subtracting the best-fit emission line model from the observed spectrum; we will use this stellar continuum spectrum in Section~\ref{sec:stellar_abund_measurement} to estimate stellar abundances.

\begin{deluxetable}{cccc}[htb!]
\tablecaption{Emission-Line Parameters\label{tab:emline_list}}
\tablehead{\colhead{ID} & \colhead{Line Name} & \colhead{$\lambda_{\rm rest}^a$} & \colhead{Fixed Ratio} \\
\colhead{} & \colhead{} & \colhead{(\AA\@)} & \colhead{}}
\startdata
1 & H16 & 3704.913 & ... \\
2 & H15 & 3713.034 & ... \\
3 & H14 & 3723.003 & ... \\
4 & [O$\,${\sc ii}]$\lambda$3726 & 3727.092 & ... \\
5 & [O$\,${\sc ii}]$\lambda$3729 & 3729.875 & 0.28--1.47 [O$\,${\sc ii}]$\lambda$3726$^b$ \\
6 & H13 & 3735.436 & ... \\
7 & H12 & 3751.217 & ... \\
8 & H11 & 3771.701 & ... \\
9 & H$\theta$ & 3798.976 & ... \\
10 & He$\,${\sc i}$\lambda$3820 & 3820.691 & ... \\
11 & H$\eta$ & 3836.472 & ... \\
12 & [S$\,${\sc iii}]$\lambda$3856 & 3857.111 & ... \\
13 & [Ne$\,${\sc iii}]$\lambda$3869 & 3869.860 & ... \\
14 & He$\,${\sc i}$\lambda$3889 & 3889.749 & ... \\
15 & H$\zeta$ & 3890.151 & ... \\
16 & [Ne$\,${\sc iii}]$\lambda$3967 & 3968.590 & 0.31 [Ne$\,${\sc iii}]$\lambda$3869 \\
17 & H$\epsilon$ & 3971.195 & ... \\
18 & He$\,${\sc i}$\lambda$4026 & 4027.328 & ... \\
19 & [S$\,${\sc ii}]$\lambda$4069 & 4069.749 & ... \\
20 & [S$\,${\sc ii}]$\lambda$4076 & 4077.500 & 0.40 [S$\,${\sc ii}]$\lambda$4069 \\
21 & H$\delta$ & 4102.892 & ... \\
22 & H$\gamma$ & 4341.684 & ... \\
23 & [O$\,${\sc iii}]$\lambda$4363 & 4364.436 & ... \\
24 & He$\,${\sc i}$\lambda$4471 & 4472.734 & ... \\
25 & [Fe$\,${\sc iii}]$\lambda$4658 & 4659.414 & ... \\
26 & He$\,${\sc ii}$\lambda$4686 & 4687.015 & ... \\
27 & He$\,${\sc i}$\lambda$4713 & 4714.466 & ... \\
28 & H$\beta$ & 4862.683 & ... \\
29 & He$\,${\sc i}$\lambda$4922 & 4923.305 & ... \\
30 & [O$\,${\sc iii}]$\lambda$4959 & 4960.295 & 0.34 [O$\,${\sc iii}]$\lambda$5007 \\
31 & [Fe$\,${\sc iii}]$\lambda$4985 & 4985.900 & ... \\
32 & [Fe$\,${\sc iii}]$\lambda$4986 & 4987.200 & ... \\
33 & [O$\,${\sc iii}]$\lambda$5007 & 5008.240 & ... \\
34 & He$\,${\sc i}$\lambda$5016 & 5017.077 & ... \\
35 & [N$\,${\sc i}]$\lambda$5198 & 5199.349 & ... \\
36 & [N$\,${\sc i}]$\lambda$5200 & 5201.705 & 0.37 [N$\,${\sc i}]$\lambda$5198
\enddata
\tablenotetext{a}{Ritz wavelengths in vacuum from the National Institute of Standards and Technology (NIST) Atomic Spectra database (\url{https://physics.nist.gov/PhysRefData/ASD/lines_form.html}).}
\tablenotetext{b}{The line ratio of [O$\,${\sc ii}]$\lambda$3726/[O$\,${\sc ii}]$\lambda$3729 is only allowed to vary in the range of 0.28--1.47, as required by atomic physics \citep{Osterbrock06}. }
\end{deluxetable}

\subsection{Gas-phase oxygen abundance}\label{subsec:gas_met}
The intrinsic dust attenuation of each galaxy was calculated from the Balmer decrement H$\gamma$/H$\beta$, because the spectral range of the KCWI-blue data did not include H$\alpha$.  Assuming Case B recombination, $T_e = 10,000$~K and $n_e = 100$~cm$^{-3}$, the theoretical ratio of H$\gamma$/H$\beta$ is 0.47 \citep{Osterbrock06}. The intrinsic $E(B-V)$, derived from the theoretical ratio of  H$\gamma$/H$\beta$ along with the \citet{Cardelli89} dust extinction law, was applied to the emission-line measurements to yield the dereddened line fluxes for nebular oxygen abundance determination. 

Ideally, nebular oxygen abundances would be measured from the reliable direct $T_e$ method, which directly determines the electron temperature of ionized gas. However, the direct $T_e$ method relies on the detection of auroral lines, such as [O$\,${\sc iii}]$\lambda$4363, which are exceedingly faint, especially for relatively metal-rich systems. We detected [O$\,${\sc iii}]$\lambda$4363 in fewer than $1/3$ of the sample. We therefore used an empirical strong-line calibration derived from $T_e$-based measurements to estimate the gas-phase oxygen abundance for constructing an unbiased MZR\@. We will discuss the gas-phase oxygen abundances derived from different approaches in more detail in future work focusing on the KCWI-red data (Zhuang et al. in prep.). 

For the present, we use the ``$R$ calibration'' \citep{Pilyugin16}, defined as:

\begin{equation} \label{eq:gasZ_upper}
\begin{split}
{\rm (O/H)}_{R,U}  = \, &\, 8.589 + 0.022 \, \log (R_{3}/R_{2}) + 0.399 \, \log N_{2}   \\  
& +  (0.164 \, \log (R_{3}/R_{2}) + 0.589 \log N_{2} - 0.137)   \\ 
& \times  \log R_{2},
\end{split}
\end{equation}
when $\log N_2 \geq -0.6$; and

\begin{equation} \label{eq:gasZ_lower}
\begin{split}
 {\rm (O/H)}_{R,L}   =\,  &\,  7.932 + 0.944 \, \log (R_{3}/R_{2}) + 0.695 \, \log N_{2}   \\  
& +   (0.970 - 0.291 \, \log (R_{3}/R_{2}) - 0.019 \log N_{2})   \\ 
& \times \log R_{2} ,
\end{split}
\end{equation}
when $\log N_2 < -0.6$, where $R_2 = I_{\rm [O\,II] \lambda 3727+ \lambda 3729} /I_{{\rm H}\beta }$, $R_3  = I_{{\rm [O\,III]} \lambda 4959+ \lambda 5007} /I_{{\rm H}\beta }$, and $N_2  = I_{\rm [N\,II] \lambda 6548+ \lambda 6584} /I_{{\rm H}\beta }$\footnote{The $N_2$ ratio defined by \citet{Pilyugin16} is different from the commonly used ``N2'' index used by other strong line calibrations \citep[e.g.,][]{Pettini04}, where $N_2 = \log{({\rm [N\,II]\lambda 6584} /{{\rm H}\alpha })}$ instead.}.

Because our full sample of the KCWI data does not cover the spectral region of [N$\,${\sc ii}]$\lambda\lambda$6548,6584, we use the emission line fluxes of H$\alpha$ and [N$\,${\sc ii}] from the SDSS MPA-JHU catalog \citep[\textit{galSpecLine}, SDSS DR17;][]{Brinchmann04, Tremonti04}. Assuming Case B recombination at $T_e = 10,000$~K and $n_e = 100$~cm$^{-3}$, the $N_2$ ratio is calculated assuming that 

\begin{equation} 
\begin{split}
N_2  = \, &\, I_{\rm [N\,II] \lambda 6548+ \lambda 6584} /I_{{\rm H}\beta }   \\  
= &I_{\rm [N\,II] \lambda 6548+ \lambda 6584} /I_{{\rm H}\alpha } \times 2.86 
\end{split}
\end{equation}

Combining $R_2$ and $R_3$ based on the KCWI measurements with the SDSS results for $N_2$, we derive the gas-phase oxygen abundance of each galaxy using Equations~(\ref{eq:gasZ_upper}) and (\ref{eq:gasZ_lower}),  with results listed in Table~\ref{tab:measurements}. The reported uncertainties are the square root of the quadrature sum of the systematic uncertainty of the $R$ calibration \citep[$0.1$~dex,][]{Pilyugin16, Pilyugin18} and the random errors from the flux measurement uncertainties. Our measurements from KCWI+SDSS data are  consistent within $1\sigma$ of the values derived from using only the SDSS emission line fluxes.

\section{Stellar abundance determination}\label{sec:stellar_abund_measurement}
Stellar abundances of iron and magnesium can be measured from spectral indices of stellar absorption lines sensitive to the stellar population parameters \citep[e.g.,][]{Thomas05, Thomas10}, or full-spectrum fitting that utilizes the information from the entire spectrum \citep[e.g.,][]{Choi14, Conroy18}. In this work, we use the full-spectrum fitting method that relies on galaxy templates with variable [Mg/Fe] to constrain [Mg/Fe] in our star-forming galaxies. 
Existing models for measuring the detailed abundances of individual $\alpha$ elements \citep[\texttt{alf},][]{Conroy18}, can only be applied to quiescent galaxies because the models exist only for older populations, usually $>1$ Gyr in age.   

In order to fit galaxies with a wide range of ages, we made use of the latest FSPS SSP models with variable [$\alpha$/Fe] (Park et al., in prep.), as detailed in Section~\ref{subsec:ssp_afe}, in order to estimate [Mg/Fe] in the star-forming galaxies. The updated model was designed to capture variation in [$\alpha$/Fe] over the full range of possible stellar ages, from 0.1~Myr to 20 Gyr.  This was achieved by constructing a new set of stellar spectral libraries and isochrones (see Park et al. in prep., for details).

Similar to existing SSP models with variable [$\alpha$/Fe] \citep{Vazdekis15, Knowles23}, our models assume that all $\alpha$ elements (O, Ne, Mg, Si, S, Ca, and Ti) vary in lock-step (i.e., [Mg/Fe] = [$\alpha$/Fe]). One must be cautious when interpreting an inferred [$\alpha$/Fe] as identical to [Mg/Fe]; as addressed in \citet{Choi14} and \citet{Beverage23}, individual $\alpha$ elements may not track each other in quiescent galaxies when each $\alpha$ element is fitted separately using \Alf\@. To mitigate this issue, we developed our own fitting algorithm (Section~\ref{subsec:full_spec_fitting}) and show in Section~\ref{subsec:validate_fitting} that our measured [$\alpha$/Fe] represents [Mg/Fe] consistent with the \Alf\ results in the sample galaxies. 

\subsection{SSP models with variable [$\alpha$/Fe]}\label{subsec:ssp_afe}
To construct SSP models with variable [$\alpha$/Fe], we followed the methodology described by \citet{Conroy12} and \citet{Conroy18}. At a given age and metallicity, the SSP spectrum ($f$) of arbitrary [$\alpha$/Fe] can be calculated as:
\begin{equation}
\label{eq:ssp_afe}
    f({\rm[\alpha/ Fe]}) = f_{\rm base}({\rm[\alpha/ Fe]_{lib}})\frac{f_{\rm theo}({\rm[\alpha/ Fe]})}{f_{\rm theo}({\rm[\alpha/ Fe]_{\rm lib}})},
\end{equation}
where $f_{\rm base}$ is the base empirical SSP model and $f_{\rm theo}$ is the synthetic SSP template with arbitrary abundance pattern from theoretical models. The ratio $f_{\rm theo}({\rm[\alpha/ Fe]}) / f_{\rm theo}({\rm[\alpha/ Fe]_{\rm lib}})$ is the ``response function'' which indicates the relative change in the spectrum due to a change in [$\alpha$/Fe] from the base model.  

For the base empirical model, we adopted the FSPS SSP models generated using the MILES stellar libraries (hereafter FSPS-MILES SSPs) described in Section~\ref{subsec:stellar_gas_fit}. Because all the available empirical SSP models including FSPS-MILES SSPs do not have $\alpha$ variation, we have to resort to theoretical models.  The theoretical SSP models (hereafter FSPS-C3K SSPs) were generated with FSPS using the Kroupa IMF \citep{Kroupa01}, $\alpha$-enhanced MIST isochrones (Dotter, et al., in prep), and the theoretical C3K stellar model grids (Park et al., in prep). The FSPS-C3K SSPs have the same grid in age and metallicity as the FSPS-MILES SSP models and span a wide range of [$\alpha$/Fe] ([-0.2, 0.0, 0.2, 0.4, 0.6]). 

Because our goal is to determine [Mg/Fe], we used the relation for [Mg/Fe] as a function of [Fe/H] for MILES stars reported by \citet{Milone11} as a proxy for [$\alpha$/Fe]$_{\rm lib}$ at a given [Fe/H]. The theoretical SSP model at a given [$\alpha$/Fe]$_{\rm lib}$ is derived from linear interpolation of two adjacent grids. We therefore obtain our new SSP models with variable age, [Fe/H], and [$\alpha$/Fe] using Equation~\ref{eq:ssp_afe}. The final models have 27 grid steps in age ($\rm \log(Age)[yr] = 5.0$ to 10.2 at an interval of 0.2~dex), 11 grid steps in [Fe/H] ($\rm [Fe/H] = -2.0$ to $0.5$ at an interval of 0.25~dex), and 5 grid setps in [$\alpha$/Fe] (${\rm [\alpha/Fe]} = -0.2$ to $0.6$ at an interval of 0.2~dex).

\subsection{Model fitting}\label{subsec:full_spec_fitting}
We developed a two-step fitting approach to measure both [Fe/H] and [Mg/Fe] of the star-forming galaxies in our sample. 

\begin{figure*}[htb!]
    \centering
    \includegraphics[width=0.95\textwidth]{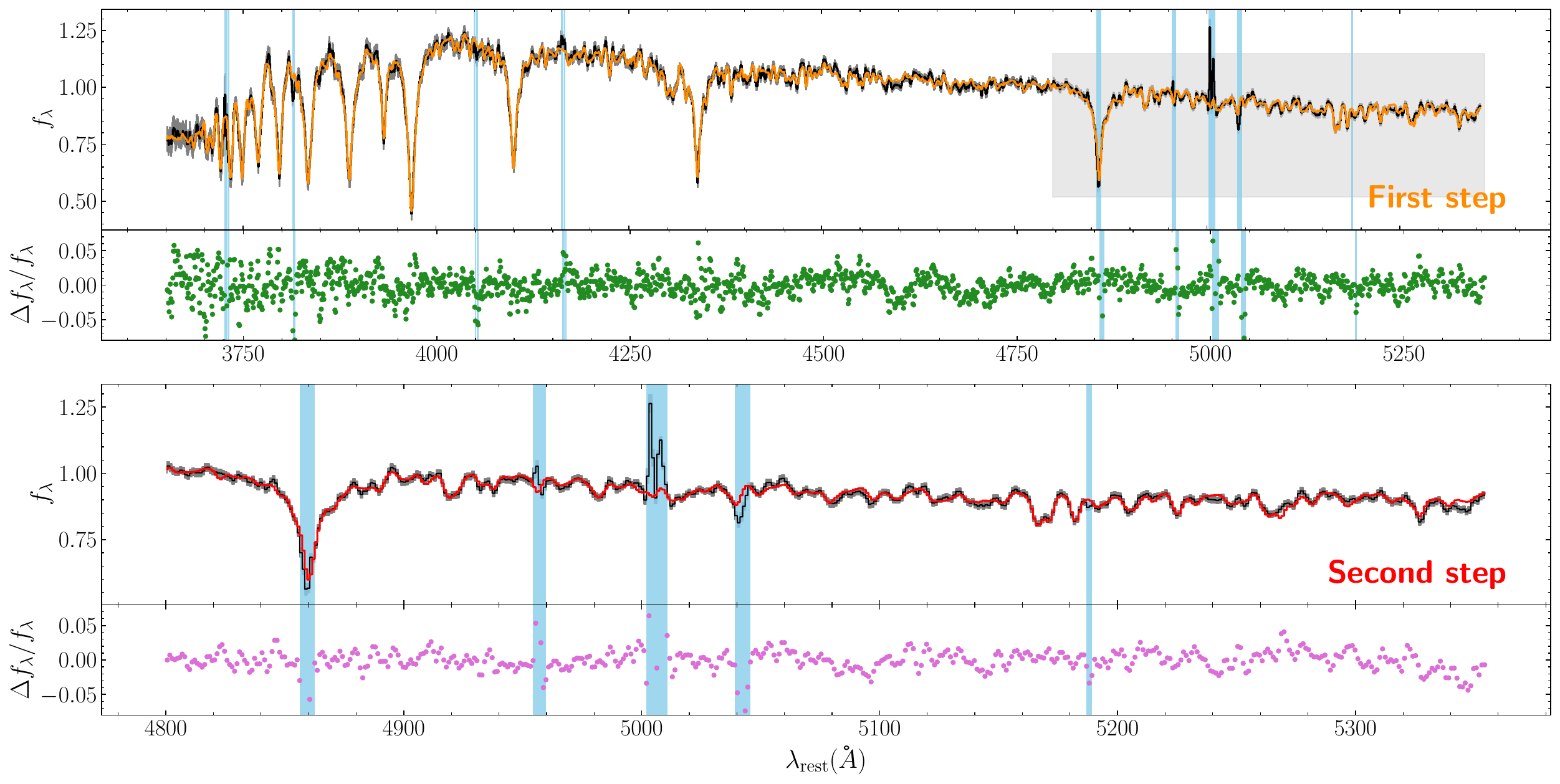}
    \caption{Illustration of the two-step approach for the same star-forming galaxy in Figure~\ref{fig:example_kcwi_jointfit}. The flux densities are normalized to have median in the spectral region 4800-5500~\AA\@. Note that the two panels cover different wavelength ranges of the same spectrum. \textbf{Top}: The emission-line-subtracted stellar spectrum (black), best-fit \ppxf\ model spectrum (orange), and the model residual (green) in the first fitting step, which operates on the entire KCWI spectrum. The blue shaded regions indicate the masked pixels that are $3\sigma$ outliers in the first iteration. The gray shaded region marks the spectral regions used in the second step. \textbf{Bottom}: the stellar continuum (black), best-fit model (red), and the model residual (pink) in the second fitting step of the same spectrum, which operates on the spectral region $\sim 4800-5500$~\AA\@. The blue shaded regions are the same masked pixels inherited from the first step. In this iteration, [Fe/H] and [Mg/Fe] are allowed to vary, but the SFH and the stellar kinematics are fixed to the values determined in the first step.    }
    \label{fig:stellar_cont_fit_example}
\end{figure*}

In the first step, we use \ppxf\ to fit the full emission-free stellar continuum spectrum of each galaxy from $3650$ to $5500$~\AA\ in the rest frame to infer the SFHs for each galaxy. 
It fits for a linear combination of discrete SSPs at different ages, [Fe/H] and [$\alpha$/Fe], along with the stellar velocity dispersion and offset from the initial spectroscopic redshift in the SDSS catalog to the observed spectrum.
The young galaxies in our sample could have SFHs that are more complex than can be described by simple parameterizations (i.e., an SSP).  Therefore, the non-parametric approach of \ppxf\ is appropriate for our case.  In this paper, we use the SFHs only to derive average ages (described below).  The full SFHs will be analyzed in the future.  Appendix~\ref{appendix:age} gives a complete justification of our use of non-parametric SFH fits.
Each SSP model is normalized over the spectral range 4800--5500~\AA\ for ``light-weighted'' property estimates. Similar to the fits described in Section~\ref{subsec:stellar_gas_fit}, we ran two iterations for each spectrum. The 3$\sigma$ outliers from the first iteration are masked in the second iteration to remove pixels affected by imperfect subtraction of emission lines. The error spectra are also scaled by the square root of the reduced $\chi^2$ of the first iteration to yield more reasonable estimates of the stellar population uncertainties. Following \citet{Cappellari23}, ``regularization'' was also employed in the fitting with $\texttt{regul}=30$ to recover a relatively smooth SFH. We also included a ninth-order multiplicative Legendre polynomial as a ``nuanced'' parameter  to account for internal dust extinction and/or imperfect flux calibration. 

Following \citet{Cappellari23}, the light-weighted properties are derived according to 
\begin{subequations}\label{eq:ppxf_lw_values}
\begin{equation}
  \langle \log{\rm Age} \rangle = \frac{\sum_i w_i\times \log{{\rm Age}_i}}{\sum_i w_i},
\end{equation}    
\begin{equation}\label{eq:mean_feh}
  \langle {\rm [Fe/H]} \rangle = \frac{\sum_i w_i \times {\rm [Fe/H]}_i}{\sum_i w_i},
\end{equation}
\begin{equation}\label{eq:mean_afe}
  \langle {\rm [\alpha/Fe]} \rangle = \frac{\sum_i w_i \times {\rm [\alpha/Fe]}_i}{\sum_i w_i},
\end{equation}
\end{subequations}
where $w_i$ is the weight of the $i^{\rm th}$ SSP template returned by \ppxf\@.

In the second step, we re-fit the stellar continuum limited to the rest-frame spectral region 4800--5500~\AA\@. As we will demonstrate in Section~\ref{subsec:validate_fitting}, a second fit is necessary to recover a value closer to the [Mg/Fe] consistent with the \Alf\ measurements. In the real galaxy spectra, individual $\alpha$ elements may not track each other \citep{Thomas03, Conroy14, Choi14, Beverage23}; the derived [$\alpha$/Fe] from the first iteration is sensitive not only to Mg but also to other $\alpha$ elements, such as Ca. The spectral region 4800--5500~\AA\ contains most of the Fe- and Mg- sensitive features (i.e., the Mg~b triplet, Fe5270, Fe5335) as well as the age indicator H$\beta$. For example, \citet{Vazdekis15} found that they could reproduce values of [Mg/Fe] consistent with the literature for two massive ellipticals only if the fits were limited to the stellar continuum in the range 4800--5500~\AA; the fits failed when they were performed using a larger spectral range.

To fit the spectral region 4800--5500~\AA\@, we fixed the non-parametric SFHs and stellar kinematics to the values recovered in the first step. We retained the weights of each age bin returned by the first step, but allowed [Fe/H] and [$\alpha$/Fe] to vary within each bin in the second iteration.

We followed the continuum correction approach of \Alf\ \citep{Conroy18} by including in the fitting process a fourth-order multiplicative polynomial in the form of $p(\lambda) = \sum_{i=0}^4 c_i(\lambda - \mu)^i$, where $\mu$ is the mean wavelength of the region being fit. The polynomial degree was determined by $n = (\lambda_{\rm max} - \lambda_{\rm min})/200$~\AA\ so that the correction is flexible enough to account for any mismatch in the continuum shape due to imperfect flux calibration or dust attenuation, but not so flexible that it over-fits regions containing real broad absorption features. In each iteration, we calculated the ratio of the data and model, masked the spectral regions with strong stellar absorption features and gas emission lines, and performed a least-squares polynomial fit to the unmasked pixels. The data were then compared with the altered model multiplied by the best-fit polynomial to calculate the likelihood. We experimented with different polynomial orders, including $n=0$ and $n = (\lambda_{\rm max} - \lambda_{\rm min})/100$~\AA\, and found that the changes of the recovered abundances are within the reported uncertainties.


The fits were accomplished using \texttt{dynesty} \citep{Speagle20}, a Python package for dynamic nested sampling capable of estimating the Bayesian posterior distribution. For each age bin, we adopted a uniform prior for [Fe/H] in the range $[-2.0, 0.5]$ and a uniform prior for [$\alpha$/Fe] in the range $[-0.2, 0.6]$. The SSP models at non-gridpoint values of model parameters were calculated via linear interpolation. Using the posteriors of [Fe/H] and [$\alpha$/Fe] in each age bin from \texttt{dynesty} along with Equations~(\ref{eq:mean_feh}) and (\ref{eq:mean_afe}), we obtained the posterior distributions of the light-weighted [Fe/H] and [$\alpha$/Fe] for each galaxy. All the galaxies have converged, single-peaked posteriors. The best-fit results are quoted as the median of the 1D marginalized posteriors. Because the maximum a posteriori (MAP) estimates for some galaxies fell outside the $1\sigma$ (68\%) confidence interval, but were always within the $2\sigma$ interval, the quoted uncertainties are based on the $2\sigma$ confidence interval (i.e., 2.5\% and 97.5\% percentiles).

\subsection{Verification of Magnesium Abundance Measurement}\label{subsec:validate_fitting}
As mentioned above, one must be cautious when interpreting the measured [$\alpha$/Fe] as [Mg/Fe] if the SPS models used for analysis tie all the $\alpha$ elements together (i.e., the case of our analysis).
If $\alpha$ elements do not track each other---for instance, $\rm [Ca/Fe]\sim 0$ when $\rm [Mg/Fe]> 0$ \citep{Choi14, Beverage23}---then the derived [$\alpha$/Fe] may not reflect [Mg/Fe]. In fact, \citet{Leethochawalit19} showed in their Appendix A that [Mg/Fe] was underestimated by $\sim 0.1-0.2$~dex when the SSP models were fit with the response functions of individual $\alpha$ elements fixed to the same value. Therefore, it is essential to validate that our method recovers the correct [Mg/Fe] rather than an amalgam of different elements, i.e., [$\alpha$/Fe]. 

As a result, we seek to understand how to interpret the measured [$\alpha$/Fe], age, and [Fe/H] via the two-step approach along with the new SPS models before applying the method to our KCWI sample.
To validate the two-step method, we made use of the sample of 123 SDSS quiescent galaxies compiled by \citet{Leethochawalit18}, which has existing measurements of individual [Mg/Fe], [Fe/H] and SSP age from \Alf\ by \citet{Zhuang23}. We re-measured their light-weighted age, [Fe/H] and [$\alpha$/Fe] using the two-step method described in Section~\ref{subsec:full_spec_fitting} in order to compare them with the existing results obtained from \Alf\@. 

To illustrate the need for the two-step method, we also experimented with fits using only a single iteration of \ppxf\@. We performed the fits with two different spectral regions: the spectral range of KCWI-blue overlapping with those of the empirical SSP models ($\sim 3650-5500$~\AA\@), which is essentially the first step in the two-step method, and the spectral region 4800-5500~\AA, which contains most of the Fe- and Mg- sensitive features present in the full spectrum. The \ppxf\ setup was the same as the one described in Section~\ref{subsec:full_spec_fitting}. To estimate the uncertainties in the \ppxf\ measurements, we adopted the ``wild bootstrap method'' \citep{DAVIDSON08} following the example described by \citet{Cappellari23}\footnote{\url{https://github.com/micappe/ppxf_examples/blob/main/ppxf_example_population_bootstrap.ipynb}}. In summary, after running the first \ppxf\ regularized fit with $\texttt{regul}=30$, we bootstrapped the residuals and repeated the \ppxf\ fit 100 times with no regularization. For each iteration, we calculated the light-weighted properties using Equation~\ref{eq:ppxf_lw_values}. The uncertainties were estimated as the standard deviations of the distributions for the light-weighted properties.

\begin{figure*}[htb!]
    \centering
    \includegraphics[width=0.95\textwidth]{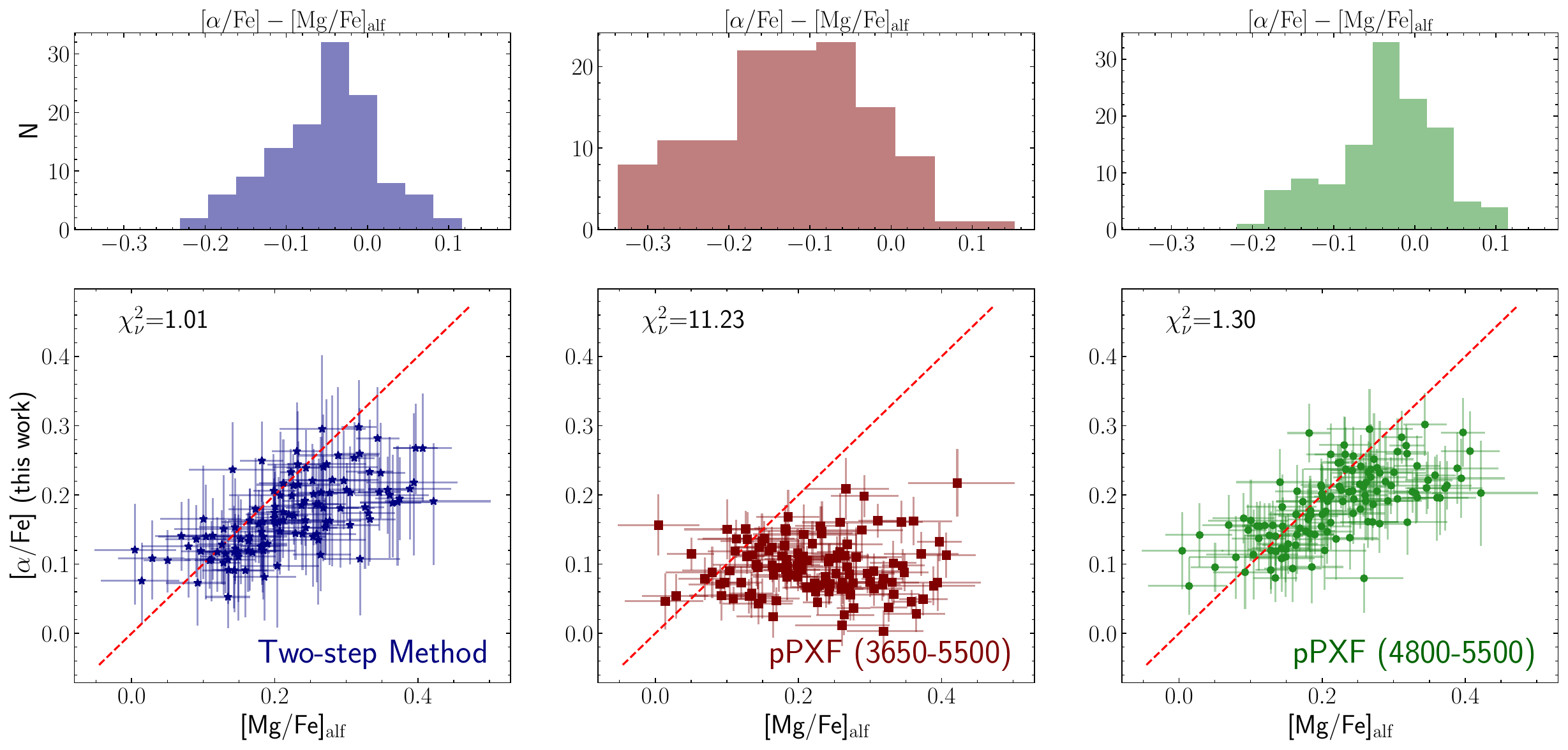}
    \caption{Comparison of [Mg/Fe] for the massive SDSS quiescent galaxies compiled by \citet{Leethochawalit19} measured from \Alf\ by \citet{Zhuang23} and from the methods presented in Section~\ref{subsec:validate_fitting}.
    Here we show the measurements of [$\alpha$/Fe] derived from the two-step method (purple stars) in the left column, those from fitting the entire spectra ($\sim 3650-5500$~\AA\@) via \ppxf\@ (red squares) in the middle column, and those from fitting the spectral regions of $\sim 4800-5500$~\AA\@ (green circles) in the right column. In each column, the top panel shows the difference between the new results and \Alf\ measurements, while the lower panel demonstrates the direct comparison between the measurements. The red dashed lines indicate equal values for both axes. The weighted RMS ($\sigma$) is also shown in each panel. }
    \label{fig:alf_comparison_afe}
\end{figure*}

\begin{figure*}[htb!]
    \centering
    \includegraphics[width=0.95\textwidth]{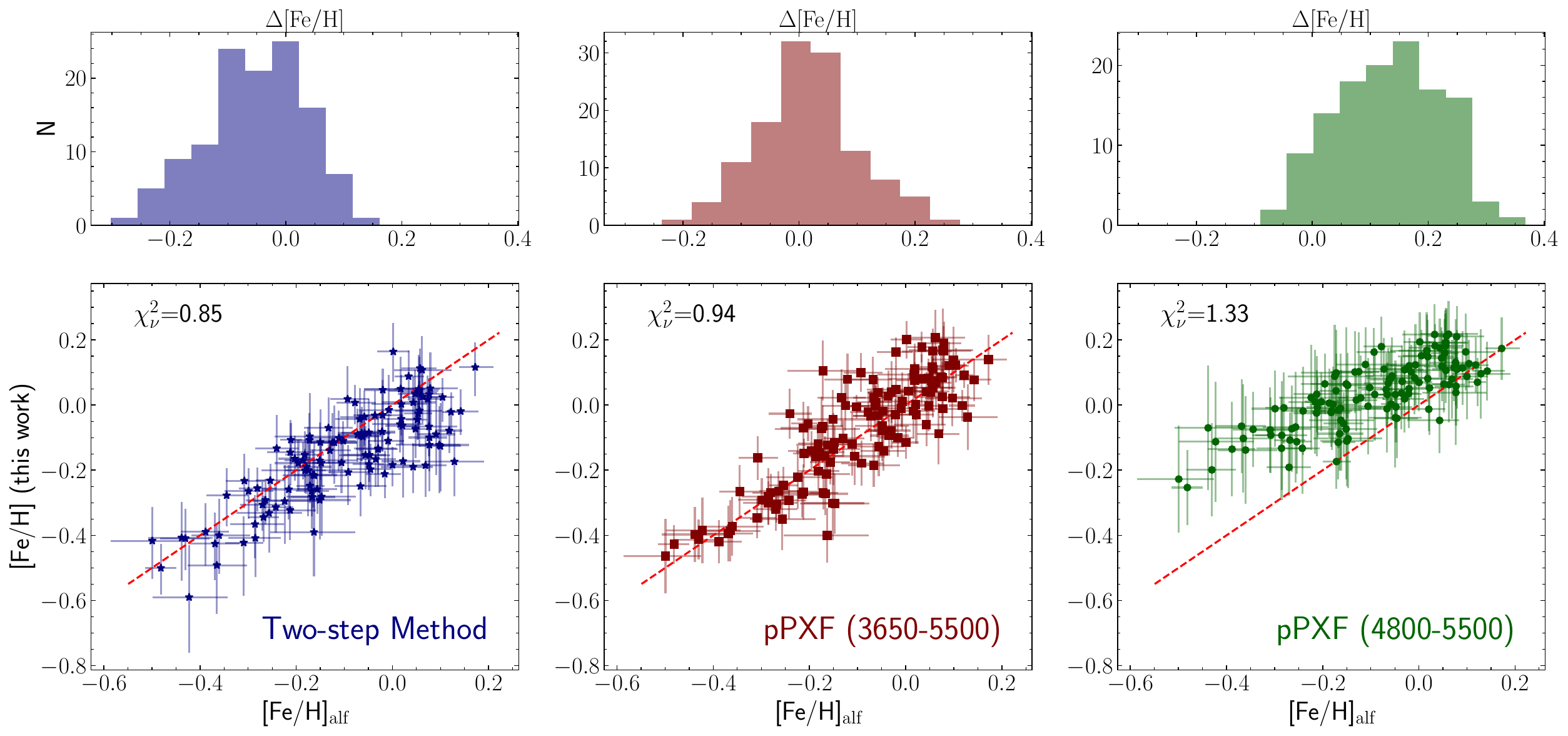}
    \caption{As for Figure~\ref{fig:alf_comparison_afe}, but comparing measurements of [Fe/H]\@.  }
    \label{fig:alf_comparison_feh}
\end{figure*}

Figure~\ref{fig:alf_comparison_afe} compares [$\alpha$/Fe] measured from the two-step and one-step methods with [Mg/Fe] derived by \Alf\ \citep{Zhuang23}; Figure~\ref{fig:alf_comparison_feh} shows the same comparison for [Fe/H]. For each method, we calculated the reduced $\chi^2$ between the \Alf\ results and our new measurements as:
\begin{equation}
    \chi_\nu^2 = \frac{1}{N} \sum_i^N \frac{(m_{i, \rm Alf} - m_{i, \rm X})^2}{\sigma_{i, \rm Alf}^2 + \sigma_{i, \rm X}^2},
\end{equation}
where $m_{i, \rm Alf}$ and $\sigma_{i, \rm Alf}$ indicate the measurements and the uncertainties from \Alf\@, while $m_{i, \rm X}$ and $\sigma_{i, \rm X}$ represent those from the three methods illustrated above.

As can be seen in the middle panels of Figure~\ref{fig:alf_comparison_afe}, [$\alpha$/Fe] derived using \ppxf\ is significantly lower than [Mg/Fe] from \Alf\ for highly $\alpha$-enhanced spectra when we fit the entire spectral range of KCWI ($\sim 3650-5500$~\AA\@), with a $\chi_{\nu}^2=11.2$. On the other hand, the two-step method as well as the \ppxf\ fit over a narrower region of $4800-5500$~\AA, give a much closer estimate of [Mg/Fe], with $\chi_{\nu}^2 = 1.01$ and 1.30, respectively.

It might seem that the single-step fit over a limited spectral range ($4800 - 5500$~\AA\@) performs as well as the two-step approach. However, Figure~\ref{fig:alf_comparison_feh} shows that fitting a narrower region between $4800-5500$~\AA\ overestimates [Fe/H] by $\sim 0.2$~dex on average compared to \Alf\@. The difference can be up to 0.5~dex for the most metal-poor galaxy in the SDSS sample. On the contrary, the two-step method and the \ppxf\ fit of the entire KCWI spectral coverage both give measurements of [Fe/H] consistent with \Alf\, yielding $\chi_{\nu}^2 = 0.85$ and 0.94, respectively. 

\begin{figure*}[htb!]
    \centering
    \includegraphics[width=0.7\textwidth]{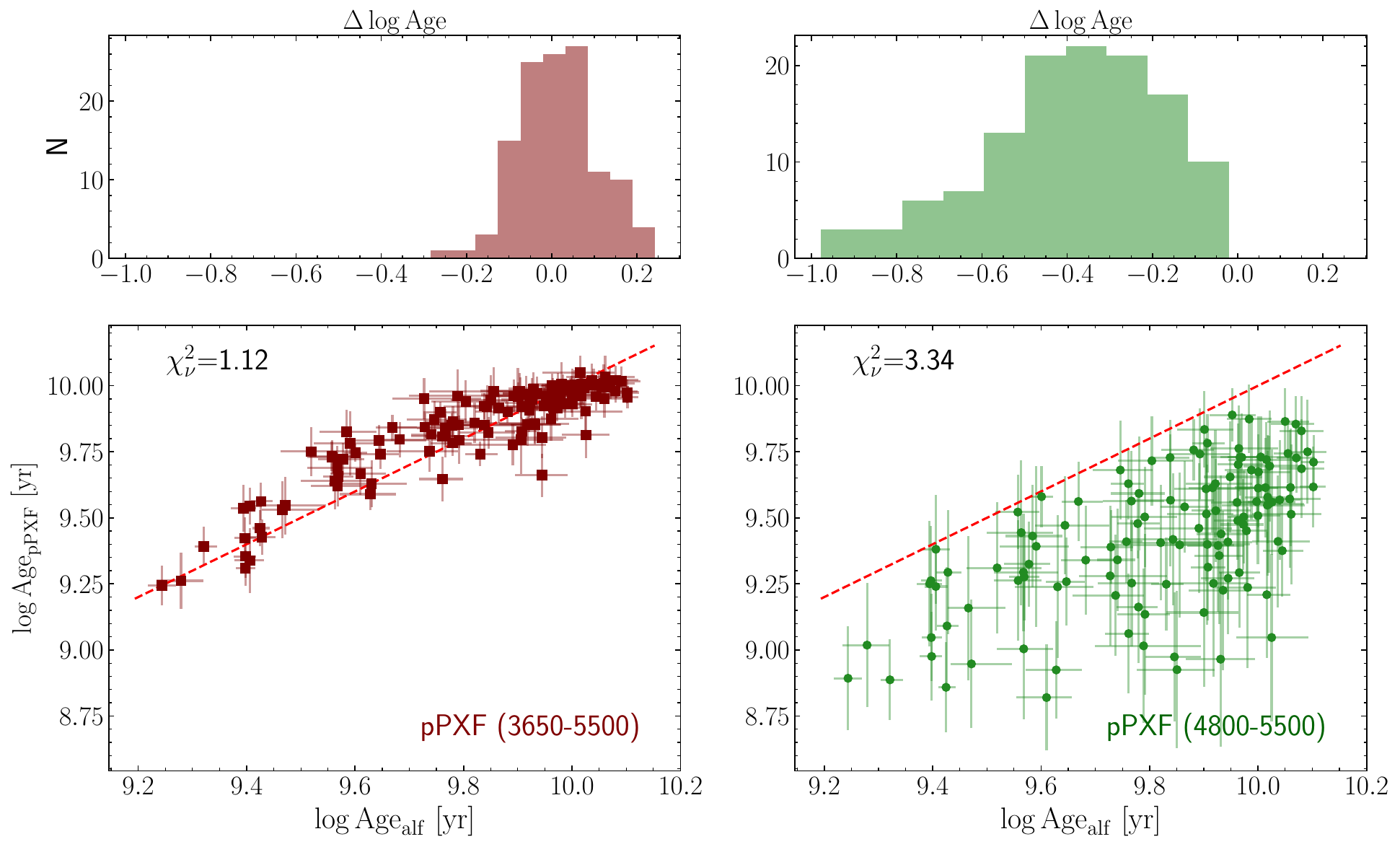}
    \caption{As for Figure~\ref{fig:alf_comparison_afe}, but comparing measurements of stellar population age.}
    \label{fig:alf_comparison_age}
\end{figure*}

The discrepancy can be explained by the fact that most of the spectral features sensitive to the stellar population age and SFH reside near the Balmer/4000~\AA\ breaks, so ignoring them inhibits the breaking of the age--metallicity degeneracy. This effect can be seen in Figure~\ref{fig:alf_comparison_age}, where we compare the light-weighted age derived from \ppxf\ using two different regions with the SSP age determined by \Alf\@. The \ppxf\ fit of the region between 4800~\AA\ and 5500~\AA\ significantly underestimates the stellar population age, while the fit of the full KCWI spectrum recovers consistent age estimates even though \ppxf\ and \Alf\ assume different forms for the SFH\@. The results of the \ppxf\ fit over different regions are in agreement with the direction of the age--metallicity degeneracy: because the models predict younger stellar populations due to a lack of age constraints from NUV and bluer optical features, they have to compensate for the deeper absorption features from older stars by increasing [Fe/H]\@. 

The tests performed above use quiescent galaxies which still have deep H$\beta$ absorption present to constrain stellar population age to some extent when we fit over a narrower range of $4800 - 5500$~\AA\@. This issue would be more severe in low-mass, star-forming galaxies in our KCWI sample because they host younger stellar populations with higher star formation activity. The shallower H$\beta$ absorption and the contamination of the H$\beta$ from the ionized gas would prohibit the single-step fit over a limited range from yielding a reasonable age estimate. We therefore have to use the information from the full spectrum to minimize the age--metallicity degeneracy.  

We note that the $\chi_{\nu}^2$ of the two-step method in the cases of [Fe/H] and [Mg/Fe] are both slightly smaller than that of the best \ppxf\ fit, which may result from larger uncertainties in the two-step method. We emphasize that we adopted the two-step because it is a compromise to recover reasonable value of [Fe/H] and [Mg/Fe] of the star-forming galaxies in our sample. We do not imply that the two-step method is significantly better in measuring [Fe/H] or [Mg/Fe] alone for quiescent galaxies than running \ppxf\ directly. 

To conclude, we have demonstrated above that our novel two-step approach can recover [Mg/Fe] and [Fe/H] consistent with the \Alf\ measurements, even using the SSP models that lock all the $\alpha$ element variation together, while a single-step fit (over either the full KCWI spectral coverage or a portion of the spectrum ($4800-5500$~\AA) carrying most of the Fe- and Mg- sensitive features) via \ppxf\ would underestimate [Mg/Fe] or overestimate [Fe/H]\@. Our two-step method preserves the age constraints from the full spectrum while reducing the contamination from other $\alpha$ elements that might not track Mg. Now, we apply the two-step approach to the KCWI star-forming galaxies to constrain their stellar [Fe/H] and [Mg/Fe]\footnote{In the following analysis, we refer the [$\alpha$/Fe] measured from the two-step method as [Mg/Fe].}. Table~\ref{tab:measurements} lists the stellar abundance measurements derived from the two-step approach described in Section~\ref{subsec:full_spec_fitting}. Table~\ref{tab:measurements} also reports the stellar velocity dispersion and the light-weighted age obtained from the first step of the two-step method, along with the uncertainties determined from 100 bootstraps of the residuals. 

\startlongtable
\begin{deluxetable*}{lccccccc}
\tablecaption{The Derived Properties of the KCWI Star-Forming Galaxies\label{tab:measurements}}
\tablehead{\colhead{Galaxy} & \colhead{$z$} & \colhead{$\log{\rm M_*}$} & \colhead{$\sigma_*$} & \colhead{$\langle \log{\rm Age} \rangle_*$} & \colhead{[Fe/H]$_*$} & \colhead{[Mg/Fe]$_*$} & \colhead{$12+\log{\rm (O/H)_g}$} \\
\colhead{} & \colhead{} & \colhead{[$M_{\odot}$]} & \colhead{km s$^{-1}$} & \colhead{[yr]} & \colhead{dex} & \colhead{dex} & \colhead{dex}
}
\startdata
2225$-$0046 & 0.016 & 8.09$\pm$0.10 & 54.8$\pm$3.5 & 8.96$\pm$0.05 & $-0.88_{-0.11}^{+0.09}$ & $0.09_{-0.06}^{+0.06}$ & 7.90$\pm$0.11 \\
0231$-$0043 & 0.009 & 8.09$\pm$0.10 & 56.7$\pm$3.1 & 8.81$\pm$0.05 & $-1.07_{-0.15}^{+0.14}$ & $0.19_{-0.09}^{+0.09}$ & 7.74$\pm$0.10 \\
0133$-$0109 & 0.016 & 8.11$\pm$0.14 & 61.7$\pm$3.4 & 8.56$\pm$0.08 & $-1.02_{-0.16}^{+0.16}$ & $0.16_{-0.09}^{+0.08}$ & 7.90$\pm$0.10 \\
2329$+$1552 & 0.009 & 8.23$\pm$0.10 & 42.3$\pm$2.2 & 9.07$\pm$0.04 & $-0.72_{-0.08}^{+0.07}$ & $0.18_{-0.05}^{+0.05}$ & 8.24$\pm$0.10 \\
2152$-$0110 & 0.016 & 8.40$\pm$0.10 & 49.4$\pm$4.6 & 9.13$\pm$0.07 & $-1.02_{-0.11}^{+0.10}$ & $0.14_{-0.08}^{+0.08}$ & 8.20$\pm$0.10 \\
0845$+$5145 & 0.014 & 8.42$\pm$0.10 & 64.5$\pm$2.5 & 8.89$\pm$0.06 & $-0.56_{-0.10}^{+0.08}$ & $0.15_{-0.05}^{+0.06}$ & 8.27$\pm$0.10 \\
2334$+$0029 & 0.024 & 8.42$\pm$0.10 & 56.9$\pm$4.6 & 8.70$\pm$0.06 & $-1.10_{-0.17}^{+0.20}$ & $0.10_{-0.09}^{+0.07}$ & 7.97$\pm$0.10 \\
0121$-$0853 & 0.018 & 8.43$\pm$0.10 & 51.0$\pm$2.8 & 8.85$\pm$0.05 & $-0.54_{-0.08}^{+0.09}$ & $0.06_{-0.05}^{+0.04}$ & 8.10$\pm$0.10 \\
0013$-$0956 & 0.013 & 8.51$\pm$0.10 & 44.7$\pm$3.0 & 9.00$\pm$0.04 & $-0.53_{-0.10}^{+0.08}$ & $0.22_{-0.05}^{+0.06}$ & 8.22$\pm$0.11 \\
0155$-$0037 & 0.023 & 8.56$\pm$0.10 & 69.0$\pm$3.3 & 8.93$\pm$0.05 & $-1.16_{-0.11}^{+0.11}$ & $0.08_{-0.07}^{+0.06}$ & 7.92$\pm$0.10 \\
0140$-$0012 & 0.018 & 8.59$\pm$0.10 & 58.0$\pm$2.8 & 8.85$\pm$0.05 & $-0.50_{-0.08}^{+0.06}$ & $0.10_{-0.05}^{+0.05}$ & 8.12$\pm$0.10 \\
0023$-$0939 & 0.02 & 8.59$\pm$0.10 & 55.2$\pm$3.9 & 8.87$\pm$0.06 & $-0.86_{-0.10}^{+0.09}$ & $0.18_{-0.06}^{+0.06}$ & 8.15$\pm$0.10 \\
0233$+$0038 & 0.024 & 8.67$\pm$0.10 & 51.7$\pm$3.3 & 8.92$\pm$0.06 & $-0.87_{-0.11}^{+0.11}$ & $0.14_{-0.07}^{+0.07}$ & 8.25$\pm$0.10 \\
0256$-$0010 & 0.029 & 8.69$\pm$0.10 & 65.0$\pm$3.7 & 8.84$\pm$0.06 & $-1.06_{-0.11}^{+0.10}$ & $0.21_{-0.08}^{+0.08}$ & 8.13$\pm$0.10 \\
2259$-$0019 & 0.016 & 8.71$\pm$0.10 & 49.2$\pm$3.0 & 9.14$\pm$0.06 & $-0.51_{-0.11}^{+0.09}$ & $0.10_{-0.06}^{+0.05}$ & 8.39$\pm$0.10 \\
2300$-$0832 & 0.025 & 8.75$\pm$0.10 & 47.6$\pm$3.1 & 9.05$\pm$0.05 & $-0.82_{-0.06}^{+0.13}$ & $0.08_{-0.07}^{+0.05}$ & 8.16$\pm$0.11 \\
0233$-$0819 & 0.019 & 8.77$\pm$0.10 & 53.7$\pm$2.1 & 8.95$\pm$0.04 & $-0.50_{-0.09}^{+0.06}$ & $0.09_{-0.04}^{+0.04}$ & 8.12$\pm$0.11 \\
0125$-$0024 & 0.016 & 8.78$\pm$0.10 & 50.5$\pm$1.9 & 9.74$\pm$0.04 & $-0.45_{-0.08}^{+0.06}$ & $0.12_{-0.07}^{+0.06}$ & -$^a$ \\
0241$-$0810 & 0.005 & 8.79$\pm$0.10 & 54.7$\pm$1.8 & 9.79$\pm$0.03 & $-0.68_{-0.03}^{+0.04}$ & $0.13_{-0.05}^{+0.05}$ & -$^a$ \\
2301$+$1451 & 0.024 & 8.81$\pm$0.11 & 47.9$\pm$3.5 & 8.87$\pm$0.06 & $-0.77_{-0.13}^{+0.13}$ & $0.09_{-0.07}^{+0.06}$ & 8.19$\pm$0.10 \\
0332$-$0600 & 0.018 & 8.82$\pm$0.10 & 52.7$\pm$2.4 & 9.04$\pm$0.06 & $-0.42_{-0.07}^{+0.06}$ & $0.09_{-0.05}^{+0.04}$ & 8.20$\pm$0.12 \\
2131$-$0613 & 0.025 & 8.82$\pm$0.10 & 57.8$\pm$3.9 & 8.78$\pm$0.07 & $-0.96_{-0.13}^{+0.13}$ & $0.20_{-0.07}^{+0.07}$ & 8.20$\pm$0.10 \\
0850$+$6145 & 0.024 & 8.88$\pm$0.10 & 64.6$\pm$4.8 & 9.06$\pm$0.08 & $-0.69_{-0.11}^{+0.09}$ & $0.10_{-0.06}^{+0.06}$ & 8.35$\pm$0.11 \\
0256$-$0707 & 0.017 & 8.96$\pm$0.10 & 60.3$\pm$2.5 & 8.93$\pm$0.04 & $-0.44_{-0.07}^{+0.06}$ & $0.12_{-0.05}^{+0.05}$ & 8.29$\pm$0.10 \\
0033$+$0012 & 0.014 & 8.97$\pm$0.10 & 61.5$\pm$2.3 & 8.72$\pm$0.04 & $-0.63_{-0.10}^{+0.10}$ & $0.17_{-0.06}^{+0.06}$ & 8.26$\pm$0.10 \\
2137$+$1243 & 0.031 & 8.98$\pm$0.11 & 56.5$\pm$4.4 & 8.95$\pm$0.07 & $-0.83_{-0.11}^{+0.09}$ & $0.19_{-0.07}^{+0.06}$ & 8.26$\pm$0.11 \\
0137$-$0942 & 0.019 & 9.02$\pm$0.10 & 60.3$\pm$2.2 & 8.71$\pm$0.04 & $-0.57_{-0.10}^{+0.07}$ & $0.10_{-0.05}^{+0.05}$ & 8.14$\pm$0.10 \\
0040$-$0017 & 0.019 & 9.02$\pm$0.10 & 50.1$\pm$2.7 & 8.72$\pm$0.04 & $-0.53_{-0.08}^{+0.08}$ & $0.08_{-0.05}^{+0.05}$ & 8.19$\pm$0.10 \\
0136$+$1356 & 0.024 & 9.03$\pm$0.10 & 58.0$\pm$3.3 & 8.60$\pm$0.05 & $-0.62_{-0.11}^{+0.09}$ & $0.09_{-0.06}^{+0.06}$ & 8.26$\pm$0.10 \\
2304$-$0933 & 0.032 & 9.04$\pm$0.10 & 60.0$\pm$2.4 & 8.84$\pm$0.05 & $-0.73_{-0.09}^{+0.08}$ & $0.21_{-0.06}^{+0.07}$ & 8.29$\pm$0.10 \\
0309$-$0041 & 0.037 & 9.05$\pm$0.10 & 66.9$\pm$3.9 & 8.72$\pm$0.06 & $-0.84_{-0.12}^{+0.11}$ & $0.12_{-0.06}^{+0.06}$ & 8.28$\pm$0.10 \\
0019$-$0004 & 0.04 & 9.15$\pm$0.10 & 69.9$\pm$3.9 & 8.74$\pm$0.05 & $-0.78_{-0.13}^{+0.13}$ & $0.09_{-0.07}^{+0.07}$ & 8.37$\pm$0.10 \\
0208$+$0048 & 0.02 & 9.15$\pm$0.10 & 46.0$\pm$3.1 & 8.93$\pm$0.05 & $-0.65_{-0.08}^{+0.08}$ & $0.15_{-0.05}^{+0.05}$ & 8.36$\pm$0.10 \\
0306$-$0033 & 0.025 & 9.16$\pm$0.10 & 49.0$\pm$2.6 & 9.04$\pm$0.04 & $-0.54_{-0.10}^{+0.08}$ & $0.17_{-0.05}^{+0.05}$ & 8.46$\pm$0.10 \\
0346$+$0023 & 0.031 & 9.18$\pm$0.10 & 64.2$\pm$3.5 & 8.81$\pm$0.06 & $-0.89_{-0.12}^{+0.11}$ & $0.21_{-0.08}^{+0.07}$ & 8.31$\pm$0.10 \\
0243$+$0032 & 0.028 & 9.34$\pm$0.10 & 49.4$\pm$2.9 & 8.75$\pm$0.06 & $-0.70_{-0.10}^{+0.10}$ & $0.20_{-0.06}^{+0.06}$ & 8.34$\pm$0.10 \\
2318$+$0105 & 0.03 & 9.35$\pm$0.10 & 65.4$\pm$3.0 & 8.82$\pm$0.05 & $-0.71_{-0.08}^{+0.08}$ & $0.22_{-0.05}^{+0.06}$ & 8.36$\pm$0.11 \\
2206$-$0041 & 0.029 & 9.37$\pm$0.16 & 65.9$\pm$2.6 & 8.43$\pm$0.06 & $-0.72_{-0.13}^{+0.13}$ & $0.22_{-0.09}^{+0.09}$ & 8.34$\pm$0.10 \\
2243$+$0040 & 0.039 & 9.37$\pm$0.10 & 57.7$\pm$4.8 & 8.76$\pm$0.05 & $-0.83_{-0.10}^{+0.09}$ & $0.13_{-0.06}^{+0.06}$ & 8.25$\pm$0.10 \\
0334$+$0106 & 0.049 & 9.44$\pm$0.10 & 77.4$\pm$4.7 & 8.13$\pm$0.07 & $-0.35_{-0.12}^{+0.10}$ & $0.12_{-0.05}^{+0.06}$ & 8.32$\pm$0.10 \\
0317$-$0004 & 0.022 & 9.46$\pm$0.10 & 55.4$\pm$2.3 & 8.62$\pm$0.05 & $-0.58_{-0.09}^{+0.09}$ & $0.15_{-0.05}^{+0.05}$ & 8.32$\pm$0.10 \\
2211$+$0006 & 0.057 & 9.58$\pm$0.10 & 50.3$\pm$4.6 & 8.43$\pm$0.09 & $-0.43_{-0.11}^{+0.12}$ & $0.17_{-0.06}^{+0.06}$ & 8.47$\pm$0.10 \\
0245$-$0045 & 0.054 & 9.60$\pm$0.10 & 55.9$\pm$3.5 & 8.49$\pm$0.06 & $-0.35_{-0.10}^{+0.10}$ & $0.13_{-0.06}^{+0.06}$ & 8.38$\pm$0.10 \\
0030$-$0042 & 0.041 & 9.63$\pm$0.10 & 63.8$\pm$2.2 & 8.63$\pm$0.04 & $-0.53_{-0.09}^{+0.09}$ & $0.17_{-0.05}^{+0.05}$ & 8.56$\pm$0.10 \\
2139$+$0019 & 0.05 & 9.72$\pm$0.10 & 66.3$\pm$3.6 & 8.58$\pm$0.08 & $-0.83_{-0.21}^{+0.20}$ & $0.18_{-0.09}^{+0.09}$ & 8.51$\pm$0.10 \\
0407$-$0634 & 0.038 & 9.88$\pm$0.10 & 59.9$\pm$3.6 & 9.13$\pm$0.07 & $-0.18_{-0.09}^{+0.08}$ & $0.09_{-0.05}^{+0.05}$ & 8.46$\pm$0.11 \\
0254$+$0103 & 0.043 & 9.90$\pm$0.10 & 71.2$\pm$3.3 & 8.61$\pm$0.06 & $-0.82_{-0.13}^{+0.13}$ & $0.20_{-0.07}^{+0.08}$ & 8.45$\pm$0.10 \\
0140$-$0013 & 0.058 & 10.02$\pm$0.10 & 76.4$\pm$2.5 & 8.57$\pm$0.04 & $-0.40_{-0.10}^{+0.09}$ & $0.21_{-0.05}^{+0.05}$ & 8.48$\pm$0.10
\enddata
\tablecomments{The columns show (1) the galaxy ID; (2-3) the spectroscopic redshift and the spectroscopic redshift from Table~\ref{tab:galaxy_prop}; (4-5) the stellar velocity dispersion and light-weighted ages measured from the \ppxf\ fit of the entire KCWI spectra (Sections~\ref{subsec:full_spec_fitting} and \ref{subsec:validate_fitting}); (6-7) the [Fe/H]$_*$ and [Mg/Fe]$_*$ estimated from the two-step method (Section~\ref{subsec:full_spec_fitting}); (8) the gas-phase oxygen abundance obtained from the $R$ calibration by \citet{Pilyugin16} described in Section~\ref{subsec:gas_met}.}
\tablenotetext{a}{The absence of emission lines in the quiescent galaxies prevent us from determining their gas-phase metallicities.}
\end{deluxetable*}

\section{Results}\label{sec:results}




\subsection{Comparison between stellar and gas-phase abundances}

We now compare the ionized gas-phase oxygen abundances with the stellar abundances of iron and magnesium for our sample. For a more direct comparison, we convert the gas-phase metallicity $\rm 12+\log{(O/H)}$ into [O/H]$_{\rm g}$ using the solar oxygen abundance of $\rm 12+\log{(O/H)_{\odot}} = 8.69$ \citep{Asplund09}. 

\begin{figure*}[htb!]
    \centering
    \includegraphics[width=0.95\textwidth]{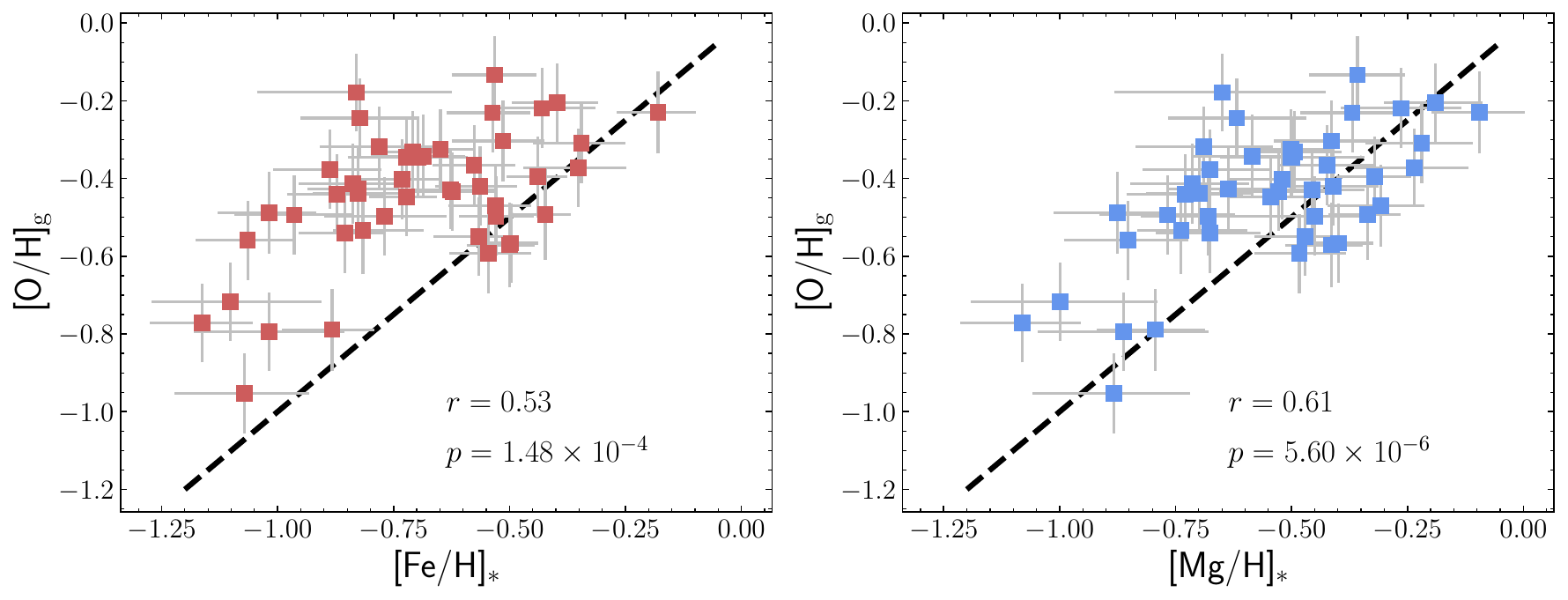}
    \caption{Comparison of [O/H]$_{\rm g}$ with [Fe/H]$_*$ (left) and [Mg/H]$_*$ (right). The dashed-dotted lines indicate the one-to-one relation (i.e., $\rm [O/H]_g=[X/H]_*$). The weighted Pearson coefficients $r$ and the $p$-values of the null hypothesis are shown in each panel. }
    \label{fig:stellar_gas_abund}
\end{figure*}

Figure~\ref{fig:stellar_gas_abund} shows the comparison between the gas-phase oxygen abundance [O/H]$_{\rm g}$ and the stellar iron abundance [Fe/H]$_*$ of our sample. We derive a inverse-variance weighted Pearson coefficient $r=0.53$ and the corresponding $p$-value $p=1.48\times 10^{-4}$, indicating that the [O/H]$_{\rm g}$ and the [Fe/H]$_*$ are correlated at the $\sim 3.8\sigma$ level. Four galaxies exhibit lower [O/H]$_{\rm g}$ than [Fe/H]$_*$, but at $< 2\sigma$ significance. The remaining galaxies in our sample exhibit higher [O/H]$_{\rm g}$ than [Fe/H]$_*$. On average, [O/H]$_{\rm g}$ is higher than [Fe/H]$_*$ by 0.25~dex, in agreement with the results of \citet{Fraser-McKelvie22} for a sample of more massive star-forming galaxies at $M* > 10^9M_{\odot}$ observed by the SAMI Galaxy Survey.
The discrepancy is not surprising, given that the two elemental abundances trace the chemical enrichment of different stellar populations. The [Fe/H]$_*$ presented here is measured from the rest-frame optical stellar continuum, so the measured [Fe/H]$_*$ indicates the chemical enrichment for older stellar populations. On the contrary, the [O/H]$_{\rm g}$ is a proxy for the chemical enrichment in regions where the youngest OB stars have just formed out of the ISM\@. In addition, the different formation timescales of oxygen and iron can also contribute to the abundance offset. Iron is primarily formed by Type Ia SNe that explode $\gtrsim 300-400$~Myr after the initial starburst, while oxygen, an $\alpha$ element, is produced by core-collapse SNe of massive stars a $\sim$ 10 Myr timescale. As a result, the chemical enrichment of iron is expected to lag behind that of oxygen, especially for galaxies that have formed the majority of their stars in the past Gyr before the Type Ia SNe began exploding, although the difference would be attenuated if galaxies have smoother and more sustained SFHs. The combined effects of the distinct stellar populations probed by oxygen and iron and the disparate formation timescales can naturally explain the lower [Fe/H]$_*$ in Figure~\ref{fig:stellar_gas_abund}.

As mentioned above, it is complicated to interpret the relation between [Fe/H]$_*$ and [O/H]$_{\rm g}$ because Fe is delayed relative to O\@. On the other hand, comparing the abundance of gas-phase oxygen and stellar magnesium is a more direct comparison between the metals in the gas and stars because magnesium is another $\alpha$ element produced by the same mechanism as oxygen.  [Mg/H]$_*$ traces the SFH-averaged stellar abundance and can be interpreted as an approximation of the gas-phase abundance averaged over the SFH of the galaxies.

In the right panel of Figure~\ref{fig:stellar_gas_abund}, we compare [O/H]$_{\rm g}$ and [Mg/H]$_*$, which is the sum of the [Fe/H]$_*$ and [Mg/Fe]$_*$ derived in Section~\ref{subsec:full_spec_fitting}. [Mg/H]$_*$ is still lower than [O/H]$_{\rm g}$, but as expected, the differences between [O/H]$_{\rm g}$ and [Mg/H]$_*$ (0.11~dex) are significantly smaller compared to [Fe/H]$_*$ (0.25~dex). The Pearson coefficient of the [O/H]$_{\rm g}$--[Mg/H]$_*$ relation ($r=0.61$) is higher than that of the [O/H]$_{\rm g}$--[Fe/H]$_*$ with a $p$-value lower by two orders of magnitude, suggesting that [O/H]$_{\rm g}$ is more linearly correlated with [Mg/H]$_*$.
Among 46 star-forming galaxies in our sample, 23 of them have consistent [O/H]$_{\rm g}$ and [Mg/H]$_*$ within $1\sigma$, whereas 19 galaxies have slightly higher [O/H]$_{\rm g}$ than [Mg/H]$_*$ at the $1-3\sigma$ level. We note that we measure [O/H]$_{\rm g}$ using strong-line calibrations based on the $T_e$ metallicity scale, which is likely to underestimate the actual $\rm (O/H)_{gas}$ due to temperature fluctuations in the H$\,${\sc ii} regions \citep{Garcia-Rojas07}. If we convert the $T_e$-based measurements to the abundance scale based on recombination lines and photoionization models by adding 0.24~dex to the measured [O/H]$_{\rm g}$ \citep{Esteban14, Blanc15, Steidel16}, all star-forming galaxies would have higher [O/H]$_{\rm g}$ than [Mg/H]$_*$.  
Most of (or all) galaxies in our sample therefore have higher [O/H]$_{\rm g}$ because the youngest populations are more chemically enriched. Nevertheless, the reduced lag between [O/H]$_{\rm g}$ and  [Mg/H]$_*$ arises from the fact that the two elements are produced by the same process, rendering them less sensitive to the shapes of the SFHs.

Even though we have demonstrated that stellar magnesium indeed tracks gas-phase oxygen more closely than stellar iron does, we emphasize that the [Mg/H]$_*$ and the [O/H]$_{\rm g}$ are not fully consistent for all galaxies in our sample as discussed above. One would always expect that [O/H]$_{\rm g}$ is higher than [Mg/H]$_*$ unless star formation is occurring in metal-poor gas from external accretion.  
Therefore, one should be cautious when using [O$_{\rm g}$/Fe$_*$] as a substitute for [$\alpha$/Fe]$_*$ in star-forming galaxies if [Fe/H]$_*$ is determined from the rest-frame optical spectrum\footnote{[Fe/H]$_*$ determined from the FUV stellar continuum is sensitive mostly to young OB stars. [O/Fe] is commonly used in high-$z$ galaxies when their rest-FUV stellar continuum can be easily obtained from ground- or space-based telescopes.}. In this case, using the difference of [O/H]$_{\rm g}$ and [Fe/H]$_*$ as a proxy for [$\alpha$/Fe]$_*$ may overestimate [Mg/Fe]$_*$.



\subsection{The gas-to-stellar abundance ratio as a function of galaxy properties }
\begin{figure*}[htb!]
        \centering
        \includegraphics[width=0.95\textwidth]{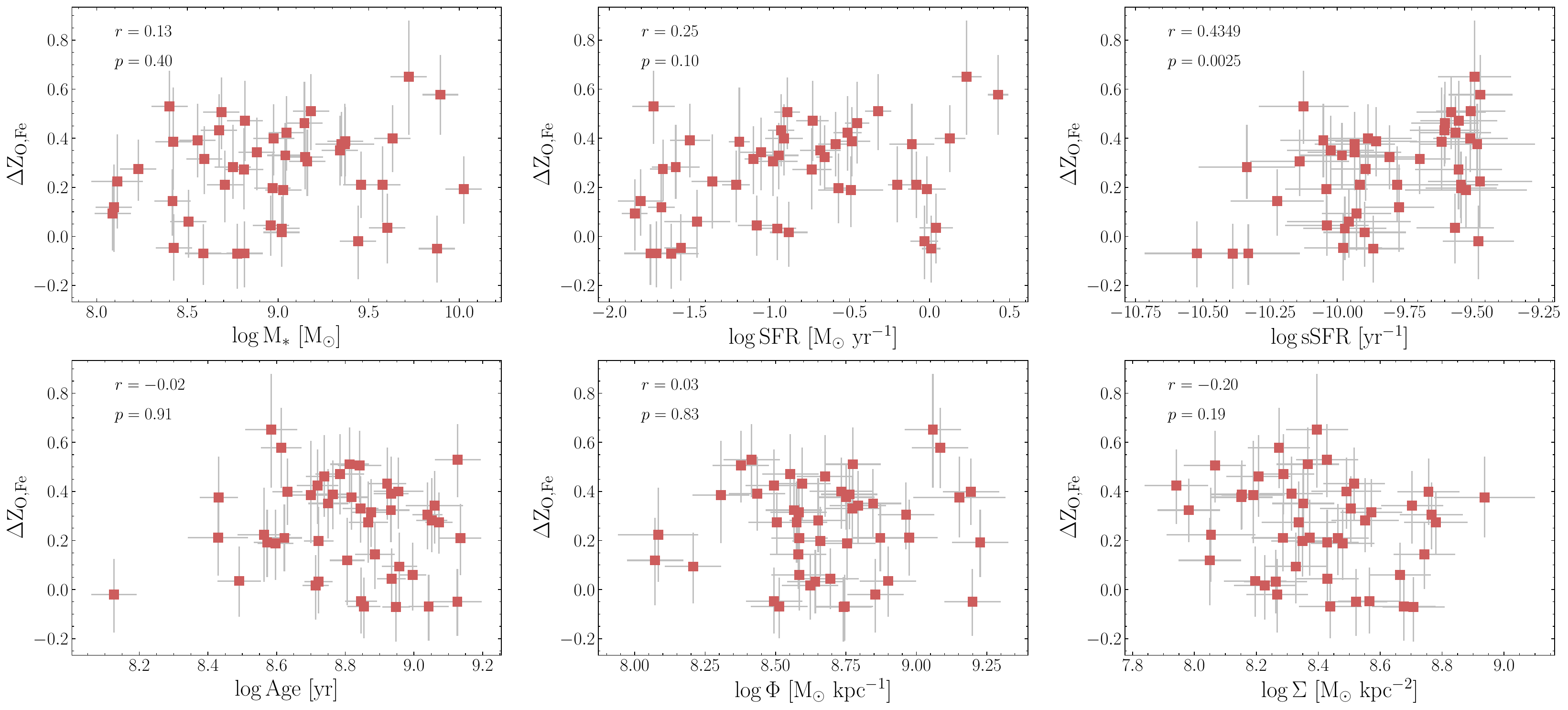}
        \caption{The differences between [O/H]$_{\rm g}$ and [Fe/H]$_*$ ($\Delta Z_{\rm O,Fe}$) as a function of different galaxy properties, including stellar mass (top left), SFR (top middle), sSFR (top right), light-weighted stellar population age (bottom left), $\log{\Phi}$ (bottom middle) and $\log{\Sigma}$ (bottom right). The weighted Pearson coefficients ($r$) and the $p$-values are shown in each panel.}
        \label{fig:ofe_offset_galprop}
\end{figure*}

\begin{figure*}[htb!]
    \centering
    \includegraphics[width=0.95\textwidth]{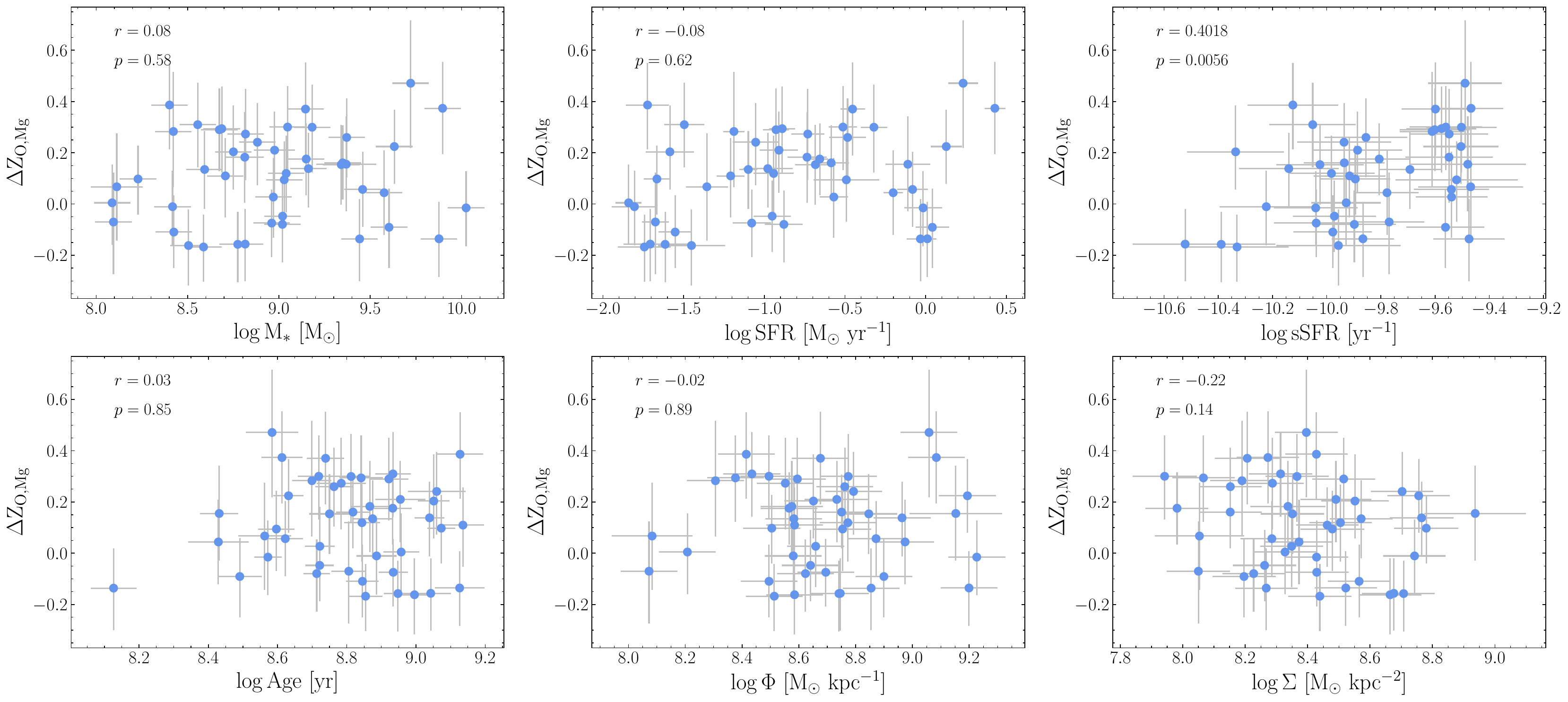}
    \caption{Similar to Figure~\ref{fig:ofe_offset_galprop}, but for the differences between [O/H]$_{\rm g}$ and [Mg/H]$_*$ ($\Delta Z_{\rm O,Mg}$). }
    \label{fig:fig2}
        \label{fig:omg_offset_galprop}
\end{figure*}

Aiming to understand the connection between metals in different phases, we explore how the gas-to-stellar abundance ratios (more accurately, their logarithmic difference), $\Delta Z_{\rm O, Fe}$ and  $\Delta Z_{\rm O, Mg}$ depend on different galaxy properties in Figures~\ref{fig:ofe_offset_galprop} and \ref{fig:omg_offset_galprop}, respectively. We define two gas-to-stellar abundance ratios as:
\begin{subequations}\label{eq:gas_to_stellar_abund}
\begin{equation}
  \Delta Z_{\rm O, Fe} = \rm [O/H]_{gas} - [Fe/H]_*
\end{equation}    
\begin{equation}
  \Delta Z_{\rm O, Mg} = \rm [O/H]_{gas} - [Mg/H]_*
\end{equation}
\end{subequations}


We explore the dependence of these quantities on six galaxy properties: (1) stellar mass, (2) SFR, (3) sSFR, (4) light-weighted age derived from the full-spectrum fitting, (5) $\log{\Phi} = \log{(M_*/R_e)}$, a proxy for the gravitational potential well, and (6) $\log{\Sigma} = \log{(M_*/R_e^2)}$, a proxy for stellar mass surface density. The properties are chosen because they are known to be correlated/anti-correlated either with stellar/gas abundances \citep[e.g.,][]{Tremonti04, Gallazzi05, Mannucci10}, or with the gas-to-stellar abundance ratios for more massive star-forming galaxies \citep[e.g.,][]{Fraser-McKelvie22}. 
To quantify whether these galaxy properties are correlated with $\Delta Z_{\rm O, Fe}$ and $\Delta Z_{\rm O, Mg}$, we calculate inverse variance-weighted Pearson coefficients ($r$) and corresponding $p$-values for each relation. 

Among all the relations investigated, we find that $\Delta Z_{\rm O, Fe}$ and $\Delta Z_{\rm O, Mg}$ are significantly correlated (i.e., $p$-value $< 0.05$) only with sSFR. A similar trend was also found by \citet{Fraser-McKelvie22}, who compared the stellar and gas-phase metallicity of the star-forming galaxies at $\log{(M_*/M_{\odot})} \gtrsim 10^{8.5}$ in the SAMI sample. Because \citet{Fraser-McKelvie22} did not measure [Mg/H]$_*$, they converted the measured gas-phase metallicity in the oxygen-based scale into the iron-based scale using the empirical abundances in the Milky Way, Large Magellanic Cloud (LMC), and Small Magellanic Cloud (SMC) by \citet{Nicholls17}.  This conversion was intended to account for the difference in the recycling time of oxygen and iron. In contrast, our analysis allows for an apples-to-apples comparison of stellar and gas-phase abundances of $\alpha$ elements, without relying on abundance patterns based on specific galaxies. 

One might expect that  $\Delta Z_{\rm O, Fe}$ and $\Delta Z_{\rm O, Mg}$ would be correlated with sSFR\@. 
$\Delta Z_{\rm O, Fe}$ and $\Delta Z_{\rm O, Mg}$ reflect the differences
in the level of chemical enrichment between that of the current ISM and that obtained when the bulk of the stars that dominate the current optical spectrum were forming, i.e., in the past, weighted roughly by light. The measured stellar abundances are therefore a proxy for the ISM abundances around $600-700$~Myr (the typical light-weighted age in this sample) ago. A galaxy with higher sSFR would have experienced a larger increase in ISM abundance over the same interval of time, leading to a larger value of $\Delta Z_{\rm O, Fe}$ and $\Delta Z_{\rm O, Mg}$.
While the correlation between $\Delta Z_{\rm O, Fe}$ and sSFR ($r=0.43$) is slightly stronger than that between $\Delta Z_{\rm O, Mg}$ and sSFR ($r=0.40$), the differences are not significant.



Perhaps surprisingly, $\Delta Z_{\rm O, Fe}$ and $\Delta Z_{\rm O, Mg}$ lack a strong dependence on SFR (see Figures~\ref{fig:ofe_offset_galprop} and \ref{fig:omg_offset_galprop}). It is well known that at a given stellar mass, galaxies with higher SFRs tend to have more metal-poor ISM \citep{Mannucci10} because their high gas fractions dilute the current gas-phase metallicity. However, [Fe/H]$_*$ and [Mg/H]$_*$, the amount of metals locked in the stars averaged over the entire SFH, are expected to be much less influenced by the current SFR because they have only a weak dependence on the current star formation. If there is a universal ``fundamental'' MZR of galaxies, where the normalization of the gas MZR decreases as SFR goes up, $\Delta Z_{\rm O, Fe}$ and $\Delta Z_{\rm O, Mg}$ should be correlated with SFR\@. The absence of dependence on SFR here implies that the stellar MZR is not universal for galaxies at different SFRs, as we will further discuss in Section~\ref{subsec:kcwi_mzr}.

We do not detect significant correlations of $\Delta Z_{\rm O, Fe}$ or $\Delta Z_{\rm O, Mg}$ with stellar mass, $\log{\Phi}$, or $\log{\Sigma}$, while \citet{Fraser-McKelvie22} found that the gas-to-stellar abundance ratio is anti-correlated with mass and $\log{\Phi}$ and uncorrelated with $\log{\Sigma}$ after correcting the abundance scale offsets between iron and oxygen. The discrepancy may be attributed to the different mass range  in the current sample as compared to that of \citet{Fraser-McKelvie22} -- the two anti-correlations found by \citet{Fraser-McKelvie22} are driven primarily by galaxies with $\log{M_*/M_{\odot}} \gtrsim 10^{9.5}$ and $\log{\Phi} \gtrsim 9.25$ (their Figure 10), whereas our sample consists of much less massive galaxies. It could be that the trends with mass and $\log{\Phi}$ have not yet been established in dwarf galaxies. 

Finally, we find that both $\Delta Z_{\rm O, Fe}$ and $\Delta Z_{\rm O, Mg}$ are uncorrelated with the light-weighted age. We suspect that this is because the star-forming galaxies in our sample are extremely young with light-weighted ages $< 1$~Gyr.  It could be that the trends with age have not yet been established.

\subsection{The Mass--Metallicity Relation}\label{subsec:kcwi_mzr}

\begin{figure*}[t]
    \centering    \includegraphics[width=0.47\textwidth]{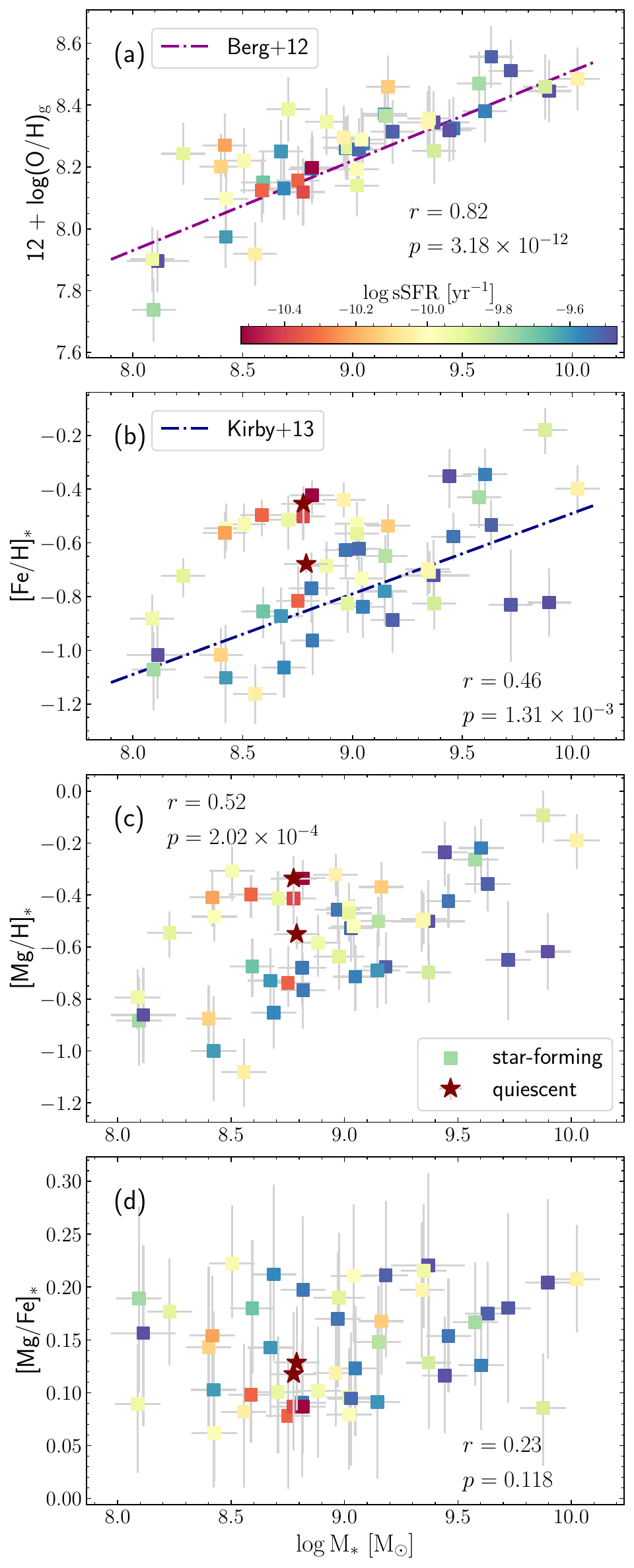}
\includegraphics[width=0.47\textwidth]{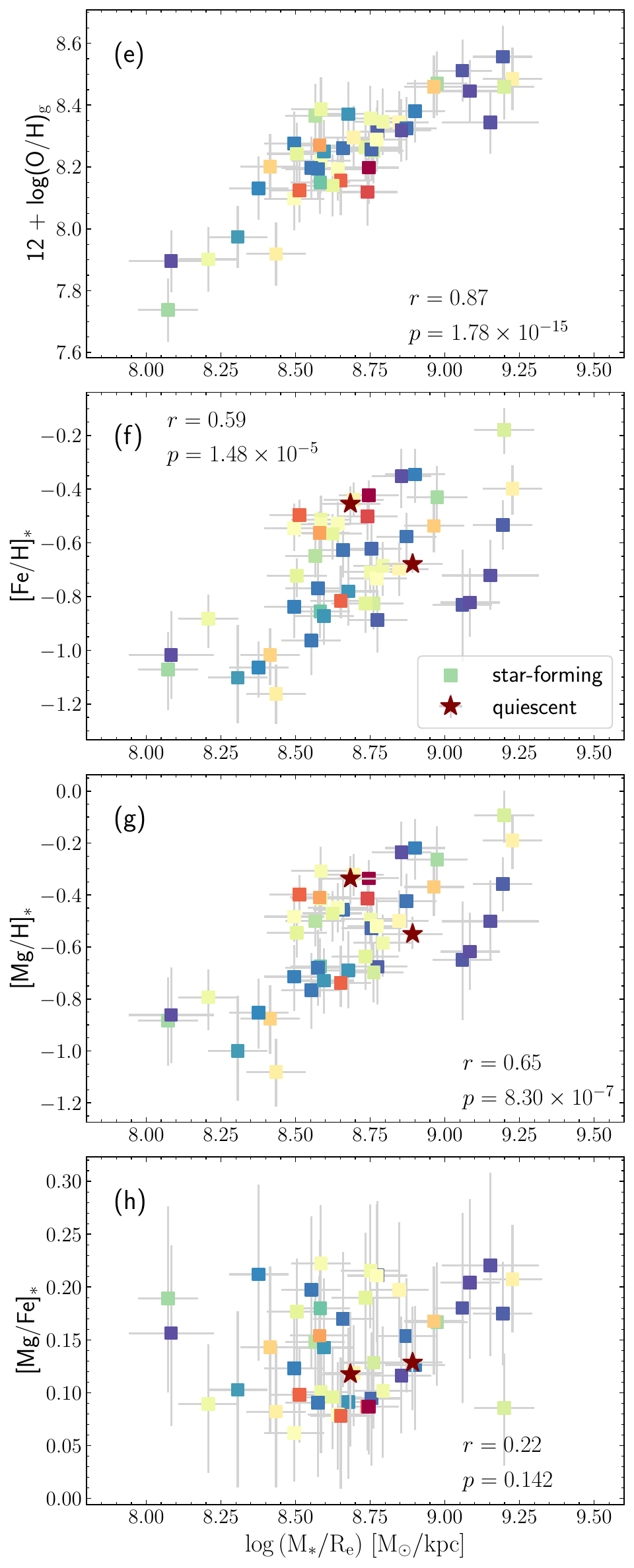}
    \caption{\textbf{Left}: The O-MZ$_{\rm g}$R (a), Fe-MZ$_*$R (b), Mg-MZ$_*$R (c), and the mass-[Mg/Fe]$_*$ relation (d) of the KCWI sample. In panel (a), the magenta dashed-dotted line shows the O-MZ$_{\rm g}$R of local dwarf galaxies derived from the direct method \citep{Berg12}. In panel (b), the blue dashed-dotted line shows the Fe-MZ$_*$R of Local Group satellites obtained from resolved stellar spectroscopy \citep{Kirby13}.
    \textbf{Right}: The gas-phase (O/H)$_{\rm g}$ (e), stellar [Fe/H]$_*$ (f), stellar [Mg/H]$_*$ (g), and [Mg/Fe]$_*$ (h) as a function of $\log{\Phi} = \log{(M_*/R_e)}$ of the KCWI sample.
    In each panel, the star-forming galaxies (squares) are color-coded by their sSFRs. The two quiescent galaxies are shown by the dark red stars. The weighted Pearson coefficients $r$ and the $p$-values of the null hypothesis are listed in each panel.    }
    \label{fig:kcwi_mzr}
\end{figure*}

The left column of Figure~\ref{fig:kcwi_mzr} shows the gas-phase O-MZ$_{\rm g}$R, stellar Fe-MZ$_*$R, and stellar Mg-MZ$_*$R along with the mass--[Mg/Fe]$_*$ relation color-coded by their sSFRs in panels (a) - (d), respectively. For each relation, we calculated the weighted Pearson coefficient and the $p$-value of the null hypothesis to quantify their linearity. Nevertheless, there is no theoretical reason to expect the MZRs should be linear across this mass range \citep[e.g.,][]{Ma16}, as simple linear relation do not capture the flatness of the MZ$_*$R as we later discuss in Section~\ref{subsec:full_mzr}. We pick the Pearson coefficient only for its simplicity to describe this small data set. Overall, we recover significant correlations between stellar mass and abundances of three elements with $p < 0.05$, but do not detect significant correlation between stellar mass and [Mg/Fe]$_*$\@. Among the three MZRs, we find that the O-MZ$_{\rm g}$R is the most linear relation with $R=0.82$. The Mg-MZ$_*$R ($R=0.52$) is more linear than Fe-MZ$_*$R ($R=0.46$). In panel (d), the dwarf quiescent galaxies as well as the star-forming galaxies with the lowest sSFRs in our sample have lower [Mg/Fe]$_*$ than the rest of our sample. The low [Mg/Fe]$_*$ in these galaxies that are already or almost quenched indicate that their past star formation may have lasted for a protracted period of time, as the star formation there would have to occur in gas whose ISM is already enriched in iron by Type Ia SNe. The [Mg/Fe]$_*$ of the star-forming galaxies with higher sSFRs show large variations in our sample. There is no reason to expect their [Mg/Fe]$_*$ to be related to stellar mass because these galaxies are still actively forming stars with evolving [Mg/Fe]$_*$.

We also compare our O-MZ$_{\rm g}$R and Fe-MZ$_*$R with the literature on the local dwarf galaxies in Figure~\ref{fig:kcwi_mzr}. The O-MZ$_{\rm g}$R of the star-forming galaxies in our KCWI sample in panel (a) exhibits a relatively tight linear relation and is consistent with the O-MZ$_{\rm g}$R derived from the direct $T_e$ method by \citet{Berg12} for local dwarf galaxies at $M_* \sim 10^{6-9.7}M_{\odot}$. Our Fe-MZ$_{*}$R in panel (b), instead, does not completely fall on the extrapolation of the low-mass Fe-MZ$_{*}$R of the Local Group satellites at $M_*\lesssim 8.5M_{\odot}$ \citep{Kirby13}. Most of the star-forming galaxies with high sSFRs fall on the extrapolation of the low-mass MZ$_*$R of Local Group dwarf satellite galaxies measured by \citet{Kirby13}, while most of the star-forming galaxies with higher [Fe/H]$_{*}$ at fixed mass that lie above the extrapolation of the low-mass MZ$_*$R have lower sSFRs. The shape of the Mg-MZ$_*$R in panel (c) is very similar to the Fe-MZ$_*$R\@. The disagreement between our Fe-MZ$_{*}$R and the relation of \citet{Kirby13} is not surprising because many physical processes such as outflows are changing rapidly with galaxy mass.  Furthermore, theoretical predictions do not necessarily predict linear MZRs \citep[e.g.,][]{Ma16}. For instance, \citet{Dave12} suggested that the slope of the MZR strongly depends on the primary feedback processes regulating galaxy growth and driving outflows. They found that the slope of the simulated MZR can be changed easily by altering the mass dependence of the mass-loading factor of the outflows alone. Because star-forming galaxies at $10^8M_{\odot} \lesssim M* \lesssim 10^{10}M_{\odot}$ are likely to different feedback processes than that of the Local Group satellites, we do not expect Fe-MZ$_{*}$Rs in this mass range to be consistent with the relation by \citet{Kirby13}.

It has been controversial whether star-forming and quiescent galaxies should exhibit distinct MZ$_*$Rs. For the dwarf satellite galaxies ($\log{(M_*/M_{\odot})} \lesssim 8.5)$ in the Local Group, \citet{Kirby13} found that star-forming and quiescent dwarf galaxies in the Local Group fall on a universal MZ$_*$R and have similar [Fe/H]$_*$ at a given mass. However, large spectroscopic surveys such as SDSS, MaNGA and SAMI of massive field galaxies ($\log{(M_*/M_{\odot})} \gtrsim 9)$ suggest that star-forming galaxies generally have more metal-poor stars than quiescent galaxies at the same masses \citep[e.g.,][]{Gallazzi21, Neumann21, Vaughan22}. When we look at the two quiescent galaxies in our sample on the Fe-MZ$_*$Rs, we find that only one of them, the more metal-poor 0241-0810 has similar [Mg/H]$_*$ and [Fe/H]$_*$ as those of heavily star-forming galaxies at the same mass. The other quiescent galaxy 0125-0024 that is more metal-rich, instead, has comparable [Fe/H]$_*$ and [Mg/H]$_*$ to the KCWI star-forming galaxies with suppressed sSFRs (the yellow to orange squares in Figure~\ref{fig:kcwi_mzr}). To confirm that the two quiescent galaxies indeed possess different metal content, we also measured their [Fe/H]$_*$ and [Mg/H]$_*$ using \Alf\ by \citet{Conroy18}, and we recovered the same results to within $1\sigma$.

The distinct [Fe/H]$_*$ and [Mg/H]$_*$ of the two quiescent galaxies may be explained by their different environments. 0241-0810 is located in the vicinity of the massive elliptical galaxy NGC 1052, and it has similar radial velocity to that of NGC 1052 \citep{Roman21}. The projected distance between 0241-0810 and NGC 1052 is around $\sim$ 50-60~kpc, while the projected virial radius of NGC 1052 is 390~kpc \citep{Forbes19}, suggesting that 0241-0810 is likely to be a satellite of NGC 1052. The overdense environment where 0241-0810 resides might explain why it is consistent with the low-mass MZ$_*$R of the Local Group dwarfs. In particular, galaxies in dense environments can be rapidly quenched by external effects, like ram pressure stripping.  Therefore, a galaxy that would have naturally evolved to a higher normalization on the MZ$_*$R may have been prevented from doing so due to a sudden cessation of star formation.  On the other hand, 0125-0024 is an isolated dwarf galaxy identified by \citet{Kado-Fong20} with no massive galaxies nearby. This galaxy does not fall on the Fe-MZ$_*$R of the Local Group dwarfs, possibly because it evolved naturally without external causes for gas removal. 

Surprisingly, compared to the O-MZ$_{\rm g}$R in panel (a), the Fe- and Mg-MZ$_*$Rs in panels (b) and (c) show much larger scatters in Figure~\ref{fig:kcwi_mzr}. The scatter in both MZ$_*$Rs appears to be driven by different sSFRs (indicated with a color bar in Figure~\ref{fig:kcwi_mzr}). Because the stellar abundances represent the metals averaged over a much longer timescale than the instantaneous metal content in the current ISM, [Fe/H]$_*$ and [Mg/H]$_*$ are expected to be less ``stochastic" and have less reason to depend on SFR/sSFR. Instead, we see the opposite behavior here. 

It has been suggested that the scatter of the MZ$_{\rm g}$R could be underestimated from the strong line estimators (see discussions in Section~7.3 of \citet{Andrews13} and Section~7.2 in Section of \citet{Steidel14}). In brief, the empirical strong line calibrations used to infer $Z_g$ are not perfect correlations between $T_e$-based measurements and strong line ratios, and thus have their own calibration errors (e.g., 0.1~dex for the $R$ calibration used in this work). 
For example, if we accounted for the $\sim0.1$ dex scatter of the "R" index-based metallicities against direct $T_{\rm e}$ measurements as a source of systematic uncertainty in our measures of oxygen abundances, the implied {\it intrinsic} scatter in the MZ${\rm g}$R  becomes uncomfortably small.
Additionally, as shown in \citet{Steidel14}, the relative intensities of the strong emission lines used in these empirical calibrations are sensitive to the changes not only in (O/H)$_{\rm g}$ but also the excitation parameter and the hardness of the stellar radiation field. These parameters are also strongly correlated with stellar mass and sSFR, and thus yield a tighter relation than the \textit{actual} MZ$_{\rm g}$R. Besides, the empirical relations are calibrated to the $T_e$-based abundance measurements of local H$\,${\sc ii} regions that only span a limited parameter space in physical properties such as excitation parameter while distant galaxies have much more diverse physical conditions, so the $Z_g$ obtained with the strong line methods may not capture the change in physical properties related to SFR as well as the direct method. 

Although the scatter of the MZ$_{\rm g}$R of local dwarf galaxies measured from the direct $T_e$ method \citep[0.15~dex,][]{Berg12} is comparable to the scatter of our O-MZ$_{\rm g}$R in Figure~\ref{fig:kcwi_mzr}, we believe it is not a fair comparison. There are only 11 galaxies in the sample of \citeauthor{Berg12} in the similar mass range of our KCWI sample, which span a narrower range of sSFR \citep[based on the measurements of the same sample by][]{Dale23} than our sample. We might expect the scatter of the MZ$_{\rm g}$R inferred from the direct $T_e$ method to be larger for a larger sample of star-forming galaxies with more diverse sSFRs. Indeed, the MZ$_{\rm g}$R of the SDSS stacked spectra\footnote{The metallicity obtained from the stacked spectra may suffer from the contamination in diffuse ionized gas ionized by the evolved, hot stars and thus also has its own systematics as we further discuss in Appendix~\ref{appendix: sdss}} measured by \citet{Andrews13} using the direct $T_e$ method has much greater scatter. At $M_*\sim 10^{8.5}M_{\odot}$, [O/H]$_{\rm g}$ can differ by up to $\sim 0.6$~dex between the most star-forming and quiescent galaxies. We therefore conclude that the scatter of our O-MZ$_{\rm g}$R is likely to be underestimated due to the strong line method used in this work. 

The scatter in the MZ$_*$R driven by sSFR/SFR has seldom been discussed in the literature. As shown in panels (b) and (c) of Figure~\ref{fig:kcwi_mzr}, the Mg-MZ$_*$R appears to be more linear than the Fe-MZ$_*$R, as the Mg-MZ$_*$R has a larger Pearson coefficient. This implies that the large scatter of the Fe-MZ$_*$R can partially attributed to the varying delay time of Type Ia SNe in galaxies. It is unclear whether the O-MZ$_{\rm g}$R is indeed tighter than the Mg-MZ$_*$R, or whether the difference in scatter is driven by the aforementioned systematics due to the strong line method. The stellar measurements of star-forming galaxies could also suffer from systematics due to the limitations of the current SPS models and the age--metallicity degeneracy, as we further discuss in Section~\ref{subsec:limitations}. However, we still believe that the large scatter in the Fe- and Mg- MZ$_*$Rs as well as our stellar abundance measurements of the star-forming galaxies should be robust, because we see that the two quiescent galaxies (marked by stars in Figure~\ref{fig:kcwi_mzr}) --- which are expected to suffer much less from the age-metallicity degeneracy --- have distinct [Fe/H]$_*$ and [Mg/H]$_*$ despite having similar stellar masses.

We therefore conclude that the MZ$_*$Rs of Fe and Mg in the mass range between the Local Group satellites and more massive field galaxies exhibit large scatter driven by their varying sSFRs and the host environments.  They should not be described by a single relation, especially not a single power law, as we further discuss in Section~\ref{subsec:full_mzr}. We also emphasize that our results do not directly imply the stellar MZRs are bimodal with a metallicity gap at $\log{(M_*/M_{\odot})} \sim 8.5$. The gap is more likely to originate from the selection bias in the sample, as we see a similar gap in the $\log{M_*}$--$\log{\rm SFR}$ plane in Figure~\ref{fig:logM_SFR}.

\subsection{Abundances and Abundance Ratio as a function of Gravitational Potential}
Recently, \citet{Vaughan22} investigated the relation between stellar metallicity and $\log{\Phi} = \log{(M_*/R_e)}$ for low-redshift galaxies from Data Release 3 of the SAMI galaxy survey \citep{Croom21}, and found that star-forming and quiescent galaxies form a single sequence on the $\Phi$--$Z$ plane, with an offset of $\Delta \rm [Z/H] \approx 0.1$~dex. They argued that the $\Phi$--$Z$ relation is more fundamental than the MZR, which they explained as a result of galaxies with deeper gravitational potentials (and thus larger $\Phi$) retaining more metals from galactic outflows. 

To investigate, we constructed the $\Phi$--$Z$ relation of (O/H)$_{\rm g}$, [Fe/H]$_*$, and [Mg/H]$_*$, as well as the $\Phi$--[Mg/Fe]$_*$ relation for our sample in the right column of Figure~\ref{fig:kcwi_mzr}. For each relation, we listed their weighted Pearson coefficient and the $p$-value of the null hypothesis to in each panel of Figure~\ref{fig:kcwi_mzr}.

Still, the scatter of the $\Phi$--$Z$ relation for (O/H)$_{\rm g}$ (panel (e) in Figure~\ref{fig:kcwi_mzr}) is significantly smaller than that of the relations for [Fe/H]$_*$ (panel (f)) and [Mg/H]$_*$ (panel (g)). As discussed above, it is possible that the scatter in the $\Phi$--$Z_{\rm g}$ relation is underestimated due to the strong line method used here.

By moving from the mass--metallicity plane to the $\Phi$--metallicity plane, we see that the relations for (O/H)$_{\rm g}$, [Fe/H]$_*$ and [Mg/H]$_*$ all become tighter and more linearly-correlated, as the Pearson coefficients (and the $p$-value) are significantly higher (smaller) in the $\Phi$--$Z$ relations. In particular, the differences in stellar abundances between galaxies with different sSFRs become smaller.

However, the dependence on sSFR is not completely eliminated when we switch to the $\Phi$--$Z$ relations (panels (e)-(g) in Figure~\ref{fig:kcwi_mzr}).
At a given $\Phi$, most star-forming galaxies with lower sSFRs still have higher [Fe/H]$_*$ and [Mg/H]$_*$\@. The scatters in the stellar $\Phi$--$Z_*$ relation measured from iron and magnesium are $\sim 0.5$~dex and 0.4~dex, respectively, which are much larger than the $\approx 0.1$~dex offset between quiescent and star-forming galaxies suggested by \citet{Vaughan22} for the SAMI sample. The disagreement between our results and those of \citet{Vaughan22} may be explained by the different SFHs of galaxies at different masses. Dwarf galaxies are known to have more bursty SFHs than massive galaxies \citep{Emami19}. The diverse SFHs of the low-mass galaxies, coupled with the delayed timescale of Type Ia SNe, can result in a wide range of stellar abundances at a given mass or $\Phi$ compared to more massive galaxies. The larger scatter we observe may originate from the dominance of dwarf galaxies in our sample. The majority of the SAMI sample of \citet{Vaughan22} instead contains more massive galaxies at $M_* > 10^{10}M_{\odot}$.  In fact, the scatter in the low-mass bins ($M_* \lesssim 10^{9.5}M_{\odot}$) of the SAMI sample (Figure 3 of \citealt{Vaughan22}) are also $> 0.3-0.5$~dex even though the mean offset is nearly zero.  

Finally, when we plot [Mg/Fe]$_*$ as a function of $\Phi$, we find that the scatter remains similar to that of the mass--[Mg/Fe]$_*$ relation. Star formation and SNe continue in the star-forming galaxies, so their [Mg/Fe]$_*$ continues to evolve. Therefore, there is no reason to expect them to be correlated with any general properties of galaxies, as we saw in the trends for massive quiescent galaxies \citep{Choi14, Leethochawalit19, Zhuang23}. 

We conclude that both the stellar and gas-phase elemental abundances of the rapidly evolving low-mass galaxies are jointly shaped by stellar mass, sSFR (or SFH), size, and the host environment, while the gas-to-stellar abundance ratios of low-mass star-forming galaxies are more related to their sSFR (or SFH)\@.

\section{Discussion}\label{sec:discussion}
\subsection{A Further Look at the Stellar MZR}\label{subsec:full_mzr}
To further investigate the shape of the MZ$_*$Rs across the full range of galaxy masses, we construct the MZRs of more massive galaxies drawn from SDSS\@. We focus the discussion on the shape of the MZ$_*$Rs only, rather than the MZ$_{\rm g}$R for the KCWI$+$SDSS sample because the strong line $R$ calibration employed in determining the oxygen abundance in the KCWI sample is more suitable for metal-poor, low-mass galaxies but inappropriate for more massive galaxies, as detailed in Appendix~\ref{appendix: sdss}\@. Because different gas metallicity calibrations are known to have significant systematics and thus affect the shape of the MZ$_{\rm g}$R, we decided not to construct the MZ$_{\rm g}$R beyond the KCWI sample in this paper. 

 As pointed out in our previous work (see Figure~7 in \citealt{Zhuang2021}), the linear extrapolation of the MZ$_*$R of the Local Group satellite galaxies \citep[$M_* \lesssim 10^8 M_{\odot}$,][]{Kirby13} appears to disagree with measurements of more massive, quiescent galaxies in the field \citep[$M_* \gtrsim 10^{9.5} M_{\odot}$,][]{Leethochawalit19}. At $M_*\sim 10^9M_{\odot}$, they predicted different [Fe/H]$_*$ by up to $\sim 0.6$~dex. By comparing the stellar metallicity of NGC 147, a dwarf elliptical satellite of Andromeda, derived from resolved stellar spectroscopy and integrated-light spectroscopy, we ruled out the possibility that the apparent discrepancy originates from systematic differences in the methods used for stellar metallicity estimates. Instead, the consistent results from the two approaches present the first evidence that it is reasonable to put the MZ$_*$Rs measured from different approaches on the same scale.  

 To give an unbiased comparison of the Fe-MZ$_*$R of the Local Group dwarfs, the KCWI sample, and the SDSS galaxies, we re-measured the stellar abundances of the SDSS quiescent galaxy sample compiled by \citet{Leethochawalit19} using the two-step method proposed in Section~\ref{subsec:full_spec_fitting}. We also selected a subsample of 124 star-forming galaxies from \citet{Gallazzi05} using criteria similar to those of \citet{Leethochawalit19}. The selection criteria, the stellar mass estimates, and the abundance measurements are detailed in Appendix~\ref{appendix: sdss}\@.

 \begin{figure*}[htb!]
    \centering
    \includegraphics[width=0.98\textwidth]{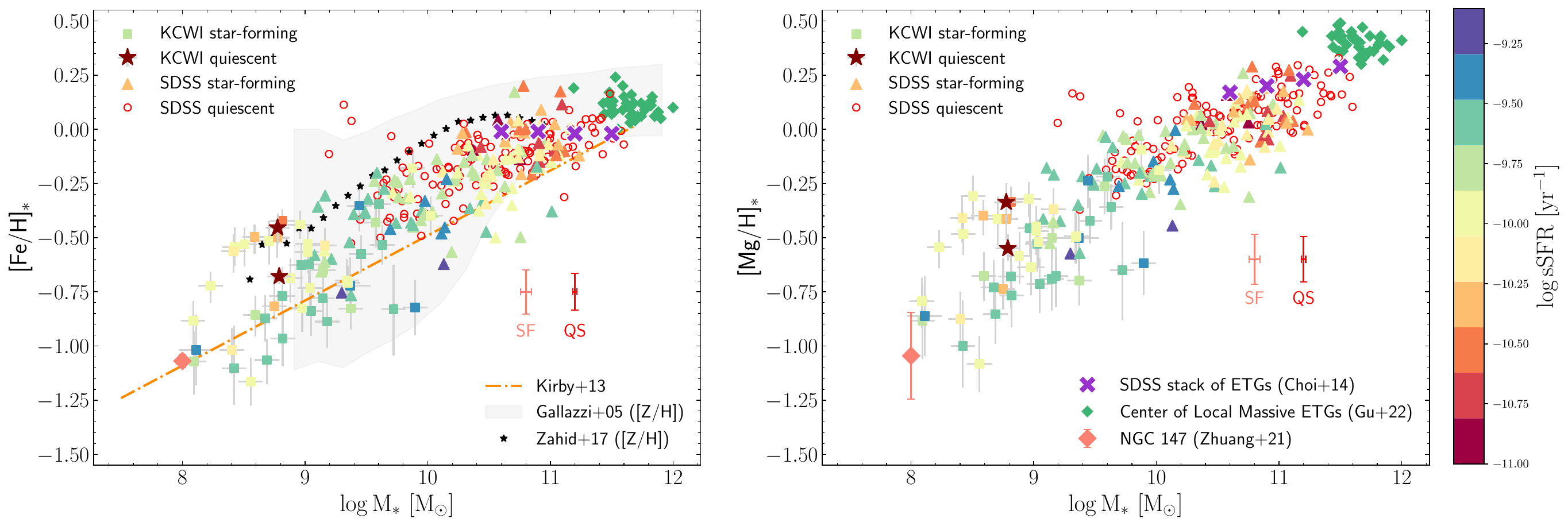}
       \caption{The MZ$_*$Rs of iron (left) and magnesium (right) for the KCWI sample (squares for the star-forming galaxies, and dark red stars for the quiescent galaxies) and the SDSS sample (triangles for the star-forming galaxies, red open circles for the quiescent galaxies). The typical measurement errors of the SDSS star-forming (SF) and quiescent (QS) galaxies are shown in each panel. The star-forming galaxies are color-coded by their SFRs.
       The local MZ$_*$Rs measured in other works \citep{Gallazzi05, Kirby13, Choi14, Zahid17} and the stellar abundance measurements of individual galaxies in the local universe \citep{Zhuang2021, Gu22} are shown. A detailed review of all references in the plot is in Section~\ref{subsec:full_mzr}. }
    \label{fig:stellar_MZR_full}
\end{figure*}

The left panel of Figure~\ref{fig:stellar_MZR_full} shows the Fe-MZ$_*$Rs of the KCWI and the SDSS samples along with the extrapolation of that of the Local Group dwarfs \citep{Kirby13}, as well as some measurements from the literature. First, our updated high-mass Fe-MZ$_*$R of the SDSS quiescent galaxies (red circles) is still inconsistent with the extrapolation of low-mass Fe-MZ$_*$R of the Local Group dwarfs (red dashed line).  They disagree by $\gtrsim 0.5$~dex in [Fe/H]$_*$ at $\log{(M_*/M_{\odot})} \sim 9$, similar to the results in our previous work \citep{Zhuang2021}, which indicates that the Fe-MZ$_*$R cannot be described a single power law across the full range of mass. Second, the Fe-MZ$_*$R of the SDSS quiescent sample (red circles) agrees with the \Alf\ measurements of the [Fe/H]$_*$ of the stacked spectra of early-type galaxies (ETGs) by \citet{Choi14} (purple crosses) and those of the center of local ETGs in the MASSIVE survey \citep{Gu22} (green diamonds). 

The left panel of Figure~\ref{fig:stellar_MZR_full} also compares our results with other existing high-mass MZ$_*$Rs measuring the total stellar metallicity [Z/H]$_*$ instead of distinguishing between [Fe/H]$_*$ and [Mg/H]$_*$\@. We interpret the measured [Z/H]$_*$ in \citet{Gallazzi05} (gray band) and \citet{Zahid17} (black stars) as [Fe/H]$_*$ for the following reason. \citet{Gallazzi05} used the spectrophotometric indices defined in the SPS models of \citet{BC03} to measure [Z/H]$_*$. These indices in \citet{BC03} were chosen to be the ones that are not sensitive to the change in [$\alpha$/Fe]$_*$ and capable of recovering solar-scaled metallicity (i.e., [Fe/H]$_*$=[Z/H]$_*$). \citet{Zahid17} determined [Z/H]$_*$ using full-spectrum fitting technique with the FSPS models derived from the solar-scaled MIST isochrones \citep{Choi16} and the MILES stellar spectral libraries \citep{Sanchez06}, similar to the FSPS-MILES SSPs models presented in Section~\ref{subsec:ssp_afe}. The metallicities of MILES stellar library were also measured in terms of [Fe/H]$_*$. Therefore, the measured [Z/H]$_*$ by \citet{Zahid17} should also best interpreted as [Fe/H]$_*$ instead of [Mg/H]$_*$. The MZ$_*$R of all SDSS galaxies in the MPA-JHU sample by \citet{Gallazzi05}, regardless of their quiescence or star-forming state, exhibits scatter larger than 1~dex at $\log{(M_*/M_{\odot})} \sim 9$ and therefore is consistent with both the low-mass MZ$_*$R of \citet{Kirby13} and our high-mass MZ$_*$R of the SDSS sample. The MZ$_*$R measured from stacked spectra of SDSS star-forming galaxies in different mass bins by \citet{Zahid17} did not quantify the scatter of the MZ$_*$R of star-forming galaxies. Their relation has overall higher normalization than our high-mass MZ$_*$R of the SDSS star-forming and quiescent sample, which might result from the different wavelength ranges, SPS models, and algorithms used in the full-spectrum fitting.

The Fe-MZ$_*$Rs of the KCWI and the SDSS samples are derived using the same approach. They are free from the systematic effects of different fitting methods and thus yield the most straightforward and interpretable relation of [Fe/H]$_*$ of galaxies at different masses. First, the SDSS star-forming galaxies exhibit large scatter in their [Fe/H]$_*$ compared to their quiescent counterparts, with the most heavily star-forming examples lying well below the high-mass MZ$_*$R of the quiescent galaxies. The trend is similar in the KCWI sample. 
The dwarf galaxies at the low mass end of the KCWI sample ($\log{(M_*/M_{\odot})} \sim 8$) have comparable [Fe/H]$_*$ as that of NGC 147 derived from \Alf\ \citep{Zhuang2021}. 
The combined results of the KCWI and SDSS samples suggest that the Fe-MZ$_*$R at $M_* > 10^{8}M_{\odot}$ is not linear. Our result reveals that the transition in the MZ$_*$R between the Local Group satellites and massive field galaxies in the SDSS is largely shaped by the change in star-formation activity indicated by sSFR and the environment.  

We also compare the Mg-MZ$_*$R of the KCWI and SDSS samples in the right panel of Figure~\ref{fig:stellar_MZR_full}. Our updated Mg-MZ$_*$R of the SDSS quiescent galaxies is consistent with the results measured from the stacked spectra of the SDSS ETGs by \citet{Choi14} but lower than the [Mg/H]$_*$ of the center of local ETGs in the MASSIVE survey \citep{Gu22}. The behavior of the Mg-MZ$_*$R is similar to Fe-MZ$_*$R, although the scatter of the Mg-MZ$_*$R of the star-forming galaxies in the KCWI sample and the SDSS sample is slightly smaller. Still, we find that at $M_* \gtrsim 10^{9} M_{\odot}$, more heavily star-forming galaxies have lower [Mg/H]$_*$ than the quiescent galaxies at similar masses.
Although there is yet no measurement of the Mg-MZ$_*$R of the Local Group satellites, we show that the dwarf elliptical galaxy NGC~147 \citep{Zhuang2021} and the quiescent satellite candidate 0241-0810 have [Mg/H]$_*$ comparable to those of the more heavily star-forming galaxies in the KCWI sample. 

To conclude, Fe- and Mg-MZ$_*$Rs beyond the Local Group do not appear to be linear and cannot be described by a relatively tight linear relation. 
 The stellar abundances of the star-forming galaxies beyond $\sim 10^8M_{\odot}$ are much more diverse and exhibit larger scatter that probably depends on factors other than stellar mass. 

\subsection{The sSFR Dependence of Stellar Abundance }\label{subsec:stellar_fmr}

Upon a closer look at Figure~\ref{fig:stellar_MZR_full}, we can tentatively see the dependence on sSFR of the Fe- and Mg-MZ$_*$Rs of star-forming galaxies in both the dwarf galaxies ($M_* \lesssim 10^9 M_{\odot}$) in the KCWI sample and the massive galaxies ($M_* \gtrsim 10^{10} M_{\odot}$) in the SDSS sample, whereas such dependence on sSFR is less obvious for galaxies at $M_*\sim 10^{9-10}M_{\odot}$. A similar behavior is seen in the MZ$_{\rm g}$R, where galaxies with higher SFRs at fixed mass are generally more metal-poor \citep[e.g.,][]{Mannucci10}, likely due to the higher gas fraction diluting the gas metallicity in the ISM. 

Recently, \citet{Looser24} reported a smooth relation between stellar mass, SFR and the light-weighted stellar metallicity in their study of 7,323 MaNGA galaxies at $\log{(M_*/M_{\odot})} > 9$ and referred to the relation as  stellar fundamental metallicity relation, suggesting a potential correlation between stellar metallicity and SFR/sSFR. \citeauthor{Looser24}\ divided the MaNGA sample into six groups based on their offsets from the SFMS (somewhat similar to sSFR) and constructed the stellar MZR within each group. They found that at $\log{(M_*/M_{\odot})} \gtrsim 9.5$ the MZ$_*$R of the galaxies below the SFMS (i.e., more quiescent) have a higher normalization than that of the galaxies above the SFMS (i.e., more starbursty). They argued that the sSFR dependence of the MZ$_*$R suggests that the metal contents of galaxies are primarily governed by long-lasting inflow of metal-poor gas accretion from the IGM/CGM instead of short-lived ``episodic'' accretion that only affects the gas metallicity, consistent with the theoretical models of IGM/CGM accretion by \citep{Forbes14, Torrey19}. 

Although our sample size is much smaller than that of \citet{Looser24}, we pushed the stellar mass limit down to $\log{(M_*/M_{\odot})} \sim 8$, bridging the gap between the Local Group satellites (in which the stellar metallicity dependence on their sSFR is absent) and the massive galaxies (where the dependence on the sSFR has already been established). In addition, the Mg-MZ$_*$R of the sample presented in this work is a better analog of O-MZ$_{\rm g}$R because the trend will be less disrupted by the delay time of Type Ia SNe.

Additionally, \citet{Garcia24} observed similar strong dependence of the normalization of the MZ$_*$R on the sSFR from the hydrodynamic simulations Illustris, TNG, and EAGLE\@. This finding is similar to the observations in this work and that of \citet{Looser24}. \citet{Garcia24} suggested that the similar sSFR dependence of the MZ$_*$R and MZ$_{\rm g}$R can be attributed to the tight correlation between stellar and gas-phase metallicities in the simulation suites, and thus the MZ$_*$R inherits the shape of the MZ$_{\rm g}$R\@. Nevertheless, they were also concerned that the sub-grid treatment of the ISM employed in these simulations can only produce smooth stellar feedback. Consequently, the SFR/sSFR dependence discovered in their work may break down if galaxies have very bursty SFHs which may significantly disrupt the processes that allows stellar metallicities to catch up to the gas-phase metallicities.

\subsection{Implications for Galaxy Growth and Quenching}\label{subsec:implications}
The interplay between stellar and gas-phase abundances we measure from the KCWI sample, as well as the stellar abundance dependence on mass, SFR, and environment, may provide us a hint as to the chemical enrichment history of galaxies as they grow and quench. 

First, Figure~\ref{fig:stellar_MZR_full} shows that dwarf quiescent galaxies in different environments can have distinct stellar abundances at similar masses. The dwarf satellite candidate 0241-0810 that is more metal-poor falls on the extrapolation of the low-mass MZ$_*$R of the Local Group satellites, implying that galaxies in the group environment would still evolve along the low-mass MZ$_*$R established in the Local Group. It is possible that galaxy 0241-0810 was quenched through ram pressure stripping (first proposed by \citealt{Gunn72}) that efficiently removed all the cold gas and resulted in fast quenching, and thus may not have had enough time to leave the low-mass MZ$_*$R before becoming quiescent \citep[e.g.,][]{Mayer06}. The isolated dwarf quiescent galaxy 0125-0041 that is more metal-rich instead was less impacted by environmental effects during its star formation. It might have  quenched more slowly than  0241-0810, so that it would have had more time to form more metal-rich stars before the cessation of star formation.

The dwarf star-forming galaxies in our sample that are on their way to quiescence in Figure~\ref{fig:stellar_MZR_full} could have experienced gas starvation \citep{Peng15, Trussler20}. When the supply of metal-poor gas from the surrounding CGM/IGM lags behind the gas consumption of the star formation, star formation would continue in the progressively more metal-enriched ISM and form more metal-rich stars, which can explain why these dwarf star-forming galaxies with low sSFRs have significantly higher stellar abundances than those that are still actively forming stars at the similar masses. The starvation scenario is also supported by the higher [O/H]$_{\rm g}$ compared to the [Mg/H]$_*$ in most galaxies in our KCWI sample (Figure~\ref{fig:stellar_gas_abund}). If the metal-poor gas inflows were dominant in driving star formation in these galaxies, one would expect [O/H]$_{\rm g}$ lower than the [Mg/H]$_*$\@. Although gas metallicity calibrations are known to have many systematic effects \citep{Kewley08} and the absolute oxygen abundance scale may be off, the $R$ calibration \citep{Pilyugin16} adopted in this work is calibrated to the $T_e$-based measurements, which are usually lower than the gas-phase metallicity measurements based on nebular recombination lines or photoionization models \citep{Esteban14, Blanc15, Steidel16, Strom18}. In other words, if we have underestimated the true [O/H]$_{\rm g}$, it only strengthens our conclusion about starvation. In addition, the prominent differences of the stellar abundances between the galaxies with enhanced and suppressed SFRs at similar masses might suggest that most of the chemical enrichment (at least for the elements that substantially affect the light-weighted stellar abundance) occurs when galaxies are approaching  quiescence. This finding is consistent with \citet{Leung24}, who found that the post-starburst galaxies in the MaNGA survey significantly increased their stellar metallicity during the recent starbursts. 

Once galaxies grow above above $\sim 10^{10}M_{\odot}$, the MZ$_*$Rs of star-forming and quiescent galaxies both become flattened in Figure~\ref{fig:stellar_MZR_full}. Massive galaxies have deeper gravitational potential well that can retain an increasingly large fraction of their metals. Theoretically, it has been suggested that energy-driven winds that are more efficient in ejecting metals dominate in low-mass galaxies \citep[e.g.,][]{Murray05}, while momentum-driven winds that drive less intense outflows dominate in high-mass galaxies \citep[e.g.][]{Hopkins12, Dave12}. Consequently, these massive galaxies retain more metals than low-mass galaxies. At this stage, massive galaxies tend to start being strongly affected by a combination of "starvation" and "cessation" of star formation due to the increasing prevalence of AGN feedback \citep[e.g.][]{Weinberger18}, and thus generate metals much less efficiently. Additionally, the ISM of the massive galaxies that have been significantly chemically enriched contains high level of metals deposited by stellar winds and SNe and are almost saturated, so they will not increase metallicity significantly even if star formation continues. All of these combined effects lead to the flattened slope we observe in Figure~\ref{fig:stellar_MZR_full}. The levels of scatter in the Fe- and Mg-MZ$_*$Rs of the SDSS star-forming galaxies are still much larger than those of the quiescent galaxies, given that the SDSS star-forming galaxies still have very diverse sSFRs and would contribute to the distinct stellar abundances at fixed masses as we discuss above.

\subsection{Limitations and Caveats of This Study}\label{subsec:limitations}
We now discuss potential limitations and caveats of this study that may affect our results. 


First, our sample is restricted to the star-forming galaxies with sSFRs in the range of $10^{-9}$--$10^{-11}$~yr$^{-1}$, so our conclusions must be limited to galaxies with sSFRs in that range. We excluded starburst galaxies due to the difficulty in determining their stellar abundances from the weak stellar absorption features in the rest-frame optical. Assuming that the dependence of stellar abundance on sSFR would exist in galaxies at all different sSFRs, it is possible that the dwarf starburst galaxies would lie even below the extrapolation of the low-mass MZ$_*$R of the Local Group satellites, resulting in even larger scatter in the Fe- and Mg-MZ$_*$Rs than the ones in Figure~\ref{fig:kcwi_mzr}. As shown in Figure~\ref{fig:logM_SFR}, the sample is also not evenly distributed in the $\log{M_*}$--$\log{\rm SFR}$ plane. The current sample lacks many star-forming galaxies with intermediate sSFRs at $M_*\sim 10^{8.5}M_{\odot}$, so the MZ$_*$Rs in Figure~\ref{fig:kcwi_mzr} are slightly bimodal. We hypothesize that denser sampling would reveal a continuous change in the normalization of MZ$_*$R with increasing sSFR\@. The scatter in the MZ$_{\rm g}$R may also increase if starburst galaxies are included because previous works have found significant correlation between sSFR and (O/H)$_{\rm g}$ \citep[e.g.,][]{Mannucci10, Andrews13}.

Second, our sample contains only two quiescent galaxies residing in distinct environments. A larger sample would be needed to take a census of the metal content of dwarf galaxies in different environments, in order to resolve whether environment is as important in shaping the MZ$_*$R as we suspect from this work. 

Third, because 40\% of the KCWI sample lacks the H$\alpha$+[N$\,${\sc ii}]$\lambda\lambda 6549,6585$ coverage\footnote{The red channel of the KCWI was not available until the summer of 2023, after some of our data were obtained.}, we determined the gas-phase oxygen abundances of our sample from a combination of line ratios measured from the KCWI spectra and SDSS spectra. Therefore, we had to assume that the [N$\,${\sc ii}]/H$\alpha$ ratio remains constant within galaxies because the KCWI FoV is much larger than the SDSS fiber aperture. We cannot construct the MZ$_{\rm g}$R using direct $T_e$ measurements because we detected [O$\,${\sc iii}]$\lambda$4363 in less than $1/3$ of the sample. The only strong line calibration that does not require the red optical line ratios is $R_{23}$ \citep[e.g.,][]{Pilyugin05} which is known as a double-branched calibration (see Figure~8 of \citealt{Pilyugin16} for an example)\footnote{The $R_{23}$ calibration is relatively insensitive to galaxies with oxygen abundances falling near turnover of the curve at $12+\log{\rm (O/H)} \sim 8.0$--$8.4$, which coincides with the gas-phase metallicity of most galaxies in our sample.}. In addition, any strong-line calibration  onto the direct $T_e$ abundance scale (like the $R$ calibration used here) may underestimate the \textit{actual} gas-phase O/H in galaxies, as discussed in Section~\ref{subsec:implications}. 

Finally, we acknowledge that the stellar abundances we measure from the two-step method might still suffer from systematic effects. Due to the lack of SPS models that directly quantify the variation in [Mg/Fe]$_*$, we still have to rely on models that vary all $\alpha$ elements together with respect to Fe.   Although we validated the method for quiescent galaxies in Section~\ref{subsec:validate_fitting}, it is possible that the two-step method may be subject to different systematic errors in star-forming galaxies due to the combined effects of the age--metallicity degeneracy and the difficulty in recovering the age of young galaxies \citep{Carnall19}. The most direct way to verify the two-step method is to apply it to a nearby star-forming galaxy where we can measure its [Mg/Fe]$_*$ from its resolved stellar population, similar to the recent work done in dwarf elliptical galaxies or globular clusters with multiple stellar populations \citep{Ruiz-Lara18,Boecker20, Zhuang2021}.

\section{Summary}\label{sec:summary}
In this work, we presented a sample of 46 star-forming and two quiescent galaxies with $M_*\sim 10^{8-10}M_{\odot}$ observed with Keck/KCWI, with the aim of understanding the shape of the MZ$_*$R in the relatively unexplored mass range between the Local Group dwarf satellite galaxies and massive field galaxies, as well as the connection between the metals in the stars and the ISM\@. 
Our findings are as follows:
\begin{enumerate}
    \item Applying SPS models with variable [$\alpha$/Fe] directly to the stellar continuum with \ppxf\ significantly underestimates [Mg/Fe]$_*$ or overestimates [Fe/H]$_*$ (Figures~\ref{fig:alf_comparison_afe} and \ref{fig:alf_comparison_feh}). To solve this problem, we developed a two-step method that is capable of recovering the [Fe/H]$_*$ and [Mg/Fe]$_*$ of the stellar population.
    
    \item We determined the gas-phase oxygen abundances via the $R$ calibration \citep{Pilyugin16} and the stellar iron and magnesium abundances via the two-step method within the same galaxies of our KCWI sample. For the first time, we can not only relate the stellar and gas-phase metallicity but also make an apples-to-apples comparison of $\alpha$ elements within the same galaxies. The correlation between [O/H]$_{\rm g}$ and [Mg/H]$_*$ is tighter than that of [O/H]$_{\rm g}$ and [Fe/H]$_*$\@. Furthermore, [O/H]$_{\rm g}$ is higher than [Fe/H]$_*$ and [Mg/H]$_*$ in most galaxies, suggesting that new stars are forming in gas that has participated in the overall chemical enrichment of the ISM, and not in metal-poor gas that has recently been accreted.

    \item The gas-to-stellar abundance ratios, $\Delta Z_{\rm O, Fe}$ and  $\Delta Z_{\rm O, Mg}$, are correlated with sSFR but do not depend on stellar mass, SFR, light-weighted age, $\log{\Phi}$ (a proxy for the  gravitational potential), or $\log{\Sigma}$ (the stellar mass surface density).

    \item We constructed the ionized O-MZ$_{\rm g}$R and the Fe-MZ$_*$R and Mg-MZ$_*$R of our KCWI sample, and we found that the scatters in the Fe- and Mg-MZ$_*$Rs are significantly larger than that of the O-MZ$_{\rm g}$R. Further work would be necessary to understand the origin of these scatters. 
    The observed scatters in the MZ$_*$Rs are primarily driven by varying sSFRs.
    At a given mass, star-forming galaxies with higher sSFRs exhibit lower [Fe/H]$_*$ and [Mg/H]$_*$. The relations become tighter in the $\Phi$--Z plane, but the dependence on sSFR is not eliminated. For the two quiescent galaxies in our sample, the dwarf satellite candidate 0241-0810 is more metal poor, has comparable stellar abundances to the heavily star-forming galaxies at similar masses, while the isolated dwarf quiescent galaxy 0125-0041 is more metal-rich and exhibits abundances similar to the star-forming galaxies with low sSFRs.

    \item We compared the Fe- and Mg-MZ$_*$Rs of our KCWI sample with those of more massive SDSS galaxies and found that they could not be described by a single linear relation. At a fixed mass, heavily star-forming galaxies still tend to have lower [Fe/H]$_*$ and [Mg/H]$_*$ than quiescent galaxies.
    This suggests that the increasing scatter in the stellar abundances at a fixed $M_*$ exhibited by the KCWI dwarfs may be driven, at least in part, by the sSFR tracing the remaining gas supply.  
    This trend also supports a physical scenario in which starvation is primarily responsible for quenching galaxies in the field. 
\end{enumerate}

This work is the first paper in a series whose goal is to understand the chemical evolution of low-mass galaxies at low redshift, especially those that remain actively star-forming, over the mass range $M_*\sim 10^{8-10}M_{\odot}$ based on IFU data cubes.  While we are limited by the sample size, the high-quality data obtained with a 10~m telescope allows us to place unprecedented constraints on the chemical enrichment of galaxies in the mass range intermediate between Local Group satellites and  massive galaxies that have been well-studied by large spectroscopic surveys. In the future, we will leverage the spatial information from the IFU data and investigate the spatially-resolved elemental abundances in the stellar and gas phases, which will provide unparalleled details to interpret the chemical evolution of star-forming dwarf galaxies. 


\section*{Acknowledgments}

The authors acknowledge the insightful and constructive feedback of the anonymous referee. We thank Yuguang Chen, Allison Strom, Adam Carnall, Claudia Maraston, and Ryan Sanders for useful discussions. We are grateful to the many people who have worked to make the Keck Telescopes and their instruments a reality and to operate and maintain the Keck Observatory, including support astronomers Greg Doppmann and Rosalie McGurk, and telescope operator Tony Connors. The authors wish to recognize and acknowledge the very significant cultural role and reverence that the summit of Maunakea has always had within the Native Hawaiian community. We are most fortunate to have the opportunity to conduct observations from this mountain.

This material is based on work supported by the National Aeronautics and Space Administration (NASA) under FINESST Grant No.\ 80NSSC22K1755 (Z.Z.\ and C.C.S.\@). E.N.K.\ acknowledges the support from the National Science Foundation (NSF) under CAREER Grant No.\ AST-2233781. C.C.S.\ has been supported in part by NSF grant No.\ AST-2009278. C.C.\ acknowledges support from NSF-AST-131547.
Most of the data presented herein were obtained at Keck Observatory, which is a private 501(c)3 non-profit organization operated as a scientific partnership among the California Institute of Technology, the University of California, and NASA. The Observatory was made possible by the generous financial support of the W. M. Keck Foundation.


\facility{Keck:II (KCWI)}
\software{\ppxf\ \citep{Cappellari17}, \texttt{dynesty} \citep{Speagle20}, BAGPIPES \citep{Carnall18, Carnall19},  Astropy \citep{Astropy2013,Astropy2018}}, 

\appendix


\section{Potential Aperture Effect}\label{appendix:aper_effect}

The main analysis in this work made use of the spectra obtained from summing the spaxels with S/N $>$ 1 in the stellar continuum. To explore whether this extraction approach would introduce any bias due to the potential metallicity radial gradients in these galaxies, we also extracted the spectra from four different circular apertures, with diameters ranging from 1\arcsec\ to 4\arcsec\@, and measured their stellar abundances using the method described in Section~\ref{subsec:full_spec_fitting}. We did not measure the gas metallicity because the scatter in gas metallicity is more likely driven by the calibration method instead of an aperture effect, as detailed in Section~\ref{subsec:kcwi_mzr}.

\begin{figure*}[htb!]
    \centering
    \includegraphics[width = 0.8\textwidth]{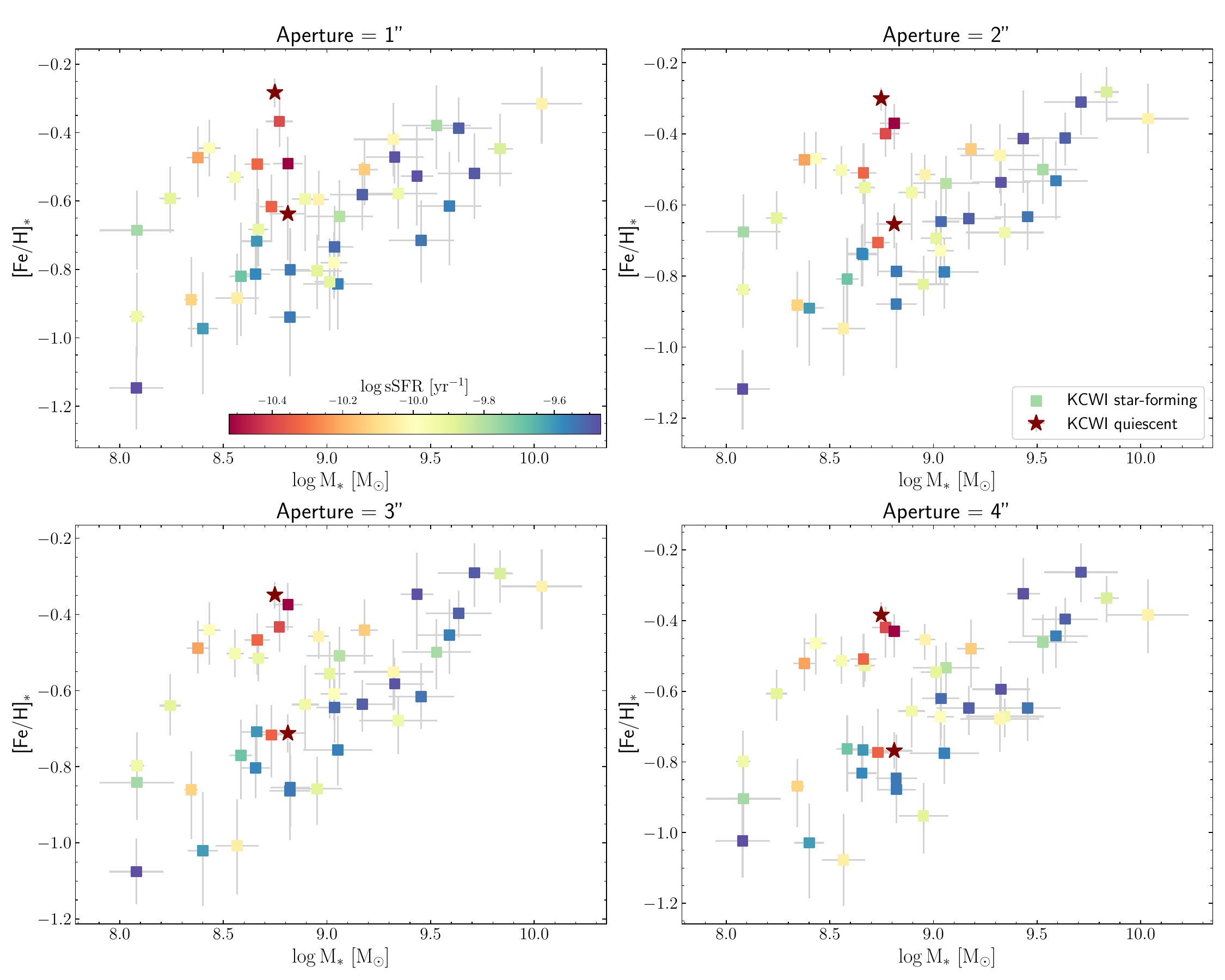}
    \caption{The Fe-MZ$_*$R of the KCWI star-forming (squares) and quiescent (stars) sample, measured from the spectra extracted from different circular apertures (1\arcsec\ -- 4\arcsec\@). The star-forming galaxies are color-coded by their sSFRs.
    The overall shape and the scatter of the Fe-MZ$_*$R remains largely unchanged across different sizes of aperture. }
    \label{fig:kcwi_mzr_aper_feh}
\end{figure*}

\begin{figure*}[t]
    \centering
    \includegraphics[width = 0.8\textwidth]{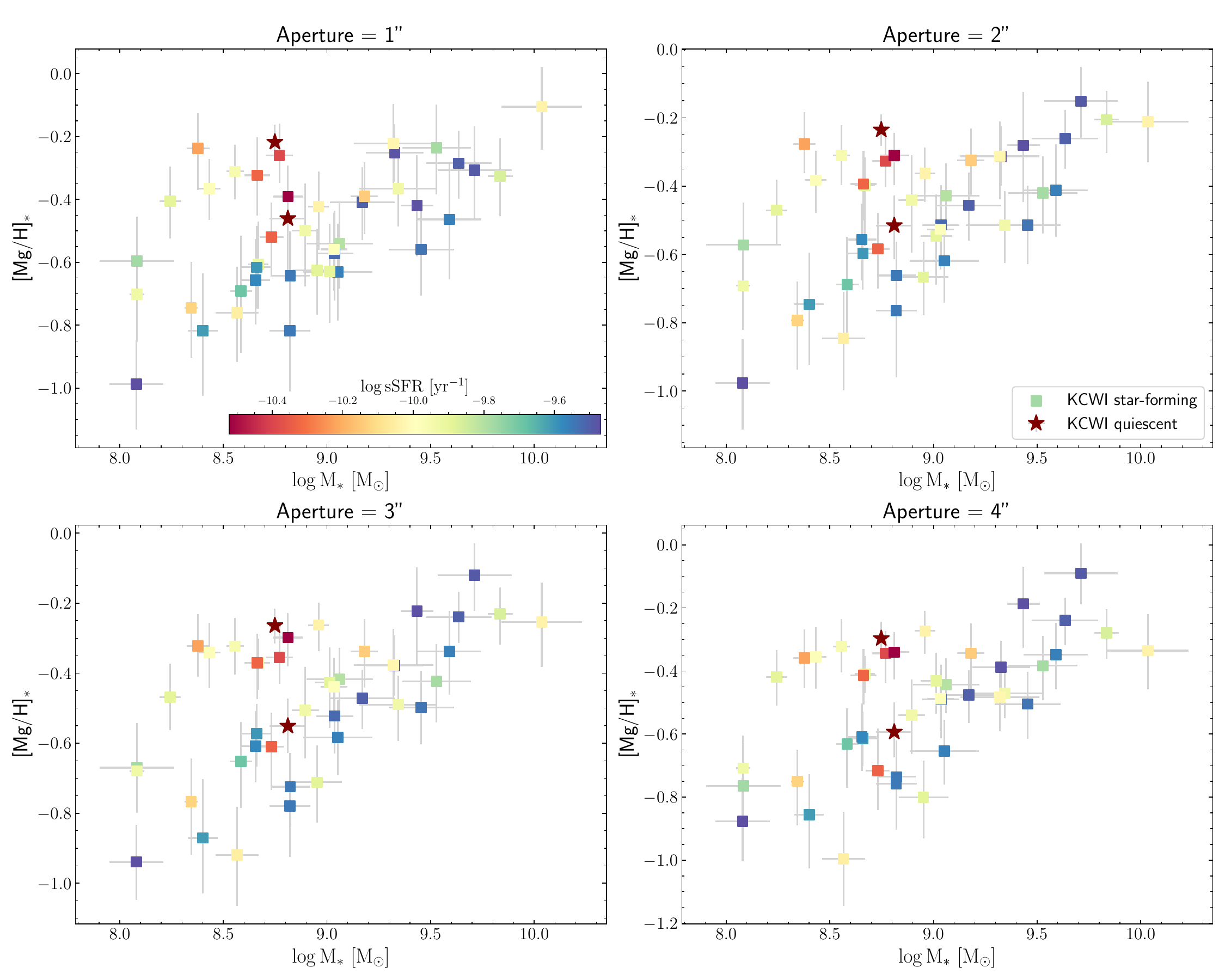}
    \caption{Similar to Figure~\ref{fig:kcwi_mzr_aper_feh}, but comparing the Mg-MZ$_*$R measured from different apertures.}
    \label{fig:kcwi_mzr_aper_mgh}
\end{figure*}

We show the Fe- and Mg-MZ$_*$Rs of the KCWI sample extracted using different aperture sizes in Figure~\ref{fig:kcwi_mzr_aper_feh} and \ref{fig:kcwi_mzr_aper_mgh}, respectively. Compared to the MZ$_*$Rs obtained from the entire KCWI data cube (roughly around 6\arcsec\ to 20\arcsec\ in diameter; Figure~\ref{fig:kcwi_mzr}), the overall shape and the scatter of the MZ$_*$Rs remained largely unchanged when smaller aperture sizes are used, indicating that potential radial gradients and aperture effects are not responsible for driving the large scatter in the MZ$_*$Rs. The dependence of the stellar abundances on sSFR remains present across all aperture sizes.

\section{Comparison of stellar population age derived from different SFH models}\label{appendix:age}

\begin{figure*}
    \centering
    \includegraphics[width = 0.8\textwidth]{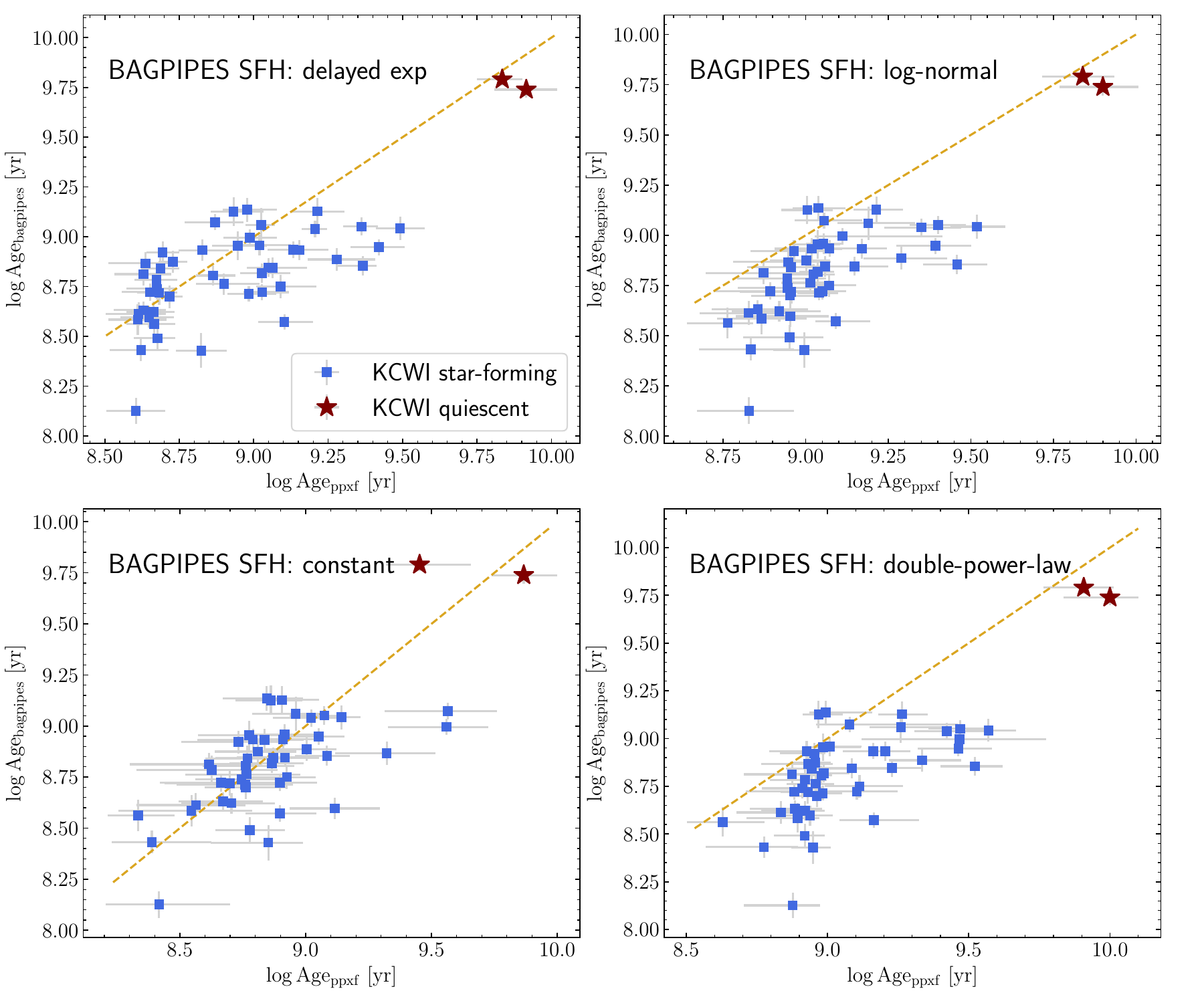}
    \caption{Comparison of light-weighted stellar population age derived from SED fitting using BAGPIPES and from full-spectrum fitting using \ppxf\@ for the star-forming (blue squares) and quiescent (red stars) galaxies in our KCWI sample. In each panel, the one-to-one line is shown by the yellow dashed line.  }
    \label{fig:age_comparison_bagpipes_ppxf}
\end{figure*}

Here we discuss the light-weighted stellar population ages of the KCWI sample derived assuming different SFH models and fitting methods. Figure~\ref{fig:age_comparison_bagpipes_ppxf} compares the light-weighted ages derived using \ppxf\ (i.e., the first step of the two-step method) assuming non-parametric SFHs with those derived from broadband SED fits using BAGPIPES assuming four different parametric SFHs: delayed exponential, log-normal, double-power-law, and constant. The light-weighted ages obtained from spectral analysis are broadly consistent with those obtained from SED fits, particularly for an assumed constant SFH\@.  The consistency indicates that invoking a non-parametric SFH that allows star formation to occur at any time and to accommodate any possible shape of the SFH does not introduce any significant bias to the light-weighted age measurement. Given that broadband photometry is more sensitive to stellar populations formed at different epochs due to its wider spectral coverage than KCWI spectroscopy, Figure~\ref{fig:age_comparison_bagpipes_ppxf} shows that the light-weighted ages derived from \ppxf\ spectral fits in the relatively narrow wavelength range 3650--5500~\AA\ are reasonable.

\section{The Selection and Abundance Measurements of the SDSS Sample}\label{appendix: sdss}
We adopt the SDSS quiescent sample selected by \citet{Leethochawalit18}. In summary, \citet{Leethochawalit18}  selected a sample of 152 quiescent galaxies with H$\alpha$ EW $<$ 1~\AA\ from 44,254 SDSS spectra from \citet{Gallazzi05}. In each bin of 0.1~dex in stellar mass spanning from $10^9$ to $10^{11.5}M_{\odot}$, they randomly selected at most eight quiescent galaxies, yielding a subsample of 152 quiescent galaxies. 

In this work, we selected the SDSS star-forming sample in a similar way to \citet{Leethochawalit18}. We restricted the galaxies to the SFMS with sSFRs in the range of $10^{-11}$ and $10^{-9}$~yr$^{-1}$ using the sSFRs from the SDSS MPA-JHU catalog \citep[\textit{galSpecExtra}, SDSS DR17;][]{Kauffmann03, Brinchmann04, Salim07}. Similarly, we randomly selected at most eight star-forming galaxies in each bin of 0.1~dex in the mass range of $10^{9-11.5}M_{\odot}$. 

\begin{figure}[htb!]
    \centering
    \includegraphics[width=0.9\columnwidth]{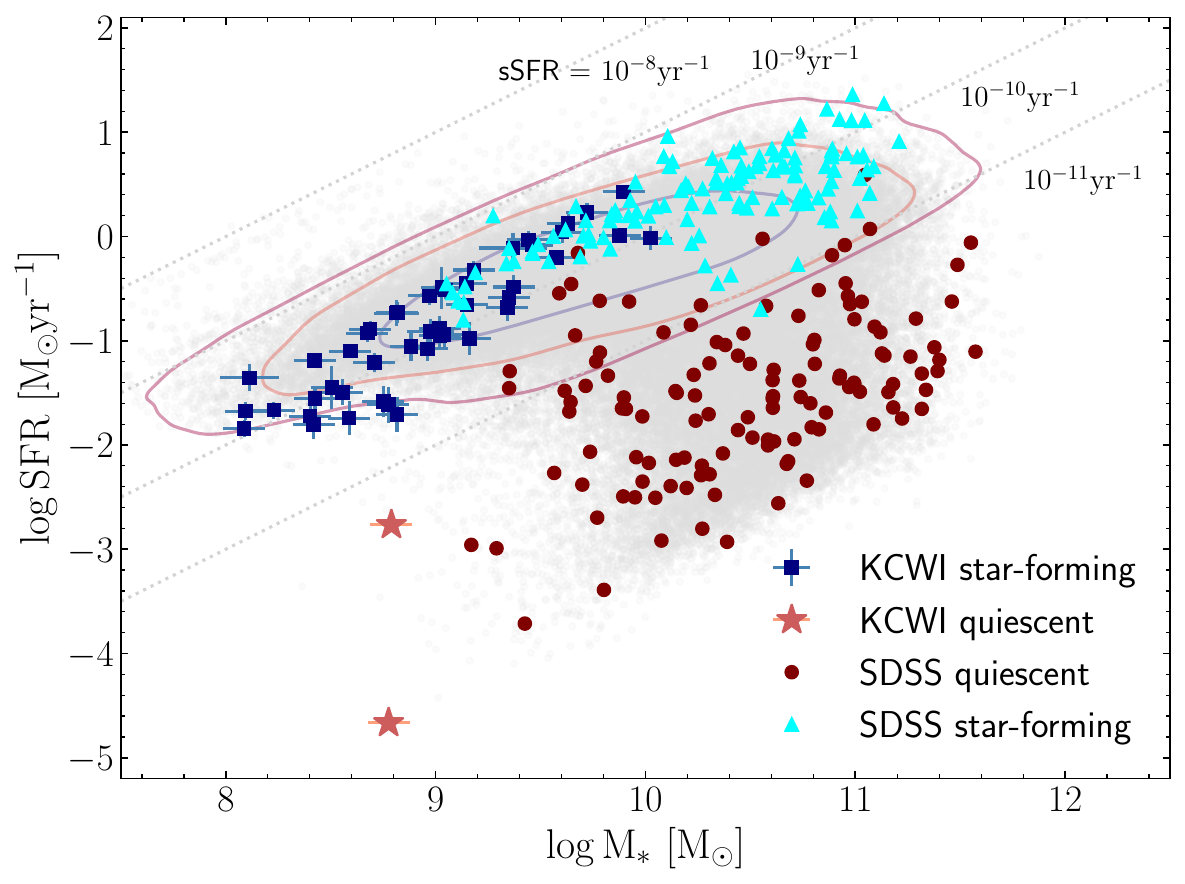}
    \caption{SFR as a function of stellar mass of the galaxies in our KCWI and SDSS samples. The blue squares and red stars represent the star-forming and the quiescent galaxies of the KCWI sample, while the cyan triangles and dark red dots indicate the star-forming and the quiescent galaxies described in Appendix~\ref{appendix: sdss}. The contours show the SDSS SFMS at a similar redshift range ($z < 0.06$) taken from the GSWLC catalog \citep{Salim16} at 1-3 $\sigma$. The grey dots indicate all the GSWLC galaxies in the same redshift range as those in our sample.  }
    \label{fig:logM_SFR_sdss}
\end{figure}

We cross-matched the subsamples with the GSWLC catalog to obtain stellar mass estimates for comparison with the KCWI sample. As the stellar masses of the KCWI sample were derived from BAGPIPES assuming the Kroupa \citep{Kroupa01} IMF, the Chabrier-based \citep{Chabrier03} stellar masses in the GSWLC catalog were adjusted to Kroupa-based values by applying an offset of $+0.025$~dex \citep{Salim07}. To be consistent with the KCWI spectra, we limited the wavelength range of the SDSS spectra to $3650-5500$~\AA\@ and measured the ages, [Fe/H]$_*$, and [Mg/Fe]$_*$ in the same manner as Section~\ref{sec:stellar_abund_measurement}. We excluded the galaxies with errors in [Fe/H]$_*$ $>$ 0.15~dex to provide a clear trend, yielding 125 quiescent and 124 star-forming galaxies in the SDSS sample. The stellar masses and SFRs of the SDSS sample are shown in Figure~\ref{fig:logM_SFR_sdss}.

\begin{figure}[htb!]
    \centering
    \includegraphics[width=0.98\columnwidth]{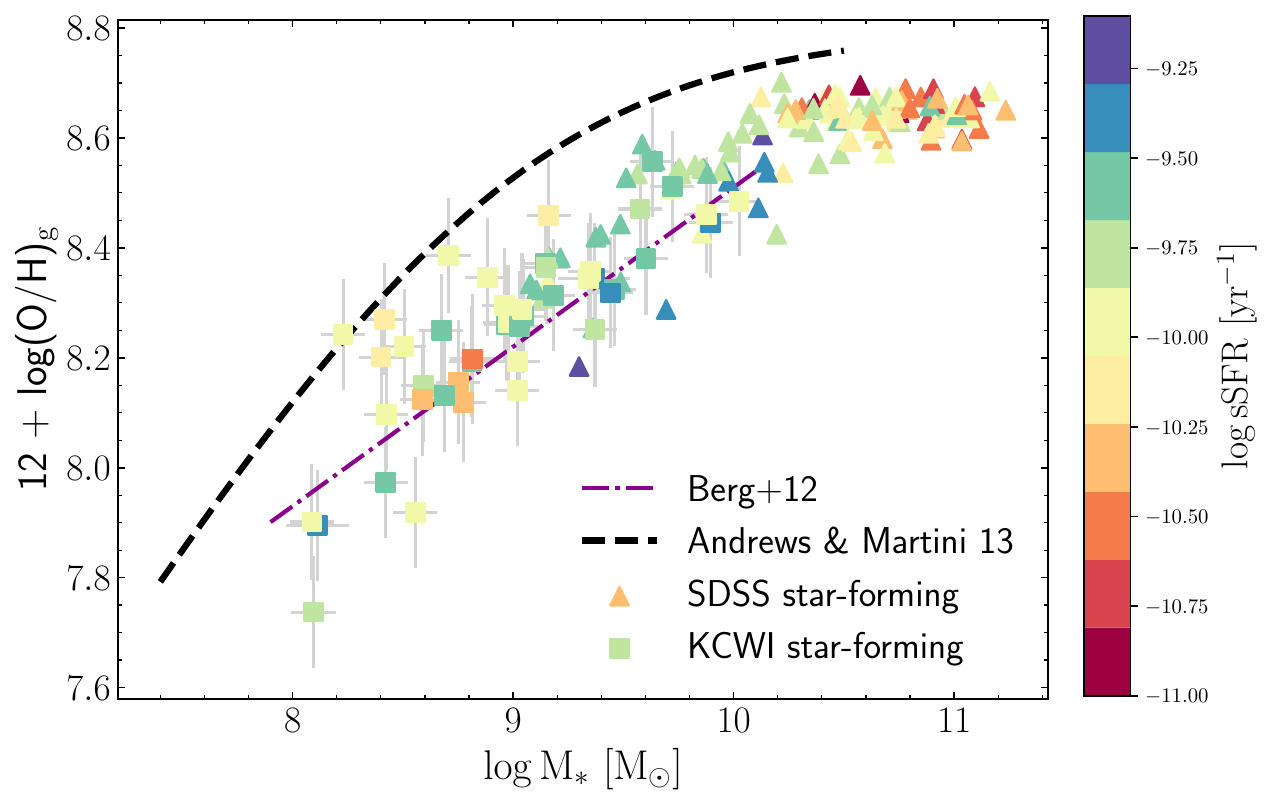}
    \caption{The O-MZ$_{\rm g}$R of the KCWI sample and the SDSS sample, color-coded by the sSFRs. The O-MZR of the local dwarf galaxies \citep[and the associated $1\sigma$ confidence interval,][]{Berg12} and the stacked spectra of the SDSS galaxies \citep{Andrews13} derived from the direct $T_e$ method are shown by the purple dashed-dotted line (and shaded region) and the black dashed line, respectively.}
    \label{fig:gas_MZR_full}
\end{figure}

We do not discuss the best-fit O-MZ$_{\rm g}$R of the SDSS star-forming galaxies in Section~\ref{sec:discussion} because it becomes flat around $\log{(M_*/M_{\odot})} \sim 10.2$ (Figure~\ref{fig:gas_MZR_full}). It is unclear to us whether the flattening of the SDSS O-MZ$_{\rm g}$R results from the underlying physical mechanisms governing the metal retention or originates from the systematics in the $R$ calibration, as \citet{Pilyugin16} only validated the $R$ calibration to the $T_e$-based measurements up to $12+\log{(\rm O/H)} \gtrsim 8.7$. Because the KCWI data cannot cover many nebular features sensitive to gas-phase metallicity in the red optical, we refrain from experimenting with other strong-line calibrators in this comparison. It is beyond the scope of this paper to investigate how the systematics of different gas metallicity indicators could affect the high-mass MZ$_{\rm g}$R of the SDSS galaxies. 

Nevertheless, we trust that the $R$ calibration is reliable for the low-mass galaxies in our KCWI sample, which all have $12+\log{(\rm O/H)} < 8.6$. The O-MZ$_{\rm g}$R of the KCWI sample also agrees with that of the local dwarf galaxies measured by \citet{Berg12} using the $T_e$ method. However, our gas-phase metallicities and those measured by \citet{Berg12} are lower than the MZ$_{\rm g}$R measured by \citet{Andrews13} for stacked SDSS spectra via the $T_e$ method. \citet{Andrews13} included a large sample of galaxies with galaxies at various sSFRs when they generated the stacked spectra. The emission from diffuse ionized gas (DIG) ionized by evolved, hot stars becomes increasingly important when the sSFR is low. As pointed out by \citet{Sanders17}, the contamination from DIG tends to affect the emission line fluxes from the low-ionization species and lower the electron temperature estimate in the low-ionization zone. Consequently, the gas metallicity derived from the stacked spectra with significant contribution from DIG would be overestimated. Although our sample also consists of star-forming galaxies with low sSFRs, we measured the gas metallicity for individual galaxies where DIG contribution should be much less significant. We therefore suspect that the contamination from DIG in the sample of \citet{Andrews13} may explain the discrepancy. 


\bibliography{ref}{}
\bibliographystyle{aasjournal}



\end{document}